\documentclass[aps,prx,showpacs,preprintnumbers,twocolumn,superscriptaddress, reprint, nofootinbib]{revtex4-2}
\usepackage{amsmath,amssymb}
\usepackage{bm}
\usepackage{tipa}
\usepackage{upgreek}
\usepackage{comment}
\usepackage{mathrsfs}
\usepackage{graphicx}
\usepackage{braket}
\usepackage{float}
\usepackage{enumitem}
\usepackage{adjustbox}
\usepackage{mathbbol}
\usepackage{booktabs}
\usepackage{gensymb}
\usepackage[normalem]{ulem}
\usepackage{xcolor}
\usepackage[colorlinks,bookmarks=true,citecolor=blue,linkcolor=red,urlcolor=blue]{hyperref}

\usepackage{pifont}
\newcommand{\cmark}{\ding{51}}
\newcommand{\xmark}{\ding{55}}

\newcommand{\nl}{{N_\ell}}
\newcommand{\bl}{{\boldsymbol{\beta}_l}}
\newcommand{\al}{{\boldsymbol{\alpha}_l}}

\newcommand{\bbeta}{{\boldsymbol{\beta}}}

\renewcommand{\comment}[1]{}
\usepackage{subfigure}
\usepackage{multirow}

\begin{document}

\title{
Three-dimensional Foliated Fractional Quantum Hall Phases
}

\author{Sahana Das}
\thanks{These authors contributed equally.}
\author{Navketan Batra}
\thanks{These authors contributed equally.}
\author{Andrea Kouta Dagnino}
\thanks{These authors contributed equally.}
\affiliation{Department of Physics, University of Z{\"u}rich, Winterthurerstrasse 190, 8057 Z{\"u}rich, Switzerland}

\author{Dan Mao}
\affiliation{Condensed Matter Theory Group, PSI Center for Scientific Computing,
Theory and Data, Paul Scherrer Institute, 5232 Villigen PSI, Switzerland}
\affiliation{Department of Physics, University of Z{\"u}rich, Winterthurerstrasse 190, 8057 Z{\"u}rich, Switzerland}

\author{Nicolas Regnault}
\affiliation{Center for Computational Quantum Physics, Flatiron Institute, 162 5th Avenue, New York, New York 10010, USA}
\affiliation{Department of Physics, Princeton University, Princeton, New Jersey 08544, USA}
\affiliation{Laboratoire de Physique de l’Ecole normale supérieure, ENS, Université PSL, CNRS, Sorbonne Université, Université Paris-Diderot, Sorbonne Paris Cité, 75005 Paris, France}

\author{Glenn Wagner}
\affiliation{Institute for Theoretical Physics, ETH Z{\"u}rich, 8093 Z{\"u}rich, Switzerland}
\author{Titus Neupert}
\affiliation{Department of Physics, University of Z{\"u}rich, Winterthurerstrasse 190, 8057 Z{\"u}rich, Switzerland}

\begin{abstract}
Foliated topological orders in three dimensions are layered systems in which anyons are free to move within a layer but cannot hop between them. A simple model with such a phase is a stack of decoupled two-dimensional electron gases in a strong magnetic field, each in the same fractional quantum Hall state. By focusing on the case of filling $\nu=1/3$ of the lowest Landau level in each layer, we show that (i) the limit of decoupled Laughlin states is stable upon introducing interlayer interactions and (ii) the system can enter a spontaneously layer-trimerized foliated non-Abelian Fibonacci phase. We support our claims by numerical exact diagonalization of up to 10 layers as well as perturbative analytical calculations. Specifically, we show that the foliated Fibonacci phase exists in the 9-layer system with pseudopotential interactions within and between neighboring layers. We identify the phase via quasihole counting and by calculating the overlap with a model wave function which we derive from the associated conformal field theory. Our numerical results suggest the possibility of realizing these phases in layered van der Waals crystals in strong magnetic fields, as well as in multilayer heterostructures. 
\end{abstract}

\maketitle
\tableofcontents

\section{Introduction}

\subsection{From fractional quantum Hall to 3D fractonic orders}

The fractional quantum Hall (FQH) effect is one of the most celebrated examples of topological order in condensed matter physics.  Since Laughlin's seminal wavefunction for the filling factor $\nu=1/3$ state \cite{Laughlin1983}, the theoretical landscape has expanded dramatically to encompass a rich hierarchy of Abelian  \cite{Haldane1983,Halperin1984statistics} and non-Abelian phases whose quasiparticles obey exotic braiding statistics \cite{Moore1991nonabelions,Nayak2008}. Among the latter, Fibonacci anyons \cite{Read1999paired,Rezayi2009non,Mong2017Fibonacci,Bonderson2012competing,Zhu2015fractional} occupy a privileged position since they support universal topological quantum computation \cite{Freedman2002a,Freedman2002b,Nayak2008,Minev2025}.

The push to more exotic two-dimensional (2D) topologically ordered phases is complemented by the exploration of topological order in three dimensions (3D)~\cite{Dennis2002, HammaZanardiWen2005, WalkerWang2012}. This pursuit is on the one hand guided by the progression of discoveries of noninteracting topological phases starting from the integer quantum Hall states, via 2D topological insulators to 3D topological insulators. Their topologically ordered counterparts have qualitatively new features in 3D where deconfined flux tubes have nontrivial statistics~\cite{WangLevin2014,JiangMesarosRan2014, WangQiGuCheng2019}. On the other hand, fracton phases~\cite{Chamon2005,Haah2011, BravyiHaah2013, Vijay2015, Vijay2016} are a more drastic 3D departure from 2D topological orders. Their ground state degeneracies are allowed to scale exponentially in the linear system size, yielding large protected computational spaces if used as topological quantum memories. Some fracton phases host mobile topological excitations whose motion is restricted to lines or planes -- subdimensional particles~\cite{Vijay2015,Pretko2017,Nandkishore2019, Pretko2020,Shirley2018}. So far, the exploration of fracton phases has largely focused on exotic exactly soluble models or field theories, with few proposals for how to realize them in materials~\cite{SlagleKim2017}. 

Curiously, relevant features of fracton topological order, namely subdimensional mobile excitations and exponentially scaling topological degeneracy, can already be realized in a setting where layers of ``conventional'' 2D topological orders are stacked. The result is foliated phases, which have emerged as a unifying organizing principle \cite{Shirley2020twisted,Shirley2020foliated,Shirley2019universal,SHIRLEY2019fractional}. Their defining feature is that anyons can only move within layers, even in the
presence of strong interlayer interactions.  
A natural but largely unexplored direction is to realize fracton topological order as 3D \emph{foliated fractional quantum Hall} (FFQH) phases, where each layer hosts a well-understood topological Hall liquid. 

There has been some theoretical work looking at the limit of infinitely many layers, including stacks of Laughlin \cite{Balents1996,Levin2009} and Jain \cite{banerjee2022bilayers} states, generalizations of Halperin states \cite{Qiu1989,Qiu1990,Naud2000,NAUD2001}, and studies of the surface states of the layered system \cite{Chalker1995,Betouras2000}. However, many of these studies do not consider the energetics of the resulting states, which is crucial for their realization. Numerical studies for large numbers of layers are therefore indispensable.

Here, we explore the potential of layered 3D systems to realize FFQH phases, combining analytically tractable limits, conformal field theory constructions of representative wave functions, and extensive numerical exact diagonalization (ED) of model systems. We deduce the principal conditions under which these phases can appear in layered materials. 

\subsection{Prior work on quantum Hall multilayers}

Our study builds on the rich literature which explores coupling of $N_\ell$ layers of FQH liquids as a route to engineering different types of topological order. The foundational single-layer states are the Laughlin states~\cite{Laughlin1983}, which minimize intralayer repulsion by attaching vortices to each electron. The simplest bilayer generalization is a decoupled stack of two such states, where each layer independently minimizes its intralayer repulsion with no interlayer correlations. Turning on interlayer interactions requires going beyond this picture: the Halperin~$(mmn)$ states~\cite{Halperin1983} provide the natural framework. They describe two-component quantum Hall systems in which correlations between layers can be encoded in the number $n$ of interlayer vortices attached to each electron position~\cite{Halperin1983}. Larger $n$ corresponds to stronger interlayer interactions, and the limiting case $n=0$ recovers the decoupled Laughlin states. 

The transitions between decoupled layers and Halperin states have been studied theoretically in great detail \cite{Wang2004,
ED1,ED3,ED0,Park1,Park2,papicThesis,Wagner2021,Simon1,Simon2,Simon3,Milovanovic,Hu2024,Hu2024,Sodemann,Halperin2020,ruegg2024dualities,Bonesteel,Pwave2,Cipri_thesis,CipriBonesteel,p_wave,Ruegg2023,lotric2024chernsimons,ShouCheng,Wagner2024,He1993,Milovanovic2010}. Beyond the Halperin states, which are Abelian, obtaining non-Abelian order in layered FQH systems has attracted sustained theoretical interest \cite{Papic2010,Peterson2010,Zhu2016,voinea2025criticalmajoranafermiontopological,Goldman2019,Goldman2021,Crepel2019,CAPPELLI2001499,Cabra2001,Liu2016,Nomura2004,Peterson2010b,Repellin2015}. It was shown by Vaezi and Barkeshli that a bilayer system of two $\nu = 1/3$ Laughlin states, coupled by single-particle interlayer tunneling, can undergo a phase transition into a non-Abelian Fibonacci phase with $\text{SU(3)}_2$ fusion rules of the excitations~\cite{Vaezi2014}. The Fibonacci anyon $\tau$, with quantum dimension $d_\tau = (1+\sqrt{5})/2$ and fusion rule $\tau \times \tau = \mathbf{1} + \tau$, emerges as a neutral interlayer excitation. It can be thought of as a superposition of an $e/3$ quasiparticle in one layer paired with a $-e/3$ quasiparticle in the other. The prediction of Fibonacci topological order in a $1/3+1/3$ quantum Hall bilayer was later confirmed numerically in an ED study \cite{Geraedts2015,Liu2015}, that also provides a comprehensive comparison of the competing topological orders at this filling \cite{Geraedts2015}. The Fibonacci state differs from the non-Abelian spin singlet (NASS) state~\cite{ardonne1999new}, another proposed non-Abelian state in FQH multilayers\footnote{The edge theory for the Fibonacci state consists of a $\mathbb{Z}_3$ parafermion and two $U(1)$ free chiral bosons \cite{PhysRevLett.113.236804}, which differs from the NASS state whose neutral edge modes consist of generalized Gepner parafermions. This results in different filling factors, shift quantum numbers, and topological ground state degeneracies for the two states. A more comprehensive explanation of the edge theory and model wavefunction construction for the Fibonacci state is provided in App.~\ref{app:modelWF}.}.

\begin{figure}[t]
    \centering
    \includegraphics[width=\columnwidth]{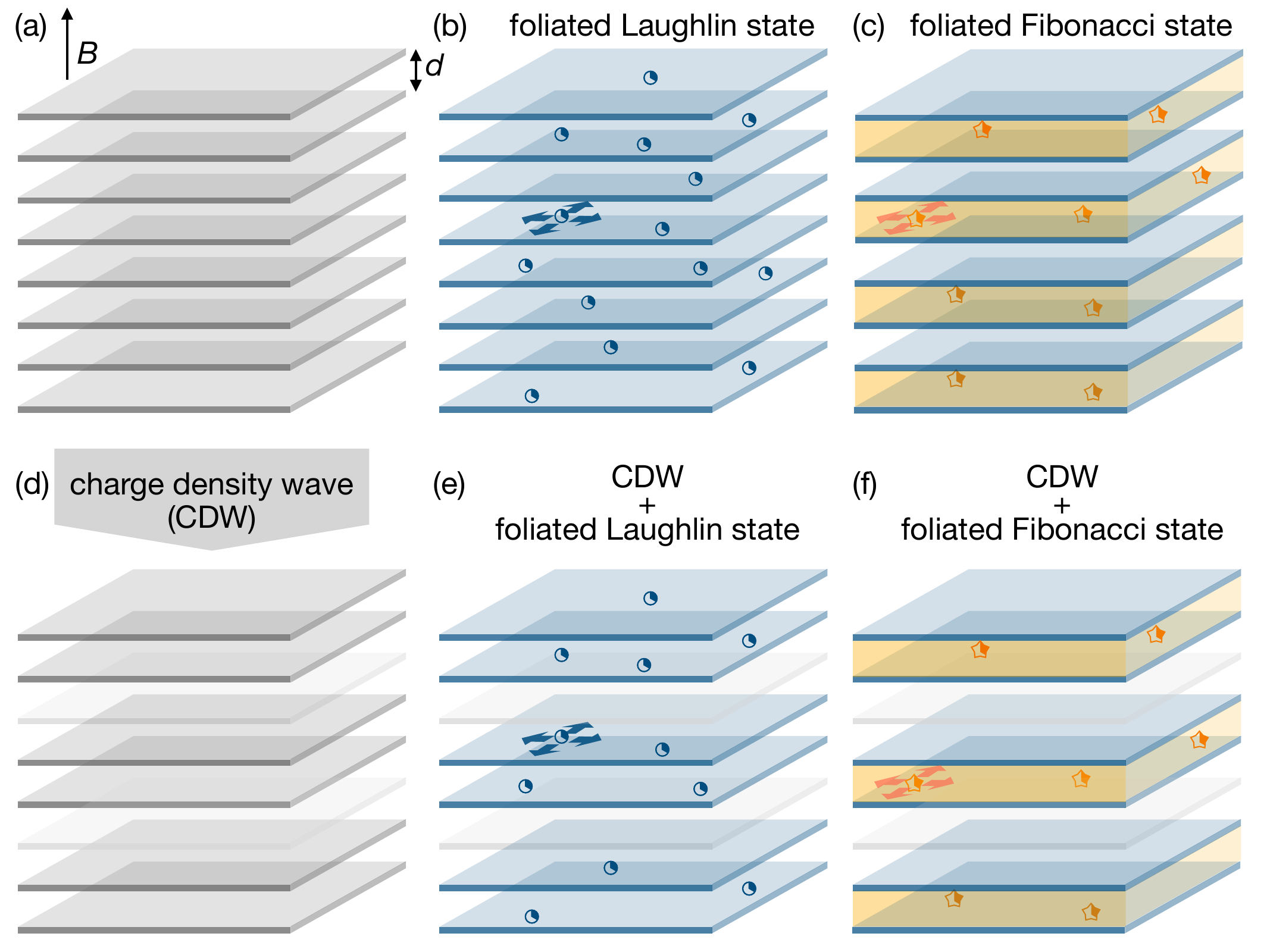}
    \caption{\textbf{Schematic of foliated fractional quantum Hall states.} (a) We consider a layered three-dimensional system in a strong magnetic field, with each layer at filling factor $\nu$ of the lowest Landau level. (b) In the absence of interlayer coupling, each layer forms a Laughlin state at appropriate filling. At $\nu=1/3$, we find that this foliated (fracton) topological order is stable up to sizable interlayer coupling. (c) Another candidate state at the same filling is a non-Abelian topological order with Fibonacci anyons, which could, however, not be energetically stabilized in our numerics. 
    (d) At $\nu=2/9$, a spontaneous charge density wave instability in the direction perpendicular to the layers triples the unit cell. For suitable interlayer interactions, in this phase we find
    (e) a foliated Laughlin state or (f) a foliated state with Fibonacci anyons. }
    \label{fig:schematic}
\end{figure}

There has been significantly less numerical work on quantum Hall multilayers for $N_\ell>2$. $N_\ell=3$ has been studied using Hartree-Fock~\cite{Hanna1996} and density matrix renormalization group (DMRG)~\cite{wang2025anyonsuperfluidtrilayerquantum}, and $N_\ell=4$ has been studied using ED due to its relevance in graphene, where the four components correspond to the four spin and valley degrees of freedom \cite{Wu2015,Toke2007,Papic2010b,Goerbig_2007,2015,DeGail2008}. To our knowledge, no numerical studies address the case of $N_\ell>4$ layers.

On the experimental side, quantum Hall multilayers have been studied since the earliest days of the FQH effect. Bilayer systems in GaAs quantum wells have revealed a wealth of correlated phases driven by interlayer interactions and tunneling, including interlayer coherent states at total filling $\nu = 1$ \cite{Experiment_Review,Eisenstein2004}. More recently, experiments on double-layer graphene systems separated by hexagonal boron nitride have opened new avenues for studying interlayer FQH physics \cite{liu2020crossover,Zhang:2025aa,nguyen2024bilayer,Liu:2019aa,Li2017}. The rapid development of van der Waals heterostructures has allowed for the experimental realization of fractional Chern insulator phases at zero magnetic field~\cite{Neupert2011fractional,Regnault2011fractional,Sheng2011fractional,Park2023}, which can potentially be stacked to form multilayer structures. These advances make the prospect of engineering quantum Hall multilayers with $N_\ell \gg 2$ increasingly realistic. A theoretical understanding of what phases can emerge in such systems, and what microscopic conditions are required to stabilize them, is therefore both timely and necessary.

\subsection{Summary of the key results}

We begin with the building blocks of the FFQH phases: few-layer FQH systems in which increasing interlayer interactions drives a transition from decoupled Laughlin states to non-Abelian topological orders. We show that stacking $\nu=1/3$ Laughlin states yields a topological order with $\text{SU(3)}_{N_\ell}$ fusion rules, which we confirm numerically for $N_\ell=2,3,4$. We further construct explicit model wave functions for these states.

Extending these few-layer results to 3D FFQH states, we establish two central findings. First, decoupled Laughlin layers remain stable against interlayer interactions up to strengths comparable to the intralayer interactions. This makes FFQH phases in as-grown 3D crystals possible in principle. Second, spontaneous breaking of translational symmetry along the layer direction can stabilize a Fibonacci FFQH phase over an extended region of interlayer coupling. 

\subsection{Outline of the paper}

The structure and content of this article are as follows: In Sec.~\ref{sec:model} we introduce the model Hamiltonian and  numerical methods. We first apply the methods to the quantum Hall bilayer at filling $\frac13+\frac13$ in Sec.~\ref{sec:bilayer}. While decoupled Laughlin states are trivially the ground state in the limit where the two layers are not interacting, we find that the phase survives up to a surprisingly large value of the interlayer interaction, a point we will return to later. For a suitable tuning of the interactions, previous work discovered a non-Abelian Fibonacci state competing with the decoupled Laughlin state. We construct a model wavefunction for the Fibonacci state using the $\mathbb{Z}_3$ parafermion conformal field theory (CFT)~\cite{zamolodchikov1985nonlocal} and the closely related Read-Rezayi states~\cite{Read1999paired}. Our CFT construction also allows us to uniquely identify the spin topological quantum field theory (TQFT) underlying the states through a series of simple current extensions. We find it to be $\mathcal{C}_{\nl=2}\equiv\mathcal{F}_{-4}\boxtimes \overline{\text{SU(3)}_2}$ where $\mathcal{F}_m$ denotes the invertible fermionic phase with topological central charge $c_\text{top}=\frac{m}{2} \mod 8$ \cite{lan2016theory} and $\overline{\text{SU(3)}_2}$ denotes the time-reversal conjugate of the $\text{SU(3)}_2$ TQFT. Using the model wavefunction, we then compute overlaps with the ED ground state in the spherical geometry as a proxy for the Fibonacci topological order. Our phase diagram based on the overlap calculations agrees with the previous work which used other diagnostics to identify the Fibonacci state~\cite{Liu2015}. Furthermore, we study the phase transition between the decoupled Laughlin and the Fibonacci state analytically in the thin-torus limit \cite{PhysRevB.28.1142}. We show that the interlayer pseudopotential perturbatively generates a correlated interlayer hopping, and has the same effect as interlayer tunneling, studied in Ref.~\onlinecite{Vaezi2014}. The resulting effective model can be mapped to a transverse field Ising model whose Ising quantum phase transition separates the decoupled Laughlin and Fibonacci phases. 

The bilayer non-Abelian topological order can be generalized to higher numbers of layers. In Sec.~\ref{sec:non-abelian}, we use quasihole counting --- counting the degeneracy of the ground state manifold when a number of quasiparticles is inserted --- to identify non-Abelian topological order in three- and four-layer systems. To establish the quasihole counting, we explicitly derive the generalized exclusion principles for 3 and 4 layer topological order with SU(3)$_3$ and SU(3)$_4$ fusion rules from the thin-torus Hamiltonian. 
We find generally good agreement between the derived quasihole counting in the $N_\ell=3$ and $N_\ell=4$ cases and the numerically observed counting (up to finite-size effects). This generalizes the $\mathcal{C}_{\nl=2}$ topological order of the Fibonacci state to $N_\ell>2$ topological order with SU(3)$_\nl$ fusion rules. The identification of the 4-layered non-Abelian topological order is interesting insofar as we only use interactions between nearest-neighbor layers (NN), yet the resulting topological order is an intrinsically four-layer order. We also derive the wavefunctions of the $N_\ell>2$ topological order with SU(3)$_\nl$ fusion rules using Gepner parafermion CFT \cite{gepner1987new}. All states studied in this work are fermionic, so they must be described by a fermionic/spin-TQFT, rather than a normal (bosonic) TQFT like a Chern-Simons theory. We find that the relevant spin-TQFT for states at filling $\nu=\frac{\nl}{3}$ is $\mathcal{C}_\nl\equiv\mathcal{F}_{6\nl}\boxtimes \overline{\text{SU(3)}_{\nl}}$.
We note that there are other examples of states with $\text{SU}(k)_r$ fusion rules, the most apparent are the Jack polynomials whose roots obey $(k,r)$ exclusion rules \cite{bernevig2008properties,ardonne2009domain,PhysRevLett.100.246802,PhysRevB.77.184502} and thus correspond to the integrable weights of $\text{SU}(k)_r$. However, these states are thought to be gapped only when $k=r+1$ (where they reduce to the $\mathbb{Z}_k$ Read-Rezayi states). Another example, as mentioned previously, are the NASS states \cite{ardonne1999new, ardonne2001non}, which also feature $\text{SU(3)}_k$ anyonic fusion rules (but have a different underlying topological order to our states).

In Sec.~\ref{sec:N-layers}, we consider the 3D limit of infinite stacks of 2D layers. The setup is a layered 3D electronic system in a strong magnetic field perpendicular to the layers (Fig.~\ref{fig:schematic}a). We show that as in the bilayer case, the decoupled Laughlin FFQH state (Fig.~\ref{fig:schematic}b) is stable up to large interlayer coupling. With next-nearest-neighbor (NNN) interlayer interactions included, electrostatic effects can cause translational symmetry breaking where pairs of layers are occupied, separated by an empty layer (Fig.~\ref{fig:schematic}d). The pairs of filled layers can either be in the decoupled Laughlin state (Fig.~\ref{fig:schematic}e), offering another realization of an Abelian FFQH state. However, more interestingly, the pairs of filled layers can together form a bilayer Fibonacci state (Fig.~\ref{fig:schematic}f). These Fibonacci states are separated by an empty layer and therefore are only weakly coupled. This is an example of a non-Abelian FFQH state. From the thin-torus Hamiltonian, we identify that the quantum energy gain that is unique to the pair-foliated charge configuration stabilizes the Fibonacci FFQH state over the competing phases. A Fibonacci FFQH state without empty layers separating the Fibonacci bilayers (Fig.~\ref{fig:schematic}c) is in principle also possible and we develop a Ginzburg-Landau theory for this state. 
However, the energetic stabilization of this state with a local Hamiltonian is an open question that we defer to future work. We close with a discussion of possible experimental platforms where FFQH states may be realized in Sec.~\ref{sec:experiments}.

\section{Model and methods}
\label{sec:model}

In this section we introduce the Hamiltonian which we will study in different geometries and layer configurations in the remainder of the paper. The Hamiltonian is defined in terms of so-called Haldane pseudopotentials which we will review for convenience below. The Haldane pseudopotentials are written in a basis-independent form, but subsequently we will discuss how to implement this Hamiltonian in both the spherical geometry (employed for the ED) and in the torus geometry (mostly as an analytical tool in the thin-torus limit).

\subsection{Haldane pseudopotentials}

We consider $N_\mathrm{e}$ spinless (or spin polarized) electrons in $N_\ell$ layers indexed by $l=1,2,\dots N_\ell$. The electrons are subjected to $N_\phi$ flux quanta. It is often useful to parametrize the interactions in quantum Hall systems in terms of certain short-range interaction potentials known as Haldane pseudopotentials $V_L^{(a)}$ with $L$ and $a$ integers~\cite{Haldane1983}. Here, $V_L^{(l-l')}$ describes the interaction energy of a pair of electrons with relative angular momentum $L$ and in layers $l$ and $l'$. The pseudopotentials satisfy layer exchange symmetry $V_L^{(l-l')}=V_L^{(l'-l)}$ and periodic boundary conditions, i.e.~$l-l'$ is defined modulo $N_\ell$.  In terms of the Haldane pseudopotentials, the Hamiltonian is 
\begin{equation}
H
= \frac{1}{2}
\sum_{l,l'=1}^{N_\ell}\ \sum_{i,j=1}^{N_\mathrm{e}}
\ \sum_{L=0}^{N_\phi}
V_L^{(l-l')}
\mathcal{P}^{(L)}_{(i,l),(j,l')},
\label{eq:Ham_general}
\end{equation}
where $\mathcal{P}^{(L)}_{(i,l),(j,l')}$ is the projector onto the two-particle subspace where particle $i$ in $l$-th layer and $j$ in $l'$-th layer have relative angular momentum $L$.

Fermionic antisymmetry dictates that the relative angular momentum $L$ between particles in the same layer must be odd. For the long-range Coulomb interaction in the lowest Landau level, the intralayer pseudopotentials are monotonically decreasing with $L$, i.e.~$V^{(0)}_1>V^{(0)}_3>\dots$ For particles in different layers, the angular momentum can be even or odd and $V^{(a)}_0>V^{(a)}_1>\dots$. Since the interaction falls off with the distance between layers, we furthermore have $V_L^{(1)}>V_L^{(2)}>\dots$ as the typical physical situation. All phases we find in this work appear in a reasonable portion of parameter space in that they can be stabilized with interaction coefficients that satisfy this set of inequalities.

To reduce the dimensionality of the parameter space that we scan to establish phase diagrams, in the following we will truncate the pseudopotentials to include only the largest terms in Eq.~\eqref{eq:Ham_general}. We will mostly be considering the largest terms of the intralayer, NN layer and NNN layer interaction, which are $V_1^{(0)}$, $V_0^{(1)}$, $V_1^{(1)}$, and $V_0^{(2)}$, respectively, and set all other pseudopotential coefficients to 0. In the rest of the paper, we will work in units where $V_1^{(0)}=1$.

\subsection{Exact diagonalization in the spherical geometry}

As is standard in quantum Hall studies, we perform ED in the spherical geometry with respect to the in-plane coordinates of the layer~\cite{Haldane1983}. We consider a sphere whose surface is penetrated by a radial magnetic field $B$ and which has radius $R=\sqrt{q}l_B$, where $l_B=\sqrt{\hbar /eB}$ is the magnetic length. Here, $q$ can be integer or half-integer. The surface of the sphere is thus pierced by $N_\phi=2q$ flux quanta and in the lowest Landau level, there will be $2q+1$ single-particle eigenstates labelled by an angular momentum $L_z$ eigenvalue $m=-q,\dots,+q$. In our ED we will always assume that all electrons are in the lowest Landau level. The number of electrons $N_\mathrm{e}$ for filling factor $\nu$ is 
\begin{equation}
    N_\phi=\frac{N_\mathrm{e}}{\nu}-\mathcal{S},
    \label{eq:shift}
\end{equation}
where the shift $\mathcal{S}$ is a topological invariant which characterizes a given topological order on curved Riemannian manifolds \cite{WenZee1992}.

On the sphere different phases in general have different shifts $\mathcal{S}$. In order to compare different potential phases, we compute the gap for different shifts $\mathcal{S}$. The state with the largest gap is the most competitive gapped state in the thermodynamic limit, where we furthermore impose that the ground state be in the $L=0$ sector as appropriate for a translationally invariant quantum Hall phase. We support this with data on the torus geometry, where the shift does not exist and hence all competing states appear in the same particle number sector. There, the ground state degeneracy can instead be used to distinguish different phases. We show that the phase diagram in the spherical geometry agrees with the toroidal geometry for the bilayer case.

While the pseudopotentials can be defined in any geometry, their construction is particularly transparent on the sphere. Due to the rotational invariance of the sphere, the angular momentum $L$ is a good quantum number (with $L_z$ labelling states within each angular momentum multiplet). Any rotationally invariant potential on the sphere can be decomposed into Haldane pseudopotentials. The two-body matrix elements for Haldane pseudopotentials $V_{2q-L}^{(a)}$ are 
\begin{align}
&U_{m_1,m_2,m_3,m_4}^{(a)}
= \\&\qquad \sum_{L=0}^{2q} V_{2q-L}^{(a)} \sum_{M=-L}^{L}
\langle q m_1, q m_2 | L M \rangle
\langle L M | q m_3, q m_4 \rangle,\nonumber
\end{align}
where $\langle q m_1, q m_2 | L M \rangle$ are the Clebsch-Gordan coefficients. In general we will allow for different matrix elements for the inter- and intralayer interactions, i.e.~we do not impose $\text{SU}(N_\ell)$ symmetry. 
In terms of the inter- and intralayer matrix elements, the many-body Hamiltonian corresponding to Eq.~\eqref{eq:Ham_general} is 
\begin{align}
    H=&\frac12\sum_{\{m_i\},l,l'}U^{(l-l')}_{m_1,m_2,m_3,m_4}c^\dagger_{m_1,l}c^\dagger_{m_2,l'}c_{m_4,l'}^{\phantom\dagger}c_{m_3,l}^{\phantom\dagger},
    \label{eq:Ham_sphere}
\end{align}
where $c_{m,l}^\dagger$ is the creation operator for an electron in layer $l$ in the lowest Landau level with $L_z$ eigenvalue $m\in\{-q,\dots,+q\}$. 

The many-body Hamiltonian Eq.~\eqref{eq:Ham_sphere} can be projected onto the many-body Fock basis in a fixed angular momentum $L_z$ sector with a fixed number of particles in each layer and the resulting matrix can be diagonalized using the Lanczos algorithm to find the lowest eigenstates.

\subsection{Thin-torus limit}
\label{sec:thin-torus-method}

The thin-torus limit provides a complementary analytical tool that we employ at several points in this work~\cite{PhysRevB.77.155308,PhysRevB.50.17199,PhysRevB.28.1142,PhysRevLett.97.056804,PhysRevLett.101.036804,PhysRevLett.95.266405,papic2014solvable}. First, in Sec.~\ref{sec:bilayerthintorus}, we derive an effective Hamiltonian for the bilayer system and motivate how interlayer interactions generate correlated hopping terms responsible for transitions between different FQH states. Next, in Sec.~\ref{sec:thintorus_multilayer}, we analyze the energetics of competing charge configurations in a system with many layers. We briefly review the construction here and comment on its regime of validity. 

We place the system on a torus of size $L_x\times L_y$ and take $L_y\ll l_B$ while keeping the total area $L_xL_y = 2\pi l_B^2 N_{\phi}$ fixed. In the Landau gauge, the lowest Landau level single-particle orbitals are Gaussian strips of width $\sim l_B$ centered at positions $x_m=m\Delta x$, where $\Delta x= 2\pi l_B^2/L_y$ is the inter-orbital spacing and $m\in \mathbb{Z}$. As $L_y\rightarrow 0$, neighboring orbitals become exponentially decoupled, with the overlap controlled by the dimensionless parameters,
\begin{align}
    \epsilon = \exp\left( -\frac{\Delta x^2}{2l_B^2} \right)= \exp\left( -\frac{\kappa^2}{2} \right), ~~~~~\kappa\equiv \frac{2\pi l_B}{L_y}.
    \label{eq:thin-torus-para}
\end{align}
Any rotationally invariant interaction projected onto the lowest Landau level can then be expanded in powers of $\epsilon$, with successive powers corresponding to processes that transfer electrons across increasingly many orbitals. In this limit the two-dimensional quantum Hall problem maps onto a one-dimensional lattice model of interacting electrons whose ground states are charge density wave (CDW) configurations. The topological structure of the quantum Hall phase is encoded in the CDW root patterns and their degeneracies, while phase transitions manifest as level crossings in the effective one-dimensional model. 

An important advantage of the thin-torus approach is that it often correctly reproduces ground-state degeneracies and exclusion statistics of the gapped quantum Hall phases, providing an intuitive picture for the competing orders in a given system. However, as a caveat, the thin-torus limit does not faithfully capture \textit{all} topological phases.
The $S_3$ wavefunctions of Ref.~\onlinecite{simon2010s3} provide an explicit example that cannot be fully characterized by their thin-torus limit or pattern of zeros, demonstrating that distinct topological orders can share the same thin-torus root configurations. Moreover, the Gaffnian state is gapped in the thin-torus limit yet expected to be gapless in the two-dimensional bulk \cite{weerasinghe2014thin,Estienne2015correlation,Jolicoeur2014absence}, showing that the thin-torus analysis can yield qualitatively incorrect predictions for the excitation spectrum. At even-denominator fillings, the adiabatic connection between the thin-torus CDW and the bulk can also fail. For example, at $\nu=1/2$, a phase transition separates the thin-torus ground state from the gapless composite Fermi liquid \cite{bergholtz2005half,bergholtz2008quantum}.

These examples illustrate that the thin-torus limit, despite its many successes, has a limited domain of validity that warrants care in interpreting its predictions. With this in mind, we use the thin-torus analysis primarily as a guide to identify candidate phases and to build physical intuition for the mechanisms driving the phase transitions and foliation, and confirm all identifications using ED calculations on the sphere and torus at finite $L_y$.

\section{Bilayer quantum Hall states with $\nu = 1/3$ per layer}
\label{sec:bilayer}

Our discussion will now focus on the case where two layers are each at $\nu=1/3$ filling of the lowest Landau level (LLL). In the trivial decoupled limit, the two layers form decoupled Laughlin states. Once the layers are coupled (either by tunnelling or interlayer interactions), a Halperin~(112) state can arise~\cite{Haldane1996}. Although this state itself is unstable and describes phase separation~\cite{DeGail2008}, the state can be modified to obtain a stable version as shown in Refs.~\onlinecite{Haldane1996,Peterson2015}. Furthermore, tunneling can drive a transition to the Jain 2/3 state \cite{Geraedts2015,Liu2015,Peterson2015}. Most interestingly, for a short-range pseudopotential interaction an ED study \cite{Liu2015} has reported signatures of the bilayer Fibonacci state originally proposed in Ref.~\onlinecite{Vaezi2014}. For long-range Coulomb interactions, an interlayer Pfaffian or a $\mathbb{Z}_4$ parafermion state have also been proposed \cite{Barkeshli2010nonabelian,Geraedts2015,Peterson2015}. In our work we will be exclusively considering short-range pseudopotential interactions and hence we do not find the latter two phases. Furthermore, in the rest of this work we will study a model without interlayer tunneling.

\subsection{Abelian phases}

The Halperin~$(mmn)$ states were originally introduced in order to describe two-component systems, where the two components could correspond to a layer degree of freedom in a quantum Hall bilayer, or they could equally well describe a spin or valley degree of freedom \cite{Halperin1983}. The wavefunction on the disk is 
\begin{align}\label{eq:halperin_mmn}
\Psi_{mmn}&(\{z_i\},\{w_i\})\\=&\prod_{i>j}(z_i-z_j)^m\prod_{i> j}(w_i-w_j)^m\prod_{ij}(z_i-w_j)^n,\nonumber
\end{align}
where $z_i$ and $w_i$ are complex electron coordinates in the top and bottom layer respectively. The filling factors of the two layers are then $\nu_1=\nu_2=\frac{1}{m+n}$. The shift on the sphere is $\mathcal{S}=m$ and the ground state degeneracy on the torus is $|m^2-n^2|$, where $m$ must be an odd integer by fermionic antisymmetry.  For the filling factor $\nu_1=\nu_2=1/3$ this gives two solutions. The Halperin~(330) state corresponds to decoupled Laughlin states in the two layers, while the (112) state introduces correlations between the two layers (see App.~\ref{app:abelianstates} for a more detailed discussion of these states and their generalization).

However we should point out that this is a Halperin~$(mmn)$ state with $n>m$, meaning that it has the property that interlayer repulsion is stronger than intralayer repulsion and so this wavefunction describes a phase separated state \cite{DeGail2008}. We work in the particle number sector where the two layers are equally populated and interlayer tunnelling is absent. The phase separation is an indication that it is energetically favourable to empty one layer and layer-polarize the system, in which case the system avoids the energy penalty from the large $V_0^\mathrm{(1)}$. This tendency was first noted in Ref.~\onlinecite{wu1993mixed}. 

We will revisit this point in Sec.~\ref{sec:translation_symmetry_breaking}. 
To fix the wavefunction one can write
\begin{align}\label{eq:singlet112}
\Psi_{\text{singlet}}&=\mathcal{P}_{\mathrm{LLL}}\prod_{i<j}|z_i-z_j|^2|w_i-w_j|^2\,
\Psi_{112}^{*},\\
&=\mathcal{P}_{\mathrm{LLL}}\prod_{i<j}(z_i-z_j)(w_i-w_j)\,
\Psi_{222}^{*}.
\end{align}
As was shown in Ref.~\onlinecite{Peterson2015}, a very good approximation for this state can be obtained on finite-size spheres by taking the ground state of the Hamiltonian with SU(2) symmetric Coulomb interactions in the $\mathcal{S}=1$ shift sector, which is a $L=S=0$ singlet. The analogous statement on the torus was shown to hold in Ref.~\onlinecite{Geraedts2015}. 

Finally, another candidate Abelian state at total layer filling 2/3 is the particle-hole transform of the Laughlin 1/3 state, for short, PH(1/3), in which all electrons would polarize to populate one layer only (a discussion of the full set of competing states at total filling $\nu=2/3$ is given in App.~\ref{app:competingstates_2/3}). The wavefunction is 
\begin{equation}\label{eq:PH1/3}
    \Psi_{\text{PH}(1/3)}(\{z_i\})=\mathcal{P}_\text{LLL}\prod_{i>j}(z_i-z_j)^2\Phi_{\nu=-2}(\{z_i\}),
\end{equation}
where $\Phi_{\nu=-2}$ describes the wavefunction of the composite fermions which fill two Lambda-levels \cite{Jain1989composite}. The lowest Landau level projection $\mathcal{P}_\text{LLL}$ can be implemented by using Jain-Kamilla orbitals for the composite fermions \cite{Jain1997,Moeller2005,Davenport2012}. 

The $K$ matrices of the Halperin~(112) state and of the layer-polarized PH(1/3) are related by an SL$(2,\mathbb{Z})$ transformation (see App.~\ref{app:abelianstates} for the explicit $K$-matrix construction of all Abelian states discussed in this work). However, the two states have a different shift and hence describe different phases on the sphere \cite{Geraedts2015}. Again, a good approximation for this state on finite spheres can be obtained by taking the ground state of the Hamiltonian with SU(2) symmetric Coulomb interactions in the fully polarized $S_z$ sector and in the $\mathcal{S}=0$ shift sector. This state is a SU(2) ferromagnet with $S=\frac{N_\mathrm{e}}{2}$, and can thus be rotated to arbitrary $S_z$ by appropriately applying $S^-$ pseudospin-lowering operators. In the bilayer system considered here, the two layers correspond to pseudospin up and pseudospin down.

\begin{table}
\centering
\begin{tabular*}{\columnwidth}{@{\extracolsep{\fill}}ccccc}
\hline\hline
FQH phase & Wavefunction & Abelian  &GSD & $\mathcal{S}$  \\
& &  & (torus) & (sphere) \\
\hline
Decoupled Laughlin & $\Psi_{330}$ & \cmark &  9 & 3  \\ 
Halperin~(112) & $\Psi_{112}/\Psi_\mathrm{singlet}$ & \cmark & 3 &  1\\ 
Fibonacci &  $\Psi_\text{Fib}$  & \xmark & 6 & $3$ \\
Particle-hole Laughlin & $\Psi_\text{PH(1/3)}$ &\cmark &3&0\\
\hline\hline
\end{tabular*}
\caption{\textbf{Different possible ground states for bilayer system.} Ground-state degeneracy (GSD) on the torus and shift $\mathcal{S}$ on the sphere for various candidate
fractional quantum Hall phases at $\nu = \frac13$ per layer in a bilayer system. 
}
\label{tab:FQHphases1}
\end{table}
\begin{figure}[b]
    \centering
    \includegraphics[width=\columnwidth]{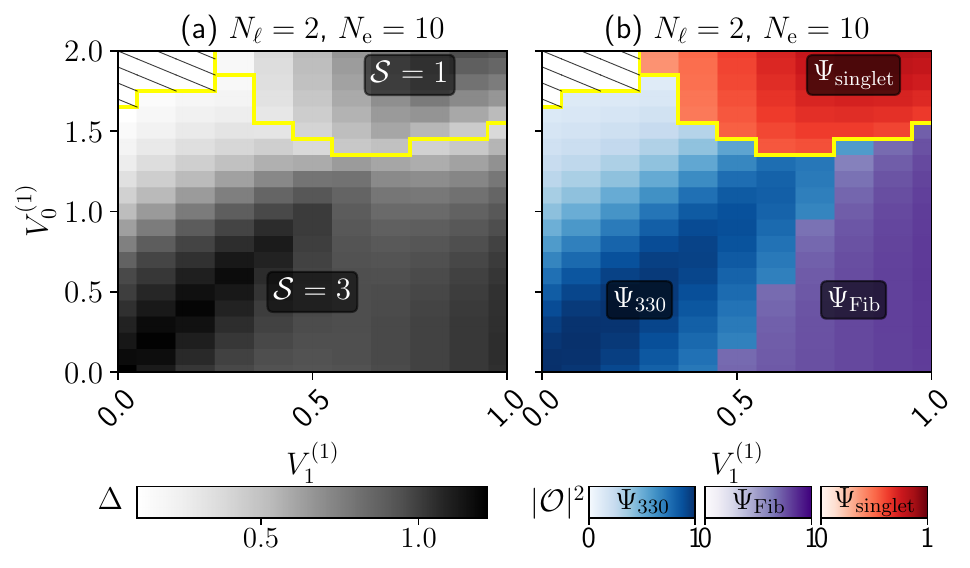}
    \caption{\textbf{Bilayer phase diagram on the sphere.} Shown are (a) the energy gap $\Delta$ above the ground state, and (b) the squared overlap $|\mathcal{O}|^2$ with model wave functions as explained in the text. For the energy gap, we take the highest gap among the shift sectors which have an $L=0$ ground state at each point in the phase diagram, and the corresponding shift sector is chosen as the ground state shift sector. The $L=0$ Hilbert space dimension is 345 for $N_\phi=12$ ($\mathcal{S}=3$ shift sector), 1087 for $N_\phi = 14$ ($\mathcal{S}=1$ shift sector), and 1794 for $N_\phi = 15$ ($\mathcal{S}=0$ shift sector). The borders between phases with different ground state shifts are highlighted in yellow. The hashed region has its ground state in the $\mathcal{S}=1$ sector, but it does not have a $L=0$ ground state consistently for different system sizes, so the nature of this phase remains elusive. 
    }
    \label{fig:bilayer}
\end{figure}

\subsection{Fibonacci state}
The bilayer Fibonacci state is a non-Abelian state 
with
$\mathcal{C}_2=\mathcal{F}_{-4}\boxtimes \overline{\text{SU(3)}_2}$ topological order (in the notation of Ref.~\onlinecite{lan2016theory}). It features a sixfold ground-state degeneracy on the torus~\cite{Vaezi2014}, which correspond to the six topologically
distinct quasiparticle species. Three of these, denoted $a_{\alpha}$ ($\alpha=0,1,2$),
carry charge $q=2\alpha e/3$ and are inherited directly from the $(330)$ state.
They are Abelian (quantum dimension $d_{a_{\alpha}}=1$) and obey the fusion rules
\begin{equation}
  a_{\alpha} \times a_{\beta} = a_{\alpha+\beta\ \mathrm{mod}\ 3}.
\end{equation}
The remaining three species are $\tau$ and $a_{\alpha}\tau$ ($\alpha=1,2$).  The neutral
excitation $\tau$ can be understood as a superposition of an $e/3$
quasiparticle in the top layer paired with a $-e/3$ quasiparticle in the
bottom layer, and vice versa.  Its quantum dimension is $d_\tau = \varphi \equiv (1+\sqrt{5})/2$ (the golden
ratio), and it obeys the fusion rule
\begin{equation}
  \tau \times \tau = \mathbf{1} + \tau,
\end{equation}
which is precisely the defining relation of the Fibonacci anyon. The quantum dimension $d_{\tau}>1$ is the hallmark of non-Abelian statistics, distinguishing $\tau$ from the Abelian sectors \cite{Freedman2002a,Freedman2002b,Nayak2008}.

There are several probes one can use to track the transition between the decoupled Laughlin phase and the Fibonacci phase, depending on the geometry of the system at hand. On the torus geometry, the two states can easily be distinguished due to their distinct ground state degeneracies. On the sphere, however, the ground state is singly-degenerate for both states. This issue is exacerbated by the fact that the two states also have the same shift [see Eq.~\eqref{eq:shift} and Tab.~\ref{tab:FQHphases1}]. An alternative probe would be to look at the quasihole excitations by introducing additional flux quanta in the system, a method which we will return to in Sec.~\ref{sec:qh_counting}. Yet another alternative probe would be to compute overlaps with a model wavefunction which is constructed to express the important aspects of the topological order of the phase. Following this logic, we have constructed a model wavefunction on the disk and sphere geometries using the underlying $\text{U(6)}_1/\text{SU(3)}_2$ conformal field theory which describes the edge excitations of this state (and through the bulk-boundary correspondence, the bulk quasihole excitations too). The wavefunction can be expressed as a function of the complex particle coordinates $\{z_i\}$ in the bottom and $\{w_i\}$ in the top layer, respectively (in the spherical geometry, these coordinates are stereographic projections of the coordinates on the sphere, as explained in App.~\ref{app:modelWF}):
\begin{align}
    \Psi_\mathrm{Fib}&(\{z_i\},\{w_i\})\label{eq:Fib_WF}\nonumber\\ = &\lim_{w_i^\pm \rightarrow w_i} \bigg[  \frac{\Psi_{k=3}^{\mathrm{RR}}({z_i},w_i^+,w_i^-)}{J_{\frac{2}{3}}(z_i,w_i^+,w_i^-)}
    \prod_i (w^+_i-w^-_i)^{\frac{2}{3}}\bigg]\nonumber\\
    & \times \prod_{i<j}(z_i-z_j)^{5/3}(w_i-w_j)^{5/3} \prod_{ij}(z_i-w_j)^{4/3},
\end{align}
where $\Psi_{k}^{\mathrm{RR}}$ is the $\mathbb{Z}_k$ Read-Rezayi wavefunction \cite{Read1999paired} and $J_\alpha$ is the Jastrow factor with exponent $\alpha$. In the above construction, we used a point-splitting procedure, where each $w_i$ coordinate is split into two coordinates $w_i^+$ and $w_i^-$, and the limit as $w_i^+\rightarrow w_i^-$ is taken. This wavefunction has a high overlap ($\sim 0.67$ for $N_\mathrm{e}=7+7$ particles where the $L=0$ Hilbert space has total dimension 101266) with the ground state in the bilayer Fibonacci phase obtained from ED. Details on the derivation of this wavefunction and its generalizations are provided in App.~\ref{app:modelWF}, and detailed overlap results including alternative model wavefunctions are provided in App.~\ref{app:overlaps}.

\subsection{Numerical results}
In Figs.~\ref{fig:bilayer} and \ref{fig:bilayer_torus}, we show the bilayer phase diagrams for $N_\mathrm{e}=10$ on the sphere and $N_\mathrm{e}=8$ on the torus, respectively. For the intralayer interaction we fix a single non-zero pseudopotential $V_1^{(0)}=1$. This guarantees that in the absence of interlayer interactions ($V_0^{(1)}=V_1^{(1)}=0$), the Laughlin state is the exact zero energy ground state (in the sector with shift $\mathcal{S}=3$ for the sphere). We then tune the interlayer pseudopotentials $V_0^{(1)}$ and $V_1^{(1)}$ to stabilize different phases. 

As shown in Tab.~\ref{tab:FQHphases1}, the decoupled Laughlin state and Fibonacci state have the same shift, while the Halperin~(112) state has a different shift. So in Fig.~\ref{fig:bilayer}(a) we first plot a phase diagram broken down by the shift of the ground state. Different shifts correspond to different system sizes and as such the ground state energies cannot be directly compared, especially in light of significant finite size effects. Instead, we compute the gap in all shift sectors and use the shift with the largest gap as a proxy for the ground state shift, as was done in Ref.~\onlinecite{Peterson2015}. In addition, we require that the ground state be an $L=0$ state, as expected from an FQH state. For each $V_0^{(1)}$ and $V_1^{(1)}$, we therefore first choose the shift based on which sectors have an $L=0$ state; if there are multiple available shift sectors we choose the one with the largest gap. In Fig.~\ref{fig:bilayer}(a) we show the thus established phase diagram. We find a large region with a ground state in the $\mathcal{S}=3$ shift sector, consistent with a decoupled Laughlin or Fibonacci state. For sufficiently large inter-layer interactions, we further find a phase in the $\mathcal{S}=1$ region, consistent with a Halperin~(112) phase. The hashed region in the top left of the phase diagram does not consistently have an $L=0$ ground state in any of the shift sectors we looked at. We note that for this system size, there is a small region with a $\mathcal{S}=1$ ground state in the $L=0$ sector (see App.~\ref{app:multilayer}), but it has a very small layer/pseudospin polarization $S_z=1$ which is likely a finite-size effect. Indeed, for $N_\mathrm{e}=8$ and $N_\mathrm{e}=12$ particles we find an $L\neq 0$ ground state with $S_z=0$ (thus likely making the ground state across the entire phase space we scanned pseudospin-unpolarized). There could also be effects due to aliasing \cite{dAmbrumenilMorf1989}. Indeed, for finite spheres, a fixed $N_\mathrm{e}, N_\phi$ could realize different FQH phases with different $\nu, \mathcal{S}$. In our case, finite-size effects could be biasing our results to show excitations of a state with $\nu\neq \frac{2}{3}$ and $\mathcal{S} \neq 0$. We are thus not able to conclusively identify a phase with its ground state in the $\mathcal{S}=0$ shift sector. Besides going to larger system sizes, one could also investigate the effect of using long-range interactions rather than short-range Haldane pseudopotentials -- we leave this to future work. In App.~\ref{app:multilayer}, we present more comprehensive phase diagrams and show that the ground state remains a pseudospin singlet along the SU(2)-symmetric line $V_1^{(1)} = V_1^{(0)}$. The corresponding energy spectra at a point on this line are shown in App.~\ref{app:su2symmetric_spectra}.

In Fig.~\ref{fig:bilayer}(b), we have plotted the overlap of the ground state on the sphere with several model wavefunctions. In the $\mathcal{S}=3$ shift sector, we computed the overlap with the Fibonacci model wavefunction from Eq.~\eqref{eq:Fib_WF} and compared it with the overlap with the decoupled Laughlin [equivalently Halperin~(330)] from Eq.~\eqref{eq:halperin_mmn}, plotting the larger of the two overlaps at each point of the phase diagram. We see that the decoupled Laughlin state does indeed transition to a state with large overlap with our Fibonacci wavefunction upon increasing $V_1^{(1)}/V_1^{(0)}$. However, the decoupled Laughlin state survives at sizeable interlayer  as long as $V_0^{(1)}\sim 2V_1^{(1)}$. For large $V_0^{(1)}$ the Halperin~(112) phase emerges, as is evident from the overlaps with the singlet wavefunction in Eq.~\eqref{eq:singlet112}. Note that here we generated a proxy for $\Psi_\mathrm{singlet}$ by taking the singlet ground state of the Hamiltonian with SU(2) symmetric Coulomb interactions. The emergence of this phase can be understood from the fact that the Halperin~(112) state is an exact zero energy state for the pseudopotential Hamiltonian where $V_0^{(1)},V_1^{(1)}>0$ and all other pseudopotentials are set to zero. This state, however, indicates a tendency of the system to phase separate or layer polarize. Indeed, for large $V_0^{(1)}$ and large $V_1^{(1)}$, the system eventually layer polarizes (the ED ground state in the layer polarized sector is lower in energy than the ED ground state in the layer balanced sector). 

Similarly, on the torus we first plot in Fig.~\ref{fig:bilayer_torus}(a) the phase diagram broken down by the degeneracy of the ground state manifold which has the largest gap above it. We find that the ground state manifold remains in the pseudospin-unpolarized sector for the range of interaction parameters considered. Moreover, we see once again that the 9-fold degenerate ground state manifold of the decoupled Laughlin state is robust to inter-layer coupling as long as $V_0^{(1)}\sim 2V_1^{(1)}$. However, for sufficiently large $V_1^{(1)}$ the ground state manifold transitions from 9-fold degenerate to 6-fold degenerate, consistent with a transition to the Fibonacci phase. We note that the location of the decoupled Laughlin and Fibonacci phases on the sphere agrees with the phase diagram on the torus (see Fig.~\ref{fig:bilayer_torus}), despite the fact that we use completely different probes for the topological order in the two geometries. We further confirm in App.~\ref{app:torus_sz_scan_330_fib} that the ground state remains in the unpolarized $S_z=0$ sector along the entire $(330)$-to-Fibonacci cut at $V_0^{(1)}=0$. In Fig.~\ref{fig:bilayer_torus}(b) we have plotted the overlap of the ground state on the sphere with several model states. The region with 9-fold degeneracy has high overlap with the decoupled Laughlin ground state manifold. For the 6-fold degenerate phase, we compute the overlap with the ground state manifold at $V_0^{(1)}=0$, $V_1^{(1)}=1$, which on the sphere corresponds to a state with large overlap with $\Psi_\text{Fib}$. We see that the overlaps do not change appreciably across the whole phase. 
Finally, in the 3-fold degenerate phase we compare two candidate states: the Halperin~(112) state and the PH(1/3) state. Proxies for each are built from the SU(2) symmetric Coulomb ground state --- taken in the $S=S_z=0$ sector for Halperin~(112), and in the $S=S_z=\frac{N_\mathrm{e}}{2}$ sector for PH(1/3). The PH(1/3) proxy is then rotated to $S_z=0$ via $S^-$ operators before computing overlaps with the actual ground state in the phase diagram. The Halperin~(112) overlap is larger throughout the 3-fold degenerate region and is shown in Fig.~\ref{fig:bilayer_torus}(b); the smaller PH(1/3) overlap is shown separately in App.~\ref{app:ph13test}.

Finally, we note that other non-Abelian states have been proposed for the $1/3+1/3$ bilayer, notably the interlayer Pfaffian \cite{Geraedts2015}. Although the overlap of the interlayer Pfaffian with the ED ground state is high in the parameter regime where we find the Fibonacci state, two strong pieces of evidence speak against this state: (i)~The ground state degeneracy on the torus is 6 for the Fibonacci state and 9 for the interlayer Pfaffian. In App.~\ref{sec:interlayer_Pfaffian} we show that the ground state degeneracy on the torus is very clearly 6-fold in the parameter regime where the Fibonacci state occurs on both geometries. (ii)~The quasihole counting in the spherical geometry is only consistent with the Fibonacci state (see Sec.~\ref{app:QHcount}).

\subsection{Effective Hamiltonian in the thin-torus limit}\label{sec:bilayerthintorus}

\begin{figure}[]
    \centering
    \includegraphics[width=\columnwidth]{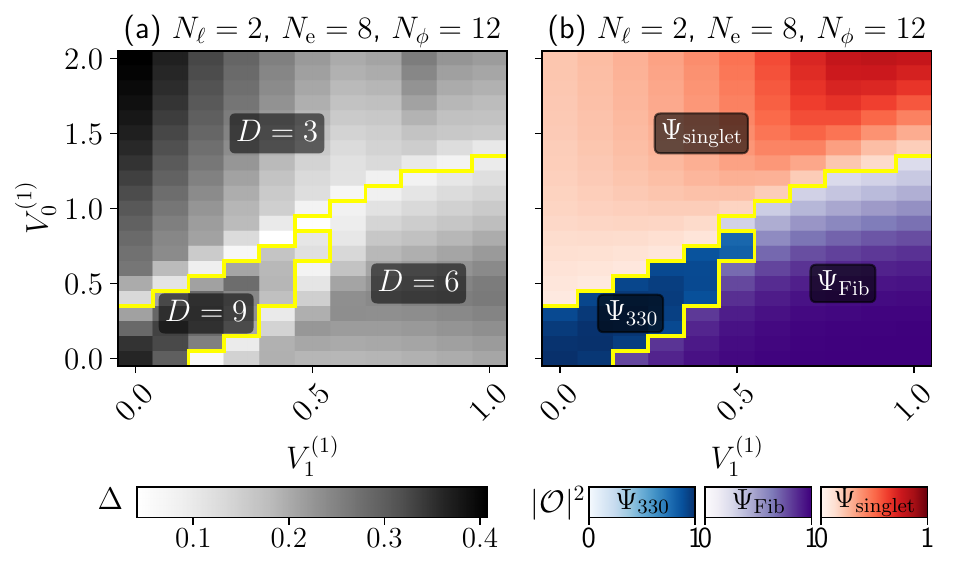}
    \caption{\textbf{Bilayer phase diagram in the torus geometry.} (a) Phase diagram for the bilayer system in the square torus geometry showing the gap $\Delta$ above the $D$ quasi-degenerate ground states. We have computed three gaps corresponding to these three degeneracies and then plotted the largest gap.  (b) Phase diagram with respect to the average of the squared overlap of the $D$ quasi-degenerate ground states with the corresponding model states. Here, we are taking ground states of different model potentials instead of analytical wave functions. For $\Psi_\mathrm{Fib}$, we have taken the ground state at $V_0^{(1)}=0.0$, $V_1^{(1)}=1.0$.}
    \label{fig:bilayer_torus}
\end{figure}
The numerical phase diagram for the bilayer system on the torus with $N_\mathrm{e}=8$ particles (Fig.~\ref{fig:bilayer_torus}) reveals a transition from the decoupled Laughlin state to the Fibonacci phase driven by interlayer interactions, in particular by increasing $V_{1}^{(1)}$ relative to $V_{0}^{(1)}$. The natural question to ask is what microscopic mechanism underlies this transition. Here we address this question from the analytically tractable limit of thin-torus geometry (see Sec.~\ref{sec:thin-torus-method} for an introduction to the method).

The projected bilayer interaction Hamiltonian takes the form up to $O(\epsilon^4)$ (see App.~\ref{appendix:thintorus} for details),
\begin{align}\label{eq:Heff_thintorus}
H &=  \sum_{r=1,2} V_r^{\parallel} \sum_{i,l} n_{i+r,l}n_{i,l} + \sum_{r=0,1,2} V_r^{\perp} \sum_{i,l}  n_{i+r,l}n_{i, \bar l}  \nonumber \\ 
 &~~ + \sum_{r=1,2} J_r \sum_{i,l} \left[ c_{i+r,l}^{\dagger} c^{\dagger}_{i,\bar l} c^{}_{i+r,\bar l} c^{}_{i,l} +\text{h.c.} \right] ,
\end{align}
where $l$ is the layer index, and $\bar l=-l$ denotes the other layer. The operator $n_{i,l}$ is the electron number operator at site $i$ and layer $l$, defined in terms of the creation $c_{i,l}^{\dagger}$ and annihilation operators $c_{i,l}$ as $n_{i,l}=c^{\dagger}_{i,l}c_{i,l}$.

There are three classes of terms in Eq.~\eqref{eq:Heff_thintorus}. 
First, the intralayer density-density interactions $V_r^\parallel$ have coefficients $V_1^{\parallel}=4\epsilon \kappa^2 V_1^{(0)}$ and 
$V_2^{\parallel}=16\epsilon^4 \kappa^2 V_1^{(0)}$. The intralayer terms are purely classical here, and the leading intralayer pair-hopping term appears only at order $O(\epsilon^5)$. 
Second, the interlayer density-density interactions $V_r^\perp$ have coefficients $V_0^\perp = V_0^{(1)}$, 
$V_1^\perp = \epsilon\left(V_0^{(1)}+\kappa^2 V_1^{(1)}\right)$, and 
$V_2^\perp = \epsilon^4\left(V_0^{(1)}+4\kappa^2 V_1^{(1)}\right)$. 
Third, the interlayer pair-hopping terms $J_r$ describe simultaneous hopping of two particles between layers while conserving the pair center of mass. 
Their amplitudes are 
$J_1 = \epsilon\left(V_0^{(1)}-\kappa^2 V_1^{(1)}\right)$ and 
$J_2 = \epsilon^4\left(V_0^{(1)}-4\kappa^2 V_1^{(1)}\right)$.

A Hamiltonian of the same form as Eq.~\eqref{eq:Heff_thintorus} was obtained in Ref.~\onlinecite{Vaezi2014}, where the interlayer pair-hopping term is generated by the single-particle interlayer tunneling. 
In our model Hamiltonian, we neglect such tunneling processes. 
Nevertheless, we find that the same interlayer pair-hopping term can be generated by expanding the interlayer pseudopotential interaction in the thin-torus limit.

We now focus on filling fraction $1/3$ in each layer and show that the ground-state degeneracies obtained from the thin-torus effective model match the torus degeneracies of the competing topological orders: the decoupled Laughlin~(330) state, the Fibonacci SU$(3)_2$ state, the Halperin~(112) state, and the layer-polarized PH(1/3) state. 
Although we do not expect the energetics from the thin-torus analysis to agree quantitatively with the numerical results, the tendencies toward these phases in the $V_{0,1}^{(1)}$ parameter space are consistent.

\textbf{Weak interlayer interactions (decoupled Laughlin phase).} With only intralayer interactions ($V_r^\parallel$ in Eq.~\ref{eq:Heff_thintorus}), the ground states are ninefold degenerate, with each layer hosting a charge-density wave of period three. 
This corresponds to the decoupled Laughlin phase, where each layer is a Laughlin state at filling $1/3$ with degeneracy three on a torus. 
When the interlayer interactions are small, i.e.~$V_r^\perp, J_r \ll V_r^\parallel$, and $V_0^\perp \sim V_1^\perp$, these nine states remain nearly degenerate. 
This gives a line in the $V_0^{(1)}$--$V_1^{(1)}$ plane along which the decoupled Laughlin phase remains stable.

\textbf{Large $V_{1}^{(1)}$, moderate $V_0^{(1)}$ (Fibonacci phase).} 
The nine ground states of the decoupled Laughlin phase separate into three diagonal states and six off-diagonal states. 
The diagonal states have the same charge-density-wave pattern in both layers, for example
\begin{equation}
\left| 
\begin{matrix}
100100\cdots \\
100100\cdots
\end{matrix}
\right\rangle ,
\label{eq:diagonal thin-torus configs}
\end{equation}
and are eigenstates of the thin-torus Hamiltonian. 
By contrast, the off-diagonal states are mixed by the interlayer pair-hopping term. 
A complementary perspective of an attractive interlayer interaction, studied in Ref.~\onlinecite{crepel2024attractive}, shows the same $6+3$ splitting with the roles of the two groups reversed.
Within each period-three unit cell, the relevant local Hilbert space is spanned by two configurations,
\begin{align}
\ket{1}\equiv  
\left| 
\begin{matrix}
100\\
010
\end{matrix}
\right\rangle,
\qquad 
\ket{2}\equiv 
\left| 
\begin{matrix}
010\\
100
\end{matrix}
\right\rangle ,
\end{align}
which may be viewed as an effective spin-$1/2$ degree of freedom. 
The thin-torus Hamiltonian projected into this subspace reduces to a transverse-field Ising model \cite{Vaezi2014} (also see App.~\ref{appendix:thintorus}); the states obtained from $\ket{1}$ and $\ket{2}$ by translating both layers by one or two sites generate two additional, disjoint subspaces.

The ferromagnetic phase of this effective Ising model is smoothly connected to the decoupled Laughlin phase. 
When the pair hopping $J_1$ dominates over the relevant interaction energy scale, set by $V_2^\parallel$, the effective Ising model enters its paramagnetic phase. 
Each of the three disjoint translation sectors then contributes one ground state, giving three off-diagonal ground states. 
Together with the three diagonal states, this yields a total ground-state degeneracy of six, consistent with the torus degeneracy of the SU$(3)_2$ Fibonacci state.

\textbf{Large $V_0^{(1)}$  (Halperin and PH-conjugate phases).} 
When $V_0^{(1)}$ is large, the on-site interlayer repulsion $V_0^\perp$ pushes the three diagonal states of the form of Eq.~\eqref{eq:diagonal thin-torus configs} far above the low-energy manifold. 
The remaining off-diagonal sector is described by the effective Ising model. 
When $|J_1|$ is large, this sector enters the paramagnetic phase, yielding one ground state in each of the three translation sectors and hence a threefold ground-state degeneracy. 
The sign of $J_1$ determines whether the paramagnet is smoothly connected to the layer-symmetric or layer-antisymmetric product state within the local two-state subspace\footnote{Here ``layer-symmetric'' and ``layer-antisymmetric'' refer to the two local thin-torus configurations $\ket{1}\pm\ket{2}$, not to the full fermionic exchange symmetry of the many-electron wavefunction.}. 
The layer-symmetric paramagnet is favored for $V_1^{(1)}>V_0^{(1)}/\kappa^2$. 
This is consistent with the Halperin~(112) wavefunction, whose interlayer Jastrow factor $\prod_{i,j}(z_i-w_j)^2$ is symmetric under exchanging $z$ and $w$ coordinates, matching the layer-symmetric combination selected in the thin-torus off-diagonal sector. 
Conversely, for $V_1^{(1)}<V_0^{(1)}/\kappa^2$, the layer-antisymmetric paramagnet is favored, consistent with the PH-conjugate Laughlin $1/3$ state, where the two layers are effectively identified into a single-component state.

\textbf{Layer polarization.} Finally, we consider what happens when the intralayer pseudopotential $V_1^{(0)}$ is no longer the dominant energy scale. 
In the thin-torus limit, when $V_0^{(1)}>4\kappa^2 V_1^{(0)}$, it becomes energetically favorable for two nearby particles to occupy the same layer rather than different layers. 
This reflects a tendency toward layer polarization driven by strong interlayer repulsion. 
If the filling in each layer is constrained to remain equal, full polarization is forbidden, and the system instead realizes a balanced state with enhanced interlayer avoidance. This is precisely the structure encoded by the Halperin~(112) and the PH(1/3) wavefunctions. When the relative layer occupation is allowed to vary in the numerical simulations, we indeed find that the energy is minimized by layer-polarized states, as shown in Fig.~\ref{fig:spin_polarization}.

\section{Non-Abelian $\text{SU(3)}_{N_\ell}$ states}
\label{sec:non-abelian}
As shown in Tables \ref{tab:FQHphases1} and \ref{tab:FQHphases2}, different topological orders can be distinguished by their shift on the sphere and by their ground state degeneracy on the torus. 

\begin{table}
\centering
\begin{tabular*}{\columnwidth}{@{\extracolsep{\fill}}cccc}
\hline\hline
FQH phase &\ \ &GSD (torus) & $\mathcal{S}$ (sphere)  \\
\hline
Decoupled Laughlin & & $3^{N_\ell}$ & 3  \\ \\
Generalized Halperin  & & $\begin{cases}
1, & N_\ell = 0 \pmod{3},\\[6pt]
3, & N_\ell \not = 0 \pmod{3}.
\end{cases}$ & 1\\ \\ Non-Abelian $\text{SU(3)}_{N_\ell}$ & & $(N_\ell+1)(N_\ell+2)/2$ & $3$ \\
\hline\hline
\end{tabular*}
\caption{\textbf{Different possible ground states for $N_\ell$-layer system.} Ground-state degeneracy (GSD) on the torus and shift $\mathcal{S}$ on the sphere for various candidate
fractional quantum Hall phases at $\nu = 1/3$ per layer in an $N_\ell$-layer system. }
\label{tab:FQHphases2}
\end{table}

In the previous section we computed overlaps with the CFT-derived wavefunction. However, this is computationally expensive for large numbers of electrons and large numbers of layers. Therefore in this section we use a different method to distinguish the decoupled Laughlin and $\text{SU(3)}_{N_\ell}$ phases, namely quasihole counting. 

Quasihole counting refers to computing the degeneracy of the ground state manifold in the presence of additional flux quanta. Adding flux quanta to the system generates quasihole excitations (e.g.~for the Fibonacci phase adding one flux quantum corresponds to adding two quasiholes) and the degeneracy of the quasihole manifold, resolved by the angular momentum quantum number, can be used to distinguish different types of topological order \cite{Read1996quasiholes,Simon2007construction,Liu2015}. Crucially, the quasihole counting is different for the case of the decoupled Laughlin state and the $\text{SU(3)}_{N_\ell}$ states. While different topological orders could have the same quasihole counting when restricted to finite-size systems \cite{PhysRevB.88.075313}, we are not aware of any such competing topological order at this shift and filling. 
\subsection{Quasihole counting}
\label{sec:qh_counting}

\begin{figure*}
    \centering
    \includegraphics[width=\linewidth]{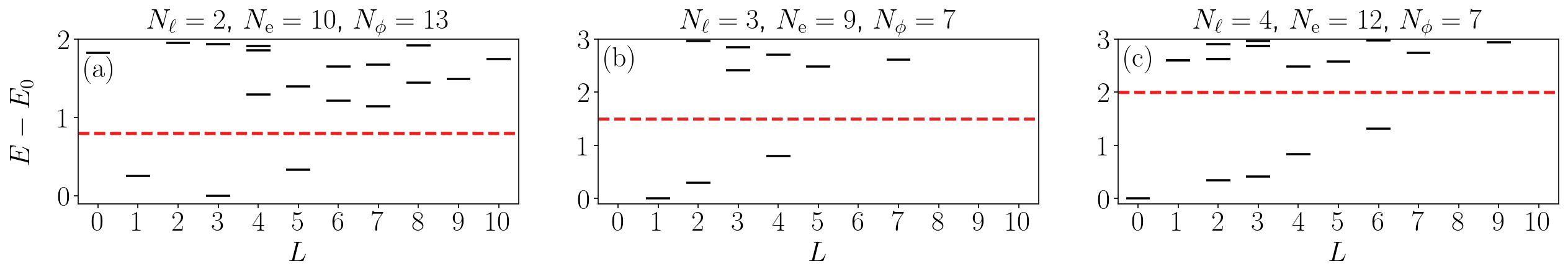}
    \caption{\textbf{Quasihole counting of the non-Abelian state for two-, three- and four-layer quantum Hall.} Energy spectra with one additional flux quantum for 2, 3, 4 layers on the sphere with  $V_1^{(0)}=V_1^{(1)}=1$ and $V_0^{(1)}=0$. The interlayer interactions in the 3 and 4 layer cases only act between NN layers (subject to periodic boundary conditions in the layer direction). For any $N_\ell$ layer case, the spectra show a clear gap, and the counting matches with the $\text{SU(3)}_{N_\ell}$ exclusion principles.} 
    \label{fig:QHspectra}
\end{figure*}

The dimension of the quasihole manifold and edge excitations for several FQH states can be analytically computed using generalized exclusion principles \cite{haldane1991fractional,Li2008}. In the case of the bilayer Fibonacci state, the exclusion principle was first derived in Ref.~\onlinecite{Liu2015} by considering the corresponding parent Hamiltonian in the thin-torus limit, which energetically favours the formation of layer-singlets on NN and NNN orbitals. We reproduce the resulting bilayer $\text{SU(3)}_2$ exclusion principles below for completeness:
\begin{enumerate}
    \item[(i)] no more than 2 particles in 3 consecutive orbitals,
    \item[(ii)] if two particles occupy NN orbitals, they will form a singlet state $\frac{1}{\sqrt{2}}(\ket{\uparrow \downarrow} - \ket{\downarrow \uparrow})$,
    \item[(iii)] if two particles occupy NNN orbitals, they will form a singlet state $\frac{1}{\sqrt{2}}(\ket{\uparrow 0 \downarrow} - \ket{\downarrow 0 \uparrow})$,
    \item[(iv)] Formation of NN singlets is prioritized over NNN singlets, unless the former increases
    the total number of singlets. 
\end{enumerate}
Here, $\ket{\uparrow}$($\ket{\downarrow}$) refers to the state with a particle in the top layer (bottom) layer, different from the spin states of the Ising model mentioned in the previous section. Notice that particles that are at least 2 orbitals apart experience no exclusion statistics and can thus form both singlet and triplet states, leading to additional multiplicity. To perform the quasihole counting, one can thus use exclusion principle (i) to write the allowed layer-summed root partitions, and then assign a multiplicity to each root following exclusion principles (ii)-(iv). 

In the spherical geometry, the rotation symmetry forces the energy spectrum to decompose into angular momentum multiplets $D_L$ labeled by the total angular momentum $L$, with degeneracy $2L+1$. The quasihole manifold can thus be fully specified by its decomposition into $L$-multiplets. For example, with $N_\mathrm{e} = 5+5$ particles and $N_\phi = 12+1$ flux quanta (one additional flux quantum above the $\mathcal{S}=3$ ground state with $N_\phi=12$), the allowed layer-summed root configurations (up to $L_z\leftrightarrow -L_z$ symmetry) are shown in Tab.~\ref{tab:su(2)roots}.
\begin{table}[t]
\centering
\begin{tabular}{c|c|c}
\hline \hline
Root & $L_z$ & \# \\
\hline
$20020020020020$ & $5$ & $1$ \\
\hline
$20020020020011$ & $4$ & $1$ \\
\hline
$20020020020002$ & \multirow{2}{*}{$3$} & \multirow{2}{*}{$2$} \\
$20020020011011$ & & \\
\hline
$20020020011002$ & \multirow{2}{*}{$2$} & \multirow{2}{*}{$2$} \\
$20020011011011$ & & \\
\hline
$20020011011002$ & \multirow{3}{*}{$1$} & \multirow{3}{*}{$3$} \\
$20020020002002$ & & \\
$20011011011011$ & & \\
\hline
$20011011011002$ & \multirow{3}{*}{$0$} & \multirow{3}{*}{$3$} \\
$20020011002002$ & & \\
$11011011011011$ & & \\
\hline \hline
\end{tabular}
\caption{\textbf{Quasihole counting for bilayer Fibonacci state, based on generalized exclusion principles.} We list the layer-summed root configurations (i.e.~the occupation numbers of each orbital, summed over both layers) that obey our $\text{SU(3)}_2$ exclusion principles, they decompose into $D_1 \oplus D_3 \oplus D_5$ angular momentum multiplets.}
\label{tab:su(2)roots}
\end{table}

Each root configuration in Table~\ref{tab:su(2)roots} has multiplicity 1, which follows from the SU(3)$_2$ exclusion principles. We thus expect the quasihole manifold to decompose as $D_1 \oplus D_3 \oplus D_5$, which matches the numerically observed counting in Fig.~\ref{fig:QHspectra}(a). We note that the interlayer Pfaffian state would have counting $D_0 \oplus 2D_1 \oplus D_2 \oplus 2D_3 \oplus D_4 \oplus D_5$ which is inconsistent with the counting in Fig.~\ref{fig:QHspectra}(a), even if one places the quasihole gap higher in energy so as to include more states in the quasihole manifold. This is further evidence against the interlayer Pfaffian occurring in this parameter regime.

The $\text{SU(3)}_{2}$ state for the bilayer case can be generalized to $N_\ell$ layers. The resulting topological order is $\mathcal{C}_\nl = \mathcal{F}_{6\nl} \boxtimes \overline{\text{SU(3)}}_{\nl}$, where $\mathcal{F}_{6\nl}=\{0,f\}$ is an invertible fermionic phase with topological central charge $3\nl \mod 8$ and $\overline{\text{SU(3)}}_{\nl}$ is the time-reversal conjugate of $\text{SU(3)}_\nl$. The total topological central charge is thus $1+\frac{3(\nl^2-1)}{\nl+3} \mod 8$ and the ground state degeneracy on the torus is ${{\nl+2} \choose {2}} = \frac{1}{2}(\nl+1)(\nl+2)$ \cite{COQUEREAUX2016119,Gannon:1994aa,Wen1998}. The topological spins of the particles are those of $\overline{\text{SU(3)}}_\nl$, up to $\frac{1}{2}$ depending on if the fermion particle is fused with the representative anyon or not. 
We have extended the generalized exclusion principles to the case of the $N_\ell$-layer $\text{SU(3)}_{N_\ell}$ state (see App.~\ref{app:QHcount} for more details) and observe excellent agreement with the numerical results in Fig.~\ref{fig:QHspectra}(b,c) for $N_\ell=3$ and $N_\ell=4$. For the trilayer $\text{SU(3)}_3$ state with $N_\mathrm{e}=9$ particles and $N_\phi = 6+1$ flux quanta, the predicted quasihole counting is $D_\frac{3}{2} \oplus D_\frac{5}{2}\oplus D_\frac{9}{2}$ while for the tetralayer $\text{SU(3)}_4$ state with $N_\mathrm{e}=12$ particles and $N_\phi = 6+1$ flux quanta, the quasihole counting is predicted to be $D_0\oplus D_2\oplus D_3\oplus D_4\oplus D_6$.  

\subsection{Wavefunctions}
\label{sec:modelWF}

We have generalized the CFT construction of the model wavefunction for the Fibonacci state to the case of $\nl$ layers, each at filling $\nu=\frac{1}{k}$, with $k$ being an odd integer. The edge theory is described by the coset $\frac{\mathrm{U}(k\nl)_1}{\text{SU}(k)_\nl}$ \cite{PhysRevLett.113.236804}. This can be understood from the fact that the fermions in layer $l$ at filling $\nu=\frac{1}{k}$ can be expressed as the product of $k$ parton fields $c_l \sim f_{1,l} ... f_{k,l}$, leading to a total of $k\nl$ complex fermions and thus to a $\text{U}(k\nl)_1$ edge theory. One must then project out the $\text{SU}(k)_\nl$ diagonal current algebra in the parton fields leading to the desired coset. The edge theory is also equivalent to $\text{SU}(\nl)_k \times \text{U}(1)_{k\nl}$ up to simple current extensions. The $[\text{U(1)}]^\nl$ layer-particle-number symmetry can be made more explicit by re-writing the neutral sector $\text{SU}(\nl)_k$ as a product of Gepner parafermions \cite{gepner1987new} and $k-1$ chiral bosons, resulting in
\begin{equation}
    \underbrace{\frac{\text{SU}(\nl)_k}{[\text{U}(1)]^{k-1}} \times [\text{U}(1)]^{k-1}}_{\text{neutral sector}} \times \underbrace{\text{U}(1)_{k\nl}}_{\substack{\text{charge}\\\text{sector}}}.
\end{equation}
We then construct electron operators for each layer in the form 
\begin{equation}
    \Psi_l(z) = \psi_\bl(z) :\text{exp}[i(\al \cdot \boldsymbol{\phi}_s(z) + \Delta \phi_c(z))]:,
\end{equation}
where $\psi_\bl(z)$ is a parafermion operator in $\frac{\text{SU}(\nl)_k}{[\text{U}(1)]^{k-1}}$ corresponding to the $\text{SU}(\nl)_k$ root $\bl$, and $\boldsymbol{\phi}_s$ and $\phi_c$ are the spin and charge bosons respectively. Using these electron operators, we can construct a model wavefunction in the form of the following conformal correlator
\begin{equation}
    \Psi(\{z_i^l\}) = \Big \langle \prod_{l=1}^\nl \prod_{i=1}^N \Psi_l(z_i^l) \mathcal{O}_{\text{bg,c}} \Big \rangle,
\end{equation}
where $z_i^l$ denotes the coordinate of the $i$th particle in layer $l$ and $\mathcal{O}_{\text{bg,c}}$ is a background charge operator that ensures charge neutrality, chosen so that it produces the terms necessary to normalize our wavefunction \cite{PhysRevB.86.245310}, which we omit in the following. For this correlator not to vanish the operators in the neutral sector must fuse to the identity, so one must have that $\sum_l \bl = 0 \mod kQ$ where $Q$ is the charge lattice of $\text{SU}(\nl)$, and $\sum_l \al = 0$. Moreover, there are further constraints arising from the filling factor and shift that we want the wavefunction to possess. In addition, one must make sure that the electron operators are all local with respect to each other. In App.~\ref{app:modelWF}, we show that these constraints are satisfied for a specific choice of $\al$, $\bl$ and $\Delta$. The model wavefunction then splits into the product of a parafermionic correlator $\bigg \langle \prod_{l=1}^\nl \prod_{i=1}^N \psi_\bl(z_i^l)\bigg \rangle$ and a vertex operator correlator for the spin and charge bosons. The vertex operator correlator results in a series of Jastrow factors, while to compute the parafermionic correlator we adopt a point-splitting procedure. Indeed we note that a very similar correlator is known already \cite{PhysRevB.95.125130,PhysRevB.87.205137,regnault2008bridge}:
\begin{align}\label{sunlk}
   \Phi_{\text{SU}(\nl)_k}^{(k)}(\{z_I^l\})  &= \Big\langle \prod_{I=1}^{kn}\prod_{l=1}^{\nl-1} J_\bl^+(z_I^l)\Big \rangle\nonumber\\
   &= S\bigg[\prod_{p=1}^k \Phi_{\text{SU}(\nl)}^\mathrm{Halp}(\{z_{i,p}^l\})\bigg],
\end{align}
where $J_\bl^+(z) =  \sqrt{k}\psi_{\bl}(z) :e^{i \bl \cdot \boldsymbol{\phi}_s(z)}:$ is a $\text{SU}(\nl)_k$ current corresponding to the $\text{SU}(\nl)_k$ root $\bl$ and $S$ denotes the symmetrization operator. Here we defined the $\text{SU}(\nl)$ Halperin states whose $K$-matrix is the Cartan matrix of $\text{SU}(\nl)$
\begin{align}
     \Phi_{\text{SU}(\nl)}^\mathrm{Halp} &= \Big\langle \prod_{I=1}^{kn}\prod_{l=1}^{\nl-1} :e^{i\bl\cdot \boldsymbol{\phi}_s(z_I^l)}:\Big \rangle\nonumber\\
     &= \prod_{l=1}^{\nl-1}\prod_{i<j}(z_i^l-z_j^l)^2  \times \prod_{\langle l,l'\rangle}\prod_{ij}(z_i^l-z_j^{l'})^{-1},
\end{align}
and $\langle l, l' \rangle$ denotes pairs of NN layers. Therefore, to compute $\Phi_{\text{SU}(\nl)_k}^{(k)}$, one must first split the $kn$ particles into $k$ clusters labelled by $p$ [so the particles initially labelled by $I=1,2,...,kn$ become labelled by $(i,p)$ with $i=1,...,n$ and $p=1,...,k$]. Then, one must take the product of the $\text{SU}(\nl)$ Halperin wavefunctions for the particles in each cluster, and finally the result must be symmetrized over all ways to group the particles into $k$ clusters. Therefore, we need to reduce our parafermionic correlator $\bigg \langle \prod_{l=1}^\nl \prod_{i=1}^N \psi_\bl(z_i^l)\bigg \rangle$ to the form in Eq.~\eqref{sunlk}, which we do by writing $\psi_{\boldsymbol{\beta}_{\nl-1}} = \psi_{-(\boldsymbol{\beta}_1+...+\boldsymbol{\beta}_{\nl-1})}$ and expanding this using the parafermionic operator product expansions. 
\begin{widetext}
The $\text{SU}(\nl)_k$ wavefunction at filling $\nu = \frac{\nl}{k}$ and shift $\mathcal{S}=k$ is thus given by 
\begin{equation}
    \Psi(\{z_i^l\}) = \lim_{\substack{{\{w_{i,a,p}\}\rightarrow z_{i,p}}\\{\{z_{i,p}\}\rightarrow z_i^\nl}}} \bigg[P_\text{split}(\{z_{i,p}\},\{w_{i,a,p}\}) \frac{ \Phi_{\text{SU}(\nl)_k}^{(k)}(\{z_I^l\}) }{\big[\Phi_{\text{SU}(\nl)}^\mathrm{Halp}(\{z_I^l\}) \big]^{1/k} } \bigg] \prod_l \prod_{i<j}(z_i^l-z_j^l)^{k-2+\frac{2}{k}}\prod_{\langle l, l'\rangle} \prod_{ij}(z_i^l-z_j^{l'})^{1-\frac{1}{k}},
\end{equation}
where $z_{(i,p)}^l = w_{i,a,l}$ for $p=1,...,k-1$ and  $z_{(i,k)}^l = z_i^k$. Here we defined the point-splitting polynomial
\begin{equation}
 P_\text{split} = \prod_{i=1}^n\bigg(\prod_{p=1}^{k-2}(z_{i,p}-z_{i,p+1})^{2p/k}\, \prod_{p=1}^{k-1}\prod_{a=1}^{\nl-2}(w_{i,a,p}-w_{i,a+1,p})^{1-\frac{1}{k}}\bigg).
\end{equation}
One can verify that the singular terms all cancel out, and that the final wavefunction is well-defined and finite with no branch-cuts or discontinuities. Further details on the construction of the model wavefunction and the identification of the corresponding (spin) TQFT are provided in App.~\ref{app:modelWF}. 
\end{widetext}

\section{From finite stacks to the 3D limit}
\label{sec:N-layers}

So far we have considered quantum Hall multilayers with two, three and four layers. In order to approach the 3D limit, we now extend our ED numerics study to up to ten layers. For interactions between NN layers we find that the decoupled Laughlin state survives up to a large value of interlayer coupling. However, once NNN layers are allowed to interact, a new phenomenon appears:  translational symmetry breaking in the layer direction.

\subsection{Abelian states: Decoupled Laughlin state and generalized Halperin states}
\begin{figure}
    \centering
    \includegraphics[width=\linewidth]{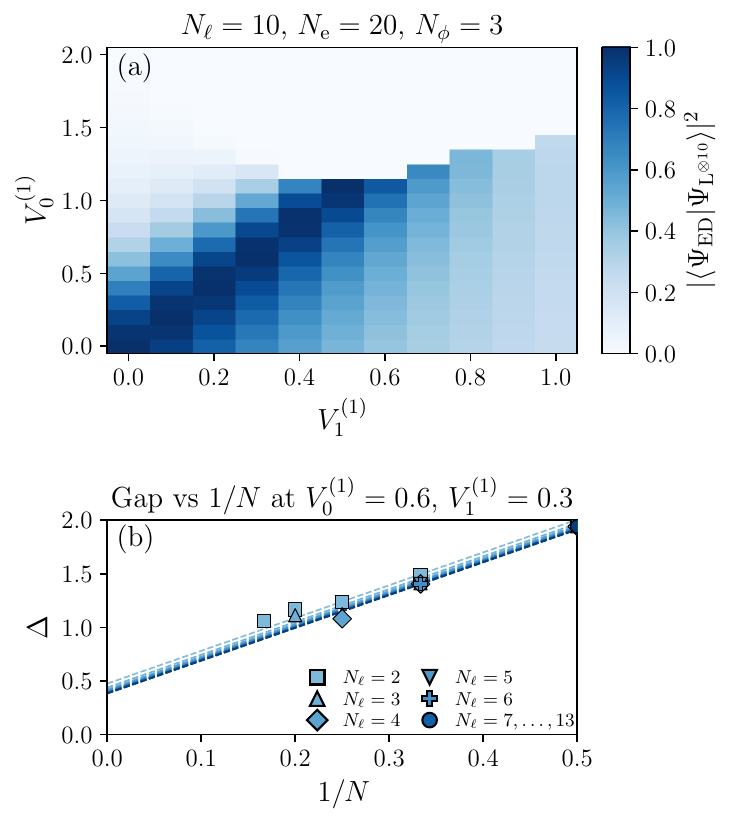}
    \caption{\textbf{Overlap and finite-size gap scaling of the decoupled Laughlin phase.} (a) We take an $N_\ell=10$ layer system in the spherical geometry with $2$ particles per layer and $N_\phi = 3$ flux quanta (equivalently $\nu=\frac{1}{3}$ per layer with shift $\mathcal{S}=3$), and compute the overlap of the ground state $\ket{\Psi_\mathrm{ED}}$ in the $L=k_z=0$ sector with the decoupled Laughlin state $\ket{\Psi_\mathrm{L}^{\otimes10}}$. Note that the region in the top half of the figure has its ground state in the $\mathcal{S}=1$ shift sector. (b) We compute the energy gap for $V_0^{(1)} = 0.6$, $V_1^{(1)} = 0.3$, filling $\frac{2}{3}$ and shift $\mathcal{S}=3$, keeping $N=\frac{N_\mathrm{e}}{\nl}$ particles in each layer fixed.}
    \label{fig:Overlap_10layer}
\end{figure}

We saw in Sec.~\ref{sec:bilayer} that the decoupled Laughlin state in the bilayer remains stable for an extended region in phase space, even for significant interlayer interactions. This robustness persists as we increase the number of layers, and we argue should extrapolate to infinitely tall stacks, i.e.~Laughlin FFQH states. Indeed, in Fig.~\ref{fig:Overlap_10layer}(a) we show the overlap of the decoupled Laughlin wavefunction
\begin{equation}
    \Psi_{\mathrm{L}^{\otimes \nl}}(\{z_i^l\}) = \prod_{l=1}^{\nl} \prod_{i<j}^N (z_i^l-z_j^l)^3
\end{equation}
with the numerically obtained ground state in a $N_\ell=10$ layer system as we vary the NN interlayer pseudopotential interactions $V_0^\textrm{(1)}$ and $V_1^\textrm{(1)}$. We observe a distinct region of stability where the ground state has near perfect overlap with the decoupled Laughlin wavefunction (the $L=k_z=0$ Hilbert space dimension is 16586). We also remark that the region of stability follows roughly a slope of $V_0^\text{(1)}/V_1^\text{(1)} \sim 2$ in the phase space, which is consistent with the phase diagrams with fewer layers (see Fig.~\ref{fig:bilayer} and App.~\ref{app:multilayer}). 

As further evidence of the robustness of the decoupled Laughlin state for large $N_\ell$, in Fig.~\ref{fig:Overlap_10layer}(b) we have  performed finite-size scaling of the neutral gap in the decoupled Laughlin phase. More precisely, we fixed $V_0^{(1)} = 0.6$, $V_1^{(1)} = 0.3$ and computed the energy gap at filling $\frac{2}{3}$ and shift $\mathcal{S}=3$, keeping $N=\frac{N_\mathrm{e}}{\nl}$ particles in each layer fixed, for several combinations of $(N,\nl)$. We find that the gap extrapolates to $\Delta_\infty \sim 0.37$ in units of $V_1^{(0)}$ for $N,\nl \rightarrow \infty$. Additional data on the finite-size scaling for a wider range of system sizes and interaction parameters are provided in App.~\ref{app:finitesize_scaling_extra}.

The region of large $V_0^{(1)}$ in  Fig.~\ref{fig:Overlap_10layer}(a) that has vanishing overlap with the decoupled Laughlin state has a ground state in the $\mathcal{S}=1$ shift sector. In the bilayer case, we saw that this corresponds to the Halperin~(112) phase describing phase separation. It is natural to ask whether also for $N_\ell >2$ a phase-separated state appears in this part of the phase diagram. The generalization of the Halperin~(112) wavefunction to $N_\ell$ components has been discussed in Refs.~\onlinecite{Goerbig_2007,DeGail2008}. The $N_\ell$-layer generalization of the Halperin state (see App.~\ref{app:abelianstates}), again assuming  ``periodic boundary conditions" such that the layers interact mod $N_\ell$, is
$K_{ij} = \delta_{ij}
+ \delta_{i,j+1 \,(\mathrm{mod}\ N_\ell)}
+ \delta_{i,j-1 \,(\mathrm{mod}\ N_\ell)}$.  The eigenvalues of $\lambda_k = 1 + 2 \cos\left(\frac{2 \pi k}{N_\ell}\right)$ for $k = 0, 1, \dots, N_\ell-1$. These generalized Halperin states have negative eigenvalues in the $K$-matrix and hence describe phase separated states (App.~\ref{app:abelianstates}). Phase separation indicates that the system is being forced to have equal occupation in each layer, even though it would be energetically favorable to form a layer imbalanced state. Instead of the electrons separating into different layers, the electrons therefore spatially separate into different parts of the system. In light of this, we should study more carefully what happens when we do not enforce equal occupation of each layer, which is the topic of the next subsection.

\subsection{Layer translational symmetry-breaking}
\label{sec:translation_symmetry_breaking}
So far, we have focused on FFQH states that preserve translation symmetry in the layer direction. However, frustrated intra- and interlayer interactions can spontaneously break this symmetry. This can occur in several ways. In a weaker form, the charge density remains uniformly distributed among the layers, while interlayer coherence develops a layer-dependent pattern. We construct a non-Abelian Chern-Simons-Ginzburg-Landau theory for such a state in App.~\ref{sec:NA_GL}, where an interlayer parton-exciton condensate spontaneously dimerizes into a foliated Fibonacci phase with no charge modulation (Fig.~\ref{fig:schematic}c) and discuss why its microscopic stabilization near the decoupled-Laughlin point remains an open problem. A stronger form of symmetry breaking occurs when the layer fillings of the ground state become non-uniform, so that different layers carry different filling fractions. We will focus on the latter possibility. Although our model does not include interlayer tunneling, the physical realization of the layer-translational symmetry-breaking state should include a small tunneling between the layers to allow charge redistribution. The FFQH phases we obtain are gapped, and hence they are stable as long as the strength of the tunneling is much smaller than the gap.

\subsubsection{Classical electrostatics}

In Sec.~\ref{sec:bilayer} we found interesting bilayer states with filling $\nu=\frac13$ per layer. We can consider the multilayer generalization of this with the goal to send the number $N_\ell$ of layers to infinity. One question is what happens when we dope holes into the system. Do the holes distribute uniformly in the layers or not? We will show that there exists a parameter regime where the lowest energy configuration has some layers remain at filling $\nu=\frac13$ while other layers are depleted. 

To gain some intuition about the spontaneous charge redistribution, we consider a classical electrostatic model for an $N_\ell$-layer system with $N_\ell$ divisible by 3. We study the model at average filling $\nu=\frac29$ per layer. The fixed total filling is $\nu_\mathrm{tot} = \sum_l \nu_l = 2 N_\ell/9$. We make this choice since this filling factor in principle allows for pairs of layers each at filling $\nu=\frac13$ separated by an empty ``buffer" layer. The electrostatic energy contribution to our Hamiltonian is modeled as:
\begin{equation}
E = \frac12\sum_{l=1}^{N_\ell} \left( V^{(0)} \, \nu_l^2+V^{(1)} \, \nu_l \nu_{l+1} + V^{(2)} \, \nu_l \nu_{l+2}   \right).
\label{eq:electrostatic}
\end{equation}
NN and NNN interactions favor charge orderings along the $z$-direction. We can show this by minimizing the energy Eq.~\eqref{eq:electrostatic} when introducing $\nu_l=\sum_qe^{iql}\nu_q$ with $\nu_q=\nu_{-q}$. Then $E=N_\ell\sum_q\lambda(q)\nu_q^2$ with
\begin{equation}
    \lambda(q)=V^{(0)}+V^{(1)}\cos q+V^{(2)}\cos 2q.
\end{equation}
We need to minimize this energy subject to the constraint $\nu_\mathrm{tot} = \sum_l \nu_l = 2 N_\ell/9$ which fixes $\nu_{q=0}$. The two competing solutions are $q=0$ and $q=2\pi/3$. For $V^{(1)}+V^{(2)}<2V^{(0)}$, the $q=0$ solution and hence equal population of all layers is favoured (denoted as  $[\frac{2}{9}\frac{2}{9}\frac{2}{9}]^{N_{\ell}/3}$, noting the filling-pattern in three consecutive layers). For $V^{(1)}+V^{(2)}>2V^{(0)}$, $\lambda(2\pi/3)<0$ and the $q=2\pi/3$ solution wins and the system forms a period 3 layer CDW ($[\frac{2}{3}00]^{N_{\ell}/3}$). The $[\frac{1}{3}\frac{1}{3}0]^{N_{\ell}/3}$ pattern can be obtained from a superposition of the $q=0$ and $q=2\pi/3$ solutions and is hence degenerate exactly at the transition between these two phases, and otherwise not part of the (classical) ground state manifold. However, a more microscopic and quantum mechanical treatment of the interactions taking into account the different angular momentum channels (specifically $V^{(1)}_1$) is expected to realize a \emph{gapped} state for $[\frac{1}{3}\frac{1}{3}0]^{N_{\ell}/3}$ (which could be a Fibonacci FFQH state or the a Laughlin FFQH state), constituting a relative energy gain of the $[\frac{1}{3}\frac{1}{3}0]^{N_{\ell}/3}$ charge configuration over $[\frac{2}{3}00]^{N_{\ell}/3}$. In that case, the charge configuration $[\frac{1}{3}\frac{1}{3}0]^{N_{\ell}/3}$ will occupy a finite region in the phase diagram, as we show in the next section. 

\subsubsection{Thin-torus limit}\label{sec:thintorus_multilayer}

\begin{figure}[t]
    \centering
    \includegraphics[width=\linewidth]{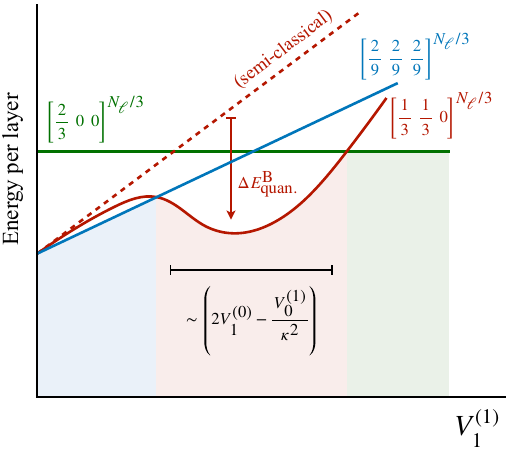}
    \caption{\textbf{Schematic illustration of the competition between different charge configurations.} From the analysis in the multilayer thin-torus framework, we show a schematic of the energy (per layer) contributions of different charge configurations as a function of the interlayer interaction $V_1^{(1)}$. The correlated interlayer hopping provides a quantum energy gain that acts exclusively on the pair-foliated configuration promoting it to a stable phase.}
    \label{fig:foliation_schematic}
\end{figure}

In this section, we compare the energies of the charge configurations discussed in the previous section in the thin-torus framework, in order to explicitly derive how their classical degeneracy is lifted. Since the interactions extend only to NNN layers, the arguments are local and hold for any $N_\ell$ divisible by 3, including the thermodynamic limit. A schematic illustration of the competition between different charge configurations is summarized in Fig.~\ref{fig:foliation_schematic}.

In the thin-torus limit, the Hamiltonian Eq.~(\ref{eq:Heff_thintorus}) generalized to $N_{\ell}$ layer system with periodic boundaries in the layer direction is,
\begin{align}\label{eq:thintorus_N6}
H &=  \sum_{r=1,2} V_r^{\parallel} \sum_{i,l} n_{i+r,l}n_{i,l} + \sum_{\substack{r=0,1,2 \\ d=1,2}} V_r^{\perp,(d)} \sum_{i,l}  n_{i+r,l}n_{i, l+d}  \nonumber \\ 
 &~~ + \sum_{\substack{r=1,2 \\ d=1,2}} J_r^{(d)} \sum_{i,l} \left[ c_{i+r,l}^{\dagger} c^{\dagger}_{i,l+d} c^{}_{i+r,l+d} c^{}_{i,l} +\text{h.c.} \right],
\end{align}
where $l\in\{0,1,\dots,N_{\ell}-1\}$ is the layer index and is defined $l\equiv l \mod N_{\ell}$. Besides the terms in the bilayer Hamiltonian, we also include density-density interaction and pair-hopping terms between the NNN layers, denoted as $V_r^{\perp, (2)}$ and $J_r^{(2)}$. The functional form of these terms is the same as the ones defined in the Sec.~\ref{sec:bilayerthintorus} using the NNN pseudopotentials $V_0^{(2)}$ and $V_1^{(2)}$ instead of the NN pseudopotentials.

We compare three charge configurations at total filling $\nu_{\rm tot}=2N_{\ell}/9$. The uniform $[\frac{2}{9}\frac{2}{9}\frac{2}{9}]^{N_{\ell}/3}$ which we label as configuration A for convenience, the pair-foliated $[\frac{1}{3}\frac{1}{3}0]^{N_{\ell}/3}$ labeled as configuration B, and $[\frac{2}{3}00]^{N_{\ell}/3}$ labeled as configuration C. Their thin-torus root configurations are $\dots 100010000\dots$ of period 9 for $\nu=2/9$, $\dots 100100100 \dots$ of period 3 for $\nu=1/3$, and $\dots 110110110\dots$ of period 3 for $\nu=2/3$. 
We split the Hamiltonian $H=H_{\rm semi.}+H_{\rm quan.}$ into its diagonal density-density part $H_{\rm semi.}$ and its off-diagonal correlated hopping part $H_{\rm quan.}$, and evaluate each on the three configurations. All energies are computed per 9-orbital unit cell. We note that the density-density contributions $H_{\rm semi.}$ differs from the classical electrostatic model Eq.~(\ref{eq:electrostatic}): it resolves variations in the density profile within each layer, whereas the electrostatic model assumes a uniform in-plane density. 

Configuration C has $N_\ell/3$ layers filled at $\nu = 2/3$, each with 3 intralayer adjacent-occupation pairs per 9-orbital cell per filled layer costing $V_1^\parallel = 4\epsilon\kappa^2 V_1^{(0)}$. The filled layers are separated by 3 layer spacings, beyond the range of both NN and NNN interactions, so all interlayer terms vanish. The semi-classical energy is $E_{\text{semi.}}^{\mathrm{C}}=N_{\ell} V_{1}^{\parallel}=4N_{\ell}\epsilon \kappa^2 V_{1}^{(0)}$. Since $V_1^{(0)} = 1$ is the dominant energy scale, configuration C is always highest in energy whenever $V_0^{(1)}, V_1^{(1)}, V_0^{(2)} < V_1^{(0)}$. We henceforth focus on the competition between A and B. 

For the remaining two configurations, the semi-classical energy at $\mathcal{O}(\epsilon)$ is determined by the number of adjacent-orbital contacts between layer pairs. This depends on the optimized root configurations through which the system minimizes its energy. This is further discussed in App.~\ref{appendix:thintorus}. The resulting semi-classical energies for configurations A and B are,

\begin{align}
    E^{\rm A}_{\rm semi.} &= N_{\ell} \left( \frac{1}{2} V_1^{\perp,(1)} + 2V_1^{\perp,(2)} \right), \\
    E^{\rm B}_{\rm semi.} &= N_{\ell} \left( V_1^{\perp, (1)} + V_1^{\perp,(2)} \right).
\end{align}
In the parameter regime $V_0^{(1)}>V_0^{(2)}$, these evaluate to $E^{\rm A}_{\rm semi.}<E^{\rm B}_{\rm semi.}$, so configuration A enjoys a classical advantage of $(E^{\rm A}_{\rm semi.}-E^{\rm B}_{\rm semi.})/N_{\ell}= - \epsilon \kappa^2 V_1^{(1)}/2$. 

This classical advantage is overcome by correlated hopping terms $J_1^{(1)}$ and $J_{1}^{(2)}$ that selectively provides enough quantum gains in configuration B. Each active pair in configuration B is precisely the bilayer system we analyzed in Sec.~\ref{sec:bilayerthintorus}. The correlated hopping splits the off-diagonal CDW states into `layer symmetric' and `layer anti-symmetric' states (see App.~\ref{appendix:thintorus}) with an energy gap $\sim |J_{1}^{(1)}|+|J_1^{(2)}|$. This leads to a quantum energy gain of
\begin{align}
    \Delta E^{\rm B}_{\text{quan.}}  \sim -\epsilon N_{\ell} \left(|V_0^{(1)}-\kappa^2 V_{1}^{(1)}|+V_0^{(2)}\right).
\end{align}
The correlated hopping operator also acts on configuration A, 
however, the resulting root lies outside the optimized CDW configuration that minimizes the density-density interaction (see App.~\ref{appendix:thintorus}). Therefore, within the low-energy projected subspace of these optimized CDW root configurations, the correlated hopping yields an energy gain for configuration A only at higher order in $\epsilon$.

\subsubsection{Exact diagonalization}
\begin{figure}[t]
\centering
\includegraphics[width=\linewidth]{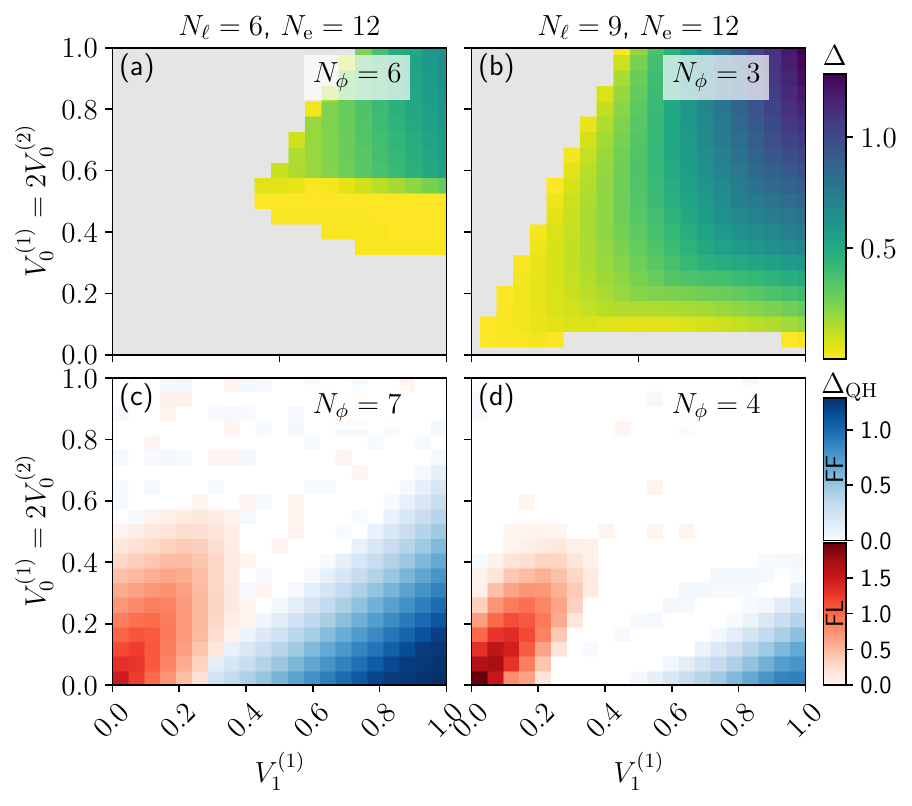}
\caption{\textbf{Phase diagrams of foliated fractional quantum Hall states.}
Phase diagram for (a) $N_\ell=6$ and (b) $N_\ell=9$ layer systems on the sphere with layer-periodic boundary condition as a function of $V_0^{\mathrm{(1)}} = 2V_0^{\mathrm{(2)}}$ and $V_1^{\mathrm{(1)}}$ keeping the intralayer interaction fixed to $V_1^\mathrm{(0)} = 1$, and setting the total number of particles to $N_\mathrm{e}=12$ with (a) $N_\phi=6$ and (b) $N_\phi=3$ flux quanta. The region in phase space where the ground state is not in the filling-per-layer sector with unit cell $\big[\frac{1}{3}\frac{1}{3}0\big]$ is masked out in gray. Where the ground state is in the desired trimerized sector, we plot the gap above the three-fold degenerate ground state manifold (the gap is taken over all possible filling per layer sectors). Quasi-hole (one added flux) gap difference between foliated Laughlin and foliated Fibonacci phases for (c) $N_\ell=6$, $N_\phi=7$, and (d) $N_\ell=9$, $N_\phi=4$. The red regioncorresponds to the foliated Laughlin phase whereas the blue region corresponds to the foliated Fibonacci phase.
}
\label{fig:6layerPhase}
\end{figure}

We use ED to characterize the possible phases for $N_\ell=6,9$ with periodic boundary conditions. At fixed $N_\mathrm{e}=12$ and $N_\phi=6$ for $N_\ell=6$ and $N_\phi=3$ for $N_\ell=9$, we scan the interaction parameter space defined by $V_0^{(1)}=2V_0^{(2)}$ and $V_1^{(1)}$, while setting the intralayer scale $V_1^{(0)}=1$. The ground state organizes into several distinct regions depending on the competition between intra- and interlayer pseudopotentials. However, we get a clear region in the phase diagram where the gapped $\big[\frac{1}{3}\frac{1}{3}0\big]^{N_\ell/3} $ type of state is favored, as shown in Fig.~\ref{fig:6layerPhase}(a) and (b) for $N_\ell=6$ and $N_\ell=9$, respectively. In the greyed-out region, the ground state is not in the $\big[\frac{1}{3}\frac{1}{3}0\big]^{N_\ell/3}$ sector. For $N_\ell=6$, the main competing state is in the $\big[\frac{2}{9}\frac{2}{9}\frac{2}{9}\big]^{N_\ell/3}$ sector. In the $N_\ell = 9$ system, the trimerized phase occupies a larger region as there is no competing state with homogeneous filling at that total particle number (since $N_\textrm{e}=12$ is not divisible by $N_\ell=9$).

Within the filling-per-layer sector $\big[\frac{1}{3}\frac{1}{3}0\big]^n$ there are two distinct possibilities for the ground state. We can either have a state with $2n$ decoupled Laughlin states $\otimes_{i=1}^{2n} L$ (with an empty layer between every other filled layer), a Laughlin FFQH state, or a state where adjacent filled layers form Fibonacci states leading to $n$ decoupled Fibonacci states $\otimes_{i=1}^{n} \text{Fib}$ (with an empty layer between each filled bilayer), a Fibonacci FFQH state. As both of these states share the same quantum numbers, we turn to their quasihole counting as a probe to distinguish them. In Fig.~\ref{fig:6layerPhase}(c) and~(d) we show the region where the quasihole counting agrees with the Laughlin FFQH state in red and the Fibonacci FFQH state in blue.

\begin{figure*}
    \centering
    \includegraphics[width=\linewidth]{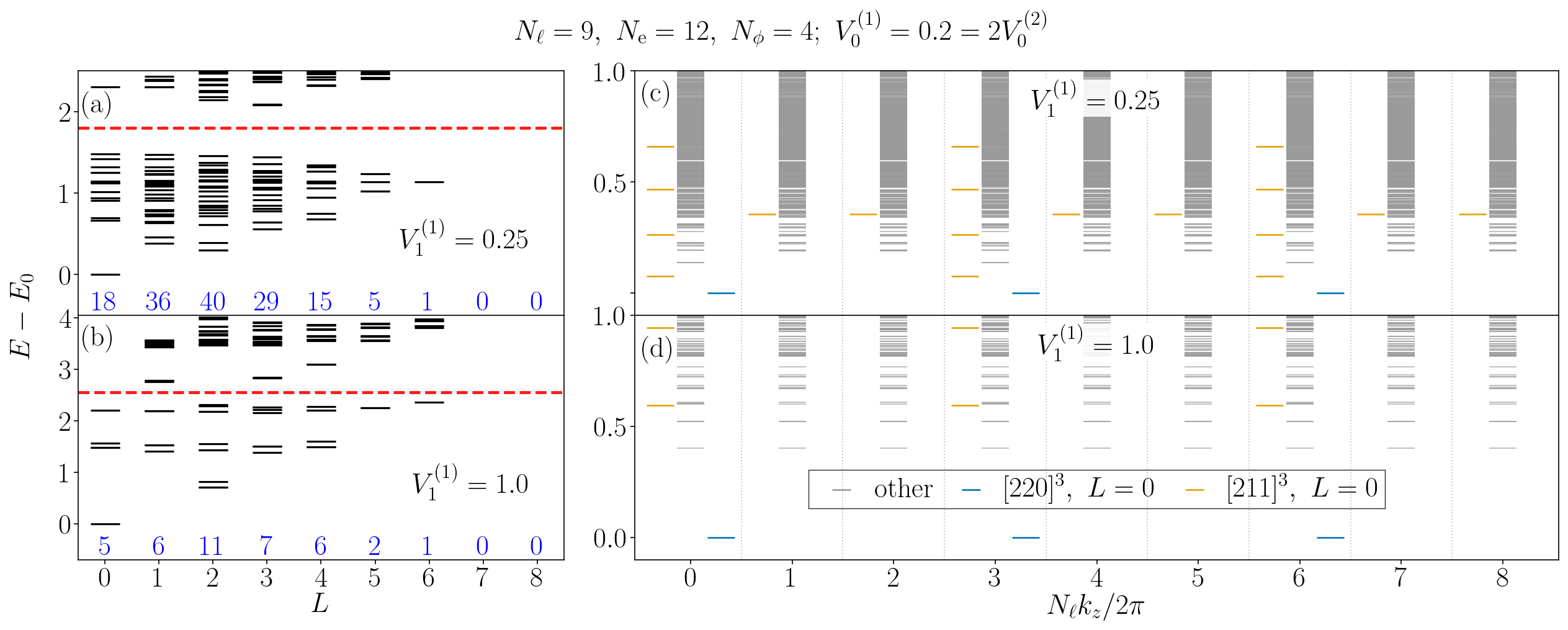}
    \caption{\textbf{Ground state and quasihole energy spectra for foliated Laughlin and foliated Fibonacci states.} $L$-resolved energy spectra for $N_\ell=9$, $N_\mathrm{e}=12$ and $N_\phi=3+1$ in the particle-number-per-layer sector $\big[220\big]^3$, fixing $V_1^{(0)}=1$, $V_0^{(1)}=0.2=2V_0^{(2)}$, (a) $V_1^{(1)}=0.25$ and (b) $V_1^{(1)}=1.0$. The dashed red line indicates the quasihole gap, and the number of states in each $L$-sector which are below the red-line in energy are shown in blue. This counting matches the predicted counting from the Laughlin and Fibonacci generalized exclusion principles. the corresponding $k_z$-resolved energy spectra at $N_\phi = 3$ (ground-state flux sector with shift $\mathcal{S}=3$) are shown in (c) and (d), highlighting the $L_z=0$ states within the $[220]^3$ particle-number-per-layer sector in blue and the $L_z=0$ states within the $[211]^3$ particle-number-per-layer sector in orange, and grouping all other sectors (including other $L_z$ sectors) in gray. The spectrum in (c) corresponds to the foliated Laughlin state, while the spectrum in (d) corresponds to the foliated Fibonacci state, both of which display a three-fold ground state degeneracy due to spontaneous trimerization.}
    \label{fig:spectra_9layer}
\end{figure*}

The quasihole counting for FFQH states can be computed by adding the angular momenta of the individual sublayers, which each obey their respective sublayer exclusion principles. For example, a Laughlin state with $N_\mathrm{e}=2$ particles and $N_\phi = 3+1$ (one additional flux quantum above the $\mathcal{S}=3$ ground state) has a $D_1$ multiplet quasihole manifold. Consequently, stacking 6 such layers leads to a quasihole manifold that decomposes into $\otimes_{i=1}^6(D_1)=D_6 \oplus 5D_5 \oplus 15D_4\oplus29D_3\oplus40D_2\oplus36D_1\oplus15D_0$. This matches the counting in Fig.~\ref{fig:spectra_9layer}(a) up to the $L=0$ sector where finite-size effects start to enter. Moreover, a bilayer Fibonacci state with $N_\mathrm{e}=4$ particles and $N_\phi = 3+1$, the quasihole manifold decomposes into $D_0 \oplus D_2$. Thus stacking three such bilayers, the total quasihole manifold decomposes as $\otimes_{i=1}^3(D_0 \oplus D_2) = D_6\oplus 2D_5\oplus6D_4\oplus7D_3\oplus11D_2\oplus6D_1\oplus5D_0$. This perfectly matches the counting in Fig.~\ref{fig:spectra_9layer}(b), where we have $N_\ell = 9$ layers with filling $\big[\frac{1}{3} \frac{1}{3} 0\big]^3$ in each layer, and thus effectively a stack of three Fibonacci bilayers. Using this quasihole counting, we can extract the quasihole gap in the foliated Laughlin and foliated Fibonacci phases shown in Fig.~\ref{fig:6layerPhase}(c) and~(d). 

In Fig.~\ref{fig:spectra_9layer}(c) and (d) we plot the momentum-resolved energy spectrum for a representative point in the foliated Laughlin and foliated Fibonacci phase in the nine layer system. In both cases we observe a three-fold degenerate ground state in the spherical geometry, which is a signature of the layer translational symmetry breaking. 

To conclude this section, let us discuss the implications for the thermodynamic limit $N_\ell\to\infty$. It is definitely the case that within the $\big[\frac{1}{3}\frac{1}{3}0\big]^{\nl/3}$ sector, there will be a transition between the foliated Fibonacci and the foliated Laughlin state. This is particularly clear since the buffer layers allow the coupling between the occupied pairs to be made arbitrarily weak, in which case we recover the bilayer scenario studied in Sec.~\ref{sec:bilayer}. Furthermore the presence of a region with $\big[\frac{1}{3}\frac{1}{3}0\big]^{\nl/3}$ charge order appears to be robust. On the other hand, whether the foliated Laughlin state itself can be realized within the $\big[\frac{1}{3}\frac{1}{3}0\big]^{\nl/3}$ sector in the thermodynamic limit [i.e.~whether there is an overlap of the red region of Fig.~\ref{fig:6layerPhase}(c) and (d) with the colored regions in Fig.~\ref{fig:6layerPhase}(a) and (b)] cannot be deduced conclusively. The evidence for the foliated Fibonacci state in the thermodynamic limit is stronger [i.e.~whether there is an overlap of the blue region of Fig.~\ref{fig:6layerPhase}(c) and (d) with the colored regions in Fig.~\ref{fig:6layerPhase}(a) and (b)]. Finally, we note that there are other competing configurations with different numbers of particles per layer, such as alternating filled and empty layers in the limit of large $V^{(1)}$ compared to $V^{(2)}$. We discuss this in App.~\ref{app:multilayer}. 

\section{Experimental realizations}
\label{sec:experiments}

To realize FFQH phases, two ingredients are required. Firstly, one requires a material platform which can host $N_\ell$ layers of quantum Hall fluid. Secondly, one needs the right type of interaction to stabilize the FFQH phase, ideally with some tunability.

\subsection{Material platform}

Different material platforms exist for realizing quantum Hall multilayers. The quantum Hall bilayer was originally experimentally studied in GaAs quantum wells \cite{Kellogg2004VanishingHall,Tutuc2004Counterflow,Eisenstein1990IndependentContacts,Eisenstein1992NewFQH,Spielman2000ResonantTunneling,Eisenstein2004}. Even triple layer quantum wells have been experimentally realized in a few cases \cite{Jo1992,LAY1996171,Lay1995}. However, larger numbers of layers remain unexplored. More recently, experiments have focused on double-layer graphene, where two graphene monolayers are separated by an insulating layer of hexagonal boron nitride \cite{zhang2024excitonsfractionalquantumhall,grapheneExp1,grapheneExp2,grapheneExp3}. This is distinct from bilayer graphene, where the graphene monolayers are directly stacked. 

In any 2D setting, the Landau level filling can be controlled by two knobs: the magnetic field strength and via gating, giving rise to Landau fan diagrams. (Nonetheless, the effects of gates for graphene multilayers can be highly non-trivial, see e.g.~Ref.~\onlinecite{kolar2025singlegatetrackingbehaviorflatband}.) In a 3D geometry relevant to FFQH states, gate tuning is not viable and one has to resort to intrinsic doping instead. Sweeping the magnetic field will thus produce a one-dimensional cut through the Landau fan diagram, which, at least in principle, offers enough control to dial into a desired Landau level filling per layer.

One route to obtain stacks of 2D electron gases are artificially designed heterostructures, where large layer separations can be achieved by including enough layers of a buffer material (such as hexagonal boron nitride) between the active layers (which should be low-doped semiconductors). This is suitable for decoupled foliated Laughlin states, but not for foliated Fibonacci states.  As-grown layered 3D materials are the preferable, more generic platform. Misfit-compounds with transition metal dichalcogenide as active layers and rocksalt `spacer' layers can be suitable candidates~\cite{Wiegers1996,
Rouxel1995,
Ng2022APL,
Atkins2014SnSeMoSe,
Zullo2023}. Recent advances in microwave imaging allow 3D imaging of van der Waals heterostructures which may help in experimentally detecting FFQH phases \cite{cao20253dmicrowaveimagingvan}.

Besides physically layered systems, FFQH physics could also be studied in multicomponent quantum Hall systems. Spin or pseudospin can lead to an SU(2) degree of freedom \cite{Abanin2010} and in the case of graphene, spin and valley provide an SU(4) degree of freedom. A distinct way to engineer a multicomponent quantum Hall system is synthetic bilayer graphene, where coherent driving is used to couple two Landau levels in a single sheet of graphene \cite{Ghazaryan2017}. 

\subsection{Form of interaction}

Once the material platform is chosen, the challenge remains to tune the interaction to a regime where the FFQH state is predicted to occur. In realistic cases, the most relevant interaction is the Coulomb interaction. Usually, for two-dimensional systems the gate screened Coulomb interaction is used. However, in the large $N_\ell$ limit the effect of the gate screening will be negligible and thus we neglect it in the following discussion. Assuming the layers are equally spaced with interlayer separation $d$, the two-dimensional Coulomb interaction between $n$th NN layers will take the form
\begin{equation}
V^{(n)}(q)=\frac{e^{-qnd}}{q}.
\label{eq:Coul}
\end{equation}
The characteristic length scale in quantum Hall problems with magnetic field $B$ is the magnetic length $l_B=\sqrt{\frac{\hbar}{eB}}$. In the lowest Landau level we then have 
\begin{equation}
V_m^{(n)}=\int \frac{d^2 q}{(2\pi)^2}\,V^{(n)}(q)\,L_m(q^2 l_B^2)\,e^{-q^2 l_B^2},
\end{equation}
where $L_m(x)$ are the Laguerre polynomials. The relevant dimensionless parameter controlling the Haldane pseudopotentials is $d/l_B$. The effect of increasing $d/l_B$ is to make the pseudopotentials $V_m^{(n)}$ a faster decaying function of $m$ and of $n$. 

The first condition we need for the realization of the foliated Fibonacci state is $V_1^{(1)}\sim V_1^{(0)}$. Hence we require $d\lesssim l_B$. At a typical magnetic field of 10T the magnetic length is around 8nm. The second condition we need to realize the foliated Fibonacci state is $V_1^{(1)}\gtrsim V_0^{(1)}$. There are several ways to change the ratio of $V_1^{(1)}/V_0^{(1)}$: (1) Considering the effect of Landau level mixing \cite{PhysRevB.87.245425,PhysRevB.80.121302},
(2) considering electrons in higher Landau Levels with a linear Dirac dispersion instead of a quadratic dispersion \cite{PhysRevLett.97.126801,PhysRevB.74.161407,PhysRevB.74.235417}, 
(3) considering the Rytova-Keldysh corrections to the interaction potential~\cite{rytova2020screenedpotentialpointcharge,Keldysh1979}. The last mentioned point, the Rytova-Keldysh interaction, can be used to model the finite layer thickness of materials. The question of exactly how large one can tune $V_1^{(1)}/V_0^{(1)}$ (ideally above 1) with the aforementioned methods remains an interesting avenue for future investigations. 

Still, $d/l_B$ gives only a single tuning parameter, while two tuning parameters are required in order to reach a precise point in our two-dimensional phase diagrams. To achieve further tunability, dielectric engineering can be used to modify the Coulomb interaction, so as to modify the ratio of certain pseudopotentials~\cite{kwan2024abelianfractionaltopologicalinsulators}. For example, using a spacer layer with specific dielectric properties will modify the interaction between the layers separated by the spacer. 

In this work we have shown ED results for short-range pseudopotential interactions. Studying the stability of the phases we have found when including a long-range tail of the interaction in Eq.~\eqref{eq:Coul} is an area which deserves future study. However, it is also possible to study synthetic systems in which the interaction takes precisely the short-range form studied in this work. In particular, this type of interaction can be engineered in synthetic bilayer graphene \cite{Ghazaryan2017}, potentially allowing for the realization of the bilayer Fibonacci state \cite{Cian2020}.

\section{Conclusion}

We have studied the $\nu=1/3$ quantum Hall system in the thermodynamic limit of many layers, $N_\ell\to\infty$. Going to this limit is what enables the qualitatively new physics we focus on: interactions can spontaneously break layer translation symmetry, enlarging the periodicity along the stacking direction $r$-fold. We focus on energetics to elucidate which of the many theoretically possible states could be realistically realized in such systems.

Concretely, we find a phase with tripled unit cell in which two consecutive layers at $\nu=1/3$ are followed by an empty layer. This produces an effective $\frac13+\frac13$ bilayer system, weakly decoupled from its neighbors by the empty layer interposed. Depending on the intra-block interactions, each block can host either decoupled Laughlin states or a Fibonacci state, and using exact diagonalization we mapped out the phase transition between these two possibilities. While prior theoretical work has largely focused on topological characterization of three-dimensional layered phases, here we have explicitly addressed the energetics required for their physical realization. 

This setup is the simplest instance of a larger family of conceivable foliated states in which the unit cell consists of $r$ layers, $p$ of which are filled at filling $\nu$ and the remaining $r-p$ are empty. A natural open question already at the level of $p=2$ is whether the foliated Fibonacci state can be stabilized energetically without intermediate buffer layers. We have formulated the corresponding non-Abelian Ginzburg--Landau theory in App.~\ref{sec:NA_GL}, but finding a microscopic Hamiltonian that realizes it as a ground state remains an open challenge.

Accessing larger $p$ requires understanding the few-layer building blocks. Using quasihole counting we have shown that, for suitable interactions, the $\frac13+\frac13+\frac13$ trilayer can host $\mathcal{C}_{N_\ell=3}$ topological order and the $\frac13+\frac13+\frac13+\frac13$ tetralayer can host $\mathcal{C}_{N_\ell=4}$ topological order. These states are the natural building blocks for $p=3$ and $p=4$ foliated phases at large $N_\ell$.

Studying layered systems with such large building blocks is, however, outside the reach of exact diagonalization. Future work will need to resort to other numerical methods. One promising avenue is variational Monte Carlo starting from the wavefunctions we have introduced for the $\mathrm{SU}(3)_\nl$ states. Once correlations are introduced between the $\mathrm{SU}(3)_\nl$ building blocks with suitable variational parameters, the competition with e.g.~the decoupled Laughlin state can be studied. Modern quantum Hall Monte Carlo can access hundreds of electrons~\cite{Gattu2025,anakru2025exploringnatureemergentgauge}, the regime needed to study large-$N_\ell$ foliated phases. The density-matrix renormalization group is another option and has been applied successfully to multicomponent quantum Hall systems~\cite{Zaletel2015infinite}, although it is already expensive in the bilayer case and may not reach significantly larger $N_\ell$ than exact diagonalization. Finally, infinite projected entangled-pair states, recently applied to chiral fermionic phases~\cite{chen2026simulatingfermionicfractionalchern}, may provide a viable alternative.

Our study maps out in which regime (parametrized in Haldane pseudopotentials) the simplest FFQH phases can be found. It calls for future work that connects these design principles to compositional and external control parameters in material systems in order to realize these specimens of fracton topological order.

\begin{acknowledgements}
Most exact diagonalization calculations were performed using DiagHam \cite{Regnault2001DiagHam}. For the $k_z$-resolved spectra, QuSpin was used \cite{QuSpin1,QuSpin2}. D.M.~is supported by SNSF Grant No.~219339. G.W.~is supported by the Swiss National Science Foundation (SNSF) via Ambizione grant number PZ00P2-216183. This work was supported by the Swiss National Science Foundation through a Consolidator Grant (iTQC, TMCG-2\_213805), a Quantum grant (20QU-1\_225225). The Flatiron Institute is a division of the Simons Foundation.
\end{acknowledgements}

\bibliography{refs}

\clearpage
\newpage

\begin{appendix}
\onecolumngrid
	\begin{center}
		\textbf{\large --- Supplementary Material ---\\Three-dimensional Foliated Fractional Quantum Hall Phases}\\
		\medskip
		\text{Sahana Das, Navketan Batra, Andrea Kouta Dagnino, Dan Mao, Nicolas Regnault, Glenn Wagner, and Titus Neupert}
	\end{center}
	
	\setcounter{equation}{0}
	\setcounter{figure}{0}
	\setcounter{table}{0}
	\makeatletter
    \@removefromreset{equation}{section}
    \makeatother
	\renewcommand{\theequation}{S\arabic{equation}}
	\renewcommand{\thefigure}{S\arabic{figure}}
    \renewcommand{\thetable}{S\arabic{table}}
	\renewcommand{\bibnumfmt}[1]{[S#1]}

\setcounter{equation}{0}
\setcounter{figure}{0}
\setcounter{table}{0}

\renewcommand{\theequation}{S\arabic{equation}}
\renewcommand{\thefigure}{S\arabic{figure}}
\renewcommand{\thetable}{S\arabic{table}}

\renewcommand{\theHequation}{S.\arabic{equation}}
\renewcommand{\theHfigure}{S.\arabic{figure}}
\renewcommand{\theHtable}{S.\arabic{table}}

\section{Abelian states}\label{app:abelianstates}

\subsection{Decoupled Laughlin state and Halperin~(112) state}

The long-wavelength physics of Abelian quantum Hall states can be described in terms of a symmetric, integer valued matrix $K_{IJ}$ and a charge vector $t_I$ \cite{WenZee1990NuclPhysSup, Read1990PRL, BlokWen1990PRB8133, BlokWen1990PRB8145, FrohlichZee1991NPB, WenZee1992PRB, Wen1995AdvPhys}. The Chern-Simons action in terms of the emergent $\text{U(1)}$ gauge fields $a^I_\mu$, with $I=1,\dots,N_{\ell}$ labelling the layers, reads
\begin{align}
    S_{\mathrm{CS}} [\{a\}]=& \frac{1}{4\pi} \int d^3x \
K_{IJ}\ \epsilon^{\mu\nu\lambda}
a^I_\mu \partial_\nu a^J_\lambda-\frac{e}{2\pi}\int d^3x \
t_I\ \epsilon^{\mu\nu\lambda}
A_\mu \partial_\nu a^I_\lambda,
\end{align}
where $A_\mu$ is the external electromagnetic gauge field. Physically, the $K$ matrix can be thought of as prescribing a flux attachment procedure. In the multilayer case the diagonal $K$-matrix entries specify the number of intralayer fluxes attached to each electron, while the off-diagonal entries specify the number of interlayer fluxes. According to the plasma analogy \cite{Laughlin1983}, we require all eigenvalues of the $K$ matrix to be positive in order to obtain a stable state \cite{DeGail2008}.   

The Halperin~$(m_1m_2n)$ states were originally introduced in order to describe two-component systems, where the two components could correspond to a layer degree of freedom in a quantum Hall bilayer, or it could equally well describe a spin or valley degree of freedom \cite{Halperin1983}. The $K$ matrix for the $(m_1m_2n)$ state is
\begin{equation}
    K = 
\begin{pmatrix}
m_1&n \\
n&m_2
\end{pmatrix}
\end{equation}
with charge vector $t=(1,1)^{\mathsf{T}}$. Assuming the two layers are equivalent, we have $m_1=m_2\equiv m$. The filling factors of the two layers are then $\nu_i=(K^{-1}t)_i=1/(m+n)$. The shift on the sphere is $\mathcal{S}=m$ and the ground state degeneracy on the torus is $|\textrm{det}(K)|=|m^2-n^2|$. The wavefunction on the disk is 
\begin{align}
\Psi_{mmn}&(\{z_i\},\{w_i\})=\prod_{i>j}(z_i-z_j)^m\prod_{i> j}(w_i-w_j)^m\prod_{i, j}(z_i-w_j)^n,
\end{align}
where $z_i$ and $w_i$ are complex electron coordinates in the top and bottom layer respectively. $m$ must be an odd integer in order to satisfy fermionic antisymmetry. For the filling factor $\nu_1=\nu_2=1/3$ this gives two solutions. The Halperin~(330) state is trivial and corresponds to decoupled Laughlin states in the two layers, while the (112) state introduces correlations between the two layers.
The Halperin~(112) state is an exact zero energy state in the case where $V_0^\mathrm{(1)},V_1^\mathrm{(1)}>0$ and all other pseudopotentials are zero. However we should point out that this is a Halperin~$(mmn)$ state with $n>m$, meaning that it has the property that interlayer repulsion is stronger than intralayer repulsion and so this wavefunction describes a phase separated state \cite{DeGail2008}. We work in the particle number sector where the two layers are equally populated and interlayer tunnelling is absent. The phase separation is an indication that it is energetically favourable to empty one layer and layer-polarize the system (in which case the system avoids the energy penalty from the large $V_0^\mathrm{(1)}$). We also discuss this point in Sec.~\ref{sec:translation_symmetry_breaking}.

The wavefunction of the (112) state can be modified to obtain a spin singlet state which has the same $K$ matrix \cite{wu1993mixed,MacDonald1996,Peterson2015}. 
The $K$ matrices of the Halperin~(112) state and of the layer polarized 2/3 state (which is the particle-hole transform of the Laughlin 1/3 state) are related by an SL$(2,\mathbb{Z})$ transformation; however, the two states have a different shift and hence describe different phases on the sphere \cite{Geraedts2015}.

\subsection{Generalized Halperin states}
\label{app:genHalperin}

For the large $V^{(1)}_0$ states, we can write down the Halperin~(112) state for $N_\ell=2$ layers as
\begin{equation}
K = 
\begin{pmatrix}
1&2 \\
2&1
\end{pmatrix}.
\end{equation}
For $N_\ell=3$ layer system with each layer at filling $\nu=1/3$, we note that the system is at a total filling of $\nu_{\rm total}=1$. To generalize, we can write down a $K$-matrix description of this state motivated by the bilayer Halperin~$(111)$ state as, 
\begin{equation}\label{eq:HalperinGen_3layers}
K = 
\begin{pmatrix}
1&1&1 \\
1&1&1 \\
1&1&1
\end{pmatrix} ~~~~~~ \text{with}~~~~~~ t=\begin{pmatrix}
    1\\1\\1
\end{pmatrix}.
\end{equation}
Indeed, it is a minimal correlation state where each electron sees a single flux from its neighboring layers. Its similarities with the bilayer Halperin~(111) state has interesting consequences for the $N_{\ell}=3$ layer case that are discussed below. Fixing the filling of each layer at $\nu=1/3$, and following the same intuition of attaching single flux with the neighboring layers gives us a general structure of the $K$-matrices,
\begin{align}
N_{\ell}=4 \text{ layers}:~~~~~~~~~~~K = 
\begin{pmatrix}
1&1&0&1 \\
1&1&1&0 \\
0&1&1&1\\
1&0&1&1
\end{pmatrix} ~~~~~~ &\text{with}~~~~~~ t=\begin{pmatrix}
    1\\1\\1\\1
\end{pmatrix},\\ 
N_{\ell}=5 \text{ layers}:~~~~~~~~ K = 
\begin{pmatrix}
1&1&0&0&1 \\
1&1&1&0&0 \\
0&1&1&1&0\\
0&0&1&1&1\\
1&0&0&1&1
\end{pmatrix}~~~~~~ &\text{with}~~~~~~ t=\begin{pmatrix}
    1\\1\\1\\1\\1
\end{pmatrix}.
\end{align}
where we incorporated periodic boundary condition in the layer direction for systems with $N_{\ell}\ge 3$. We have assumed that the interlayer interactions fall off with the separation between the layers, such that the interaction between nearest neighbour layers is largest. As such, it is the most energetically favourable to attach the fluxes to the neighbouring layers. Other valid $K$-matrices at the same filling factor are possible, however they are energetically less favourable. 
For a general $N_{\ell}$, the $K$ matrix takes the form of a circulant matrix, $K_{ij}^{(N_\ell)} = \delta_{ij}
+ \delta_{i,j+1 \,(\mathrm{mod}\ N_\ell)}
+ \delta_{i,j-1 \,(\mathrm{mod}\ N_\ell)}$. 
In particular, computing the eigenvalues we can show  
\begin{equation}
\det K^{(N_\ell)} =
\prod_{m=0}^{N_\ell-1} \left[1 + 2\cos \left(\frac{2\pi m}{N_\ell}\right)\right]
=
\begin{cases}
0, & N_\ell = 0 \pmod{3}\\[6pt]
3(-1)^{N_\ell+1}, & N_\ell \not = 0 \pmod{3}
\end{cases}
\end{equation}
and ground state degeneracy (GSD) on the torus is $|\textrm{det}(K)|$, except if $\textrm{det}(K)=0$, in which case the GSD is 1. We also have $\mathcal{S}=1$. Similar to the Halperin state for the quantum Hall bilayer, where we have an exciton condensate, this generalized Halperin state can be understood in terms of \emph{trion} formation. For the $N_\ell$-layer quantum Hall $K$-matrix
\begin{equation}
K_{ij}^{(N_\ell)} = \delta_{ij} + \delta_{i,j+1 \,(\mathrm{mod}\, N_\ell)} + \delta_{i,j-1 \,(\mathrm{mod}\, N_\ell)},
\end{equation}
the eigenvalues are
\begin{equation}
\lambda_k = 1 + 2 \cos\left(\frac{2 \pi k}{N_\ell}\right), \quad k = 0, 1, \dots, N_\ell-1.
\end{equation}
The corresponding normalized eigenvectors are
\begin{equation}
v^{(k)} = \frac{1}{\sqrt{N_\ell}}
\begin{pmatrix}
1 \\
e^{2 \pi i k / N_\ell} \\
e^{2 \pi i 2k / N_\ell} \\
\vdots \\
e^{2 \pi i (N_\ell-1) k / N_\ell}
\end{pmatrix}, \quad k = 0, 1, \dots, N_\ell-1.
\end{equation}
We can thus interpret $k$ as $k_z$. The fact that we have negative eigenvalues once again shows a tendency towards phase separation, which is a sign that it is energetically favourable to depopulate some layers in favour of other layers. The instability is towards $k_z=1/3$ layer-CDW.  We therefore need to go study translational symmetry breaking in the $z$-direction in more detail. 

We now address the specific case when $N_{\ell}\mod 3 =0$, i.e., the $K$ matrix is singular. From the eigenspectrum, the vanishing eigenvalues occur at $k=N_{\ell}/3$ and $k=2N_{\ell}/3$, so there are exactly two zero modes for every $N_{\ell}$ system. This is reminiscent of the bilayer situation described by the Halperin~$(111)$ state, where the system shows quantized Hall response in the charge sector and alongside has gapless neutral modes (density fluctuations that preserve total charge). At the level of an effective field theoretic description, a singular $K$ matrix signals the presence of zero-energy modes that are not controlled by the Chern-Simons term, i.e., they are gapless at the Gaussian level because no topological mass term appears. Note that in real systems, however, these modes can acquire a gap depending on the microscopic details which are not captured by the $K$-matrix alone.

Since we have exactly two zero eigenvalues, we can transform this singular $K$-matrix into a block diagonal structure with the null space explicitly factored out. The two-dimensional null space of $K^{(N_{\ell})}$ admits the integer basis,
\begin{align}
    n^{(1)}=(1,-1,0,1,-1,0,\dots)^{\mathsf{T}},~~~~~~n^{(2)}=(0,1,-1,0,1,-1,\dots)^{\mathsf{T}}
\end{align}
One can check that $Kn^{(a)}=0$ directly since every row of $K$ sums three consecutive entries. These vectors are primitive and can be completed to a basis of $\mathbb{Z}^{N_{\ell}}$ by a unimodular matrix $W\in GL(N_{\ell},\mathbb{Z})$. In this basis,
\begin{align}
   K\rightarrow K' = W^{\mathsf{T}}KW = \begin{pmatrix}
        0_{2\times 2} & 0_{2\times (N_{\ell}-2)} \\
        0_{(N_{\ell}-2)\times2} & K_{\text{non-sing.}}
    \end{pmatrix}~~~~~\text{with}~~~~t\rightarrow t'=W^{\mathsf{T}}t=\begin{pmatrix}
        0\\ t_{\text{non-sing.}}
    \end{pmatrix}
\end{align}
where $0_{a\times b}$ is simply a null matrix of dimensions $a\times b$. The matrix $K_{\text{non-sing.}}$ is non-singular, and therefore describes a gapped charged sector. The presence of the gapless neutral sector is indicated by the structure of the transformed charged vector $t'$. Specific to the case of $N_{\ell}=3$, this representation is equivalent to the layer polarized interpretation of the $\nu_{\rm total}=1$ state as was expected from the $K$-matrix in Eq.~(\ref{eq:HalperinGen_3layers}). When generalized to more layers (with $N_{\ell} \mod 3$), the layer polarization interpretation only holds in the sense that two of the $N_{\ell}$ layers can be thought of as being `empty' while the gapped charged sector of the $N_{\ell}$ layer FQH state sits in $N_{\ell}-2$ layers.

\section{Competing states in bilayer at $\nu=2/3$}\label{app:competingstates_2/3}
We list bilayer fractional quantum Hall phases with a total filling fraction $2/3$ proposed in previous literature in Table \ref{tab:FQHphases}.
\begin{table}[H]
\centering
\begin{tabular}{cccc}
\hline\hline
FQH Phase & \qquad GSD \qquad \quad & \qquad $\mathcal{S}$ \qquad \quad & \qquad numerically found in \qquad \\
\hline
(330) & 9 & 3 & \cite{Liu2015} \\
(112) & 3 & 1 & \cite{Geraedts2015,Peterson2015}\\
Jain 2/3 & 3 & 0 & \cite{Geraedts2015,Peterson2015}\\
$\mathbb{Z}_4$ Read--Rezayi & 15 & 3 & \cite{Peterson2015} \\
Interlayer Pfaffian/paired spin singlet (PSS) & 9 & 3 & \cite{Geraedts2015} \\
Bonderson--Slingerland & 9 & 4&- \\
Intralayer Pfaffian & 27 & 3& \cite{Peterson2015}\\
Bilayer Fibonacci & 6 & 3 &\cite{Liu2015} \\
\hline\hline
\end{tabular}
\caption{Ground-state degeneracy on the torus and shift $\mathcal{S}$ on the sphere for various candidate
fractional quantum Hall phases at $\nu = 2/3$ in a bilayer system.  The (112)
and Jain 2/3 phases are actually the same phase, in the sense that
their $K$ matrices are related by an SL(2,$\mathbb{Z}$) transformation, but they have a different shift on the sphere.}
\label{tab:FQHphases}
\end{table}

\section{CFT construction of model wavefunctions}\label{app:modelWF}

In this section we provide a detailed construction of the model wavefunctions presented in Sec.~\ref{sec:modelWF} for several FQH phases, based on the underlying conformal field theory living on their boundary edge \cite{MOORE1991362}. An introduction to CFT methods applied to FQH physics can be found in Ref.~\onlinecite{hansson2017quantum}. 
\subsection{Bilayer Fibonacci}
The edge theory of the $\text{SU(3)}_2$ Fibonacci state at $\nu=\frac{2}{3}$ is the coset theory $\text{U(6)}_1/\text{SU(3)}_2$ \cite{PhysRevLett.113.236804}. The $\text{U(6)}_1$ comes from the 6 complex fermions arising from the parton decomposition of the bare electron operator in each layer, and the $\text{SU(3)}_2$ coset comes from projecting out the unphysical degrees of freedom after Higgs-ing the difference of the $\text{SU(3)}_1$ gauge fields in the two layers. As noted in Ref.~\onlinecite{Vaezi2014}, this theory can be re-expressed as $\text{SU(2)}_3 \times \text{U(1)}_6$ by using the level-rank duality $\text{U}(nk)_1 = \text{SU}(n)_k \times \text{SU}(k)_n \times \text{\text{U(1)}}_{nk}$. However, the bilayer has two $\text{U(1)}$ symmetries, corresponding to conservation of charge and layer pseudo-spin. To make the symmetry apparent, we quotient a $\text{U(1)}_6$ boson from $\text{SU(2)}_3$ to obtain $\frac{\text{SU(2)}_3}{\text{\text{U(1)}}_6} \times \text{\text{U(1)}}_6 \times \text{\text{U(1)}}_6$. The first theory is the celebrated $\mathbb{Z}_3$ parafermion theory, which is known to describe the critical 3-state Potts model, and is the neutral sector of the edge theory of the $k=3$ Read-Rezayi state. The conformal weight and fusion rules of the primaries in this theory are shown in Table~\ref{tab:Z3Pf}.

\begin{table}[h!]
    \centering
    \begin{tabular}{c||c|c|c|c|c}
        $h$ & $2/3$ & $2/3$ & $1/15$ & $1/15$ & $2/5$\\
        \hline
        $\times$ & $\psi_1$ & $\psi_2$ & $\sigma_1$ & $\sigma_2$ & $\epsilon$ \\
        \hline \hline
        $\psi_1$ & $\psi_2$ & & & &  \\
        \hline
        $\psi_2$ & $1$ & $\psi_1$ & & &  \\
        \hline
        $\sigma_1 $ & $\epsilon$ & $\sigma_2$ & $\sigma_2 + \psi_1$ & &  \\
        \hline
        $\sigma_2$ & $\sigma_1$ & $\epsilon$ & $1+\epsilon$ & $\sigma_1+\psi_2$  &\\
        \hline
        $\epsilon$ & $\sigma_2$ & $\sigma_1$ & $\sigma_1+\psi_2$ & 
        $\sigma_2+\psi_1$ & $1+\epsilon$
    \end{tabular}
    \caption{Conformal weight and fusion rules of the chiral primaries in $\text{SU(2)}_3/\text{U(1)}_6$.}
    \label{tab:Z3Pf}
\end{table}
We list the operator product expansions (OPEs) for the parafermionic fields $\psi_1, \psi_2$ below for future reference
\begin{align}\label{OPE}
    &\psi_1(z) \psi_1(w) \sim (z-w)^{-2/3}\Big[\psi_2(w)+\frac{1}{2}(z-w)\partial \psi_2(w)+...\Big]\\
    &\psi_2(z) \psi_2(w) \sim (z-w)^{-2/3}\Big[\psi_1(w)+\frac{1}{2}(z-w)\partial \psi_1(w)+...\Big]\\
    &\psi_1(z) \psi_2(w) \sim (z-w)^{-4/3} \Big[1+\frac{5}{3}(z-w)^2T(w)+...\Big]
\end{align}
where $T(w)$ is the stress-energy tensor of the theory. The two $\text{U(1)}_6$ bosons $\phi_c$ and $\phi_s$ will describe the charge and spin degree of freedom in the system. The primaries are the vertex operators $V_{\alpha,\beta}=:e^{i(\alpha\phi_c+\beta\phi_s)/\sqrt{6}}:$ with $\alpha,\beta\in \mathbb{Z}_6$ and conformal weight $h=\frac{\alpha^2+\beta^2}{12}$ and fusion rules $V_{\alpha,\beta} \times V_{\alpha',\beta'} = V_{\alpha+\alpha',\beta+\beta'}$ where $\alpha+\alpha'$ and $\beta+\beta'$ are defined modulo 6. We proceed by constructing electron operators for each layer. Firstly, we note that $\epsilon, \sigma_1, \sigma_2$ are not simple currents and thus not suitable to construct an electron operator. Even disregarding this, their conformal weight does not allow their combination with a vertex operator to form a fermionic operator with half-odd integer conformal weight. Moreover the electron operators should contain a non-trivial field from the neutral sector for them to have any chance of describing a non-Abelian fractional quantum Hall state and not some Abelian topological order. We are thus left with the parafermionic fields $\psi_1$ and $\psi_2$, from which we can construct two simple current operators:
\begin{align}
    &\Psi_\uparrow(z) = \psi_1(z) V_{3,1}(z) = \psi_1(z) :e^{i(3\phi_c(z)+\phi_s(z))/\sqrt{6}}:\\
    &\Psi_\downarrow(z) = \psi_2(z) V_{3,-1}(z)= \psi_2(z) :e^{i(3\phi_c(z)-\phi_s(z))/\sqrt{6}}:
\end{align}
with conformal weight $h=\frac{3}{2}$. We may then construct a bilayer wavefunction for $N_\mathrm{e}=n+n$ particles by evaluating the following CFT correlator:
\begin{equation}
    \Psi(z_1,...,z_n;w_1,...,w_n) = \langle \Psi_\uparrow (z_1)...\Psi_\uparrow (z_n) \Psi_\downarrow (w_1)...\Psi_\downarrow (w_n) \mathcal{O}_\text{bg}(\infty)\rangle 
\end{equation}
where $z_i$ and $w_i$ are the coordinates of the particles in the top and bottom layer respectively. Here $\mathcal{O}_\text{bg}(\infty)$ is a background charge operator that ensures charge neutrality inside this correlator, its precise form is not important for our purposes as its contributions can be fixed by normalizing  the wavefunction on the relevant geometry (sphere/disk). This correlator splits into the product of a correlator in the parafermionic theory and a correlator in the bosonic theory. The latter amounts to some Jastrow factors and one is left with
\begin{equation}
    \Psi(z_1,...,z_n;w_1,...,w_n) = \langle \psi_1 (z_1)...\psi_1(z_n)\psi_2 (w_1)...\psi_2 (w_n)\rangle \times \prod_{i<j}(z_i-z_j)^{5/3}(w_i-w_j)^{5/3}\times \prod_{ij}(z_i-w_j)^{4/3}
\end{equation}
We see that the maximum power of any coordinate (say $z_1$) in this wavefunction is $N_\phi = \frac{5}{3}(n-1) + \frac{4}{3}n - \frac{4}{3}$ where the first two terms come from the Jastrow factors, and the last term comes the parafermionic correlator. More specifically, we take all other coordinates to $\infty$, and fuse their respective fields using the OPEs, yielding $f(z_2,w_2,...z_n,w_n)\psi_2(w_1)$ where $f$ is some holomorphic function that crucially does not depend on $z_1$. We are thus left with the OPE of $\psi_1(z_1)\psi_2(w_1) \sim (z_1-w_1)^{-4/3}+...$ as desired. We thus see that $N_\phi = \frac{3}{2}N_\mathrm{e} - 3$ indicating a total filling $\nu=\frac{2}{3}$ and shift $\mathcal{S}=3$ as expected for the bilayer Fibonacci state. While this is already promising, we want to be able to compute the parafermionic correlator in order to compare it with our ground states from exact diagonalization. Thus, we perform a point-splitting regularization
\begin{equation}
     \langle \psi_1 (z_1)...\psi_1(z_n)\psi_2 (w_1)...\psi_2 (w_n)\rangle  =  \lim_{w_i^\pm \rightarrow w_i}  \bigg[\langle \psi_1 (z_1)...\psi_1(z_n)\psi_1 (w_1^+) \psi_1(w_1^-)...\psi_1(w_n^+) \psi_1(w_n^-)\rangle\prod_i (w^+_i-w^-_i)^{\frac{2}{3}}\bigg] 
\end{equation}
where we substituted $\lim_{w^\pm \rightarrow w}\psi_1(w^+) \psi_1(w^-) (w^+-w^-)^{2/3}=\lim_{w^\pm \rightarrow w}\psi_2(w^-) [1 + o(w_1^+-w_1^-)] = \psi_2(w)$. The additional terms in $o(w_1^+-w_1^-)$ come from the higher order terms of the OPE, and they can be safely neglected in the $w^\pm \rightarrow w$ limit. We are left with the task of computing the $3n$-point correlation function of the parafermionic field $\psi_1$ (or equivalently $\psi_2$, as there is a permutation symmetry of the two fields in this CFT). This correlator was shown to be equivalent to the $k=3$ Read-Rezayi wavefunction up to a Jastrow factor \cite{Read1999paired,PhysRevB.61.5473}: 
\begin{equation}
    \langle \psi(z_1) ... \psi(z_{3n}) \rangle = \frac{\Psi_{k=3}^\text{RR}(z_1,...,z_{3n})}{\prod_{i<j}(z_i-z_j)^{\frac{2}{3}}}
\end{equation}
The $k=3$ Read-Rezayi state is constructed by splitting the particles into three clusters $A,B,C$ of $n$ particles each, and evaluating the following expression
\begin{equation}
    \Psi_{k=3}^\text{RR}(z_1,...,z_{3n})=\mathcal{S}\bigg[\prod_{\substack{i<j\\p=A,B,C}} (z_i^p-z_j^p)^2\bigg]
\end{equation}
where $z_i^p$ is the coordinate of the $i$th particle in cluster $p$, and $\mathcal{S}$ symmetrizes over the different ways to partition the particles into 3 clusters \cite{CAPPELLI2001499}. We thus obtain 
\begin{equation}
     \langle \psi_1 (z_1)...\psi_1(z_n)\psi_2 (w_1)...\psi_2 (w_n)\rangle  =  \lim_{w_i^\pm \rightarrow w_i} \bigg[  \frac{\Psi_{k=3}^\text{RR}({z_i},w_i^+,w_i^-)}{J_{\frac{2}{3}}(z_i,w_i^+,w_i^-)}\prod_i (w^+_i-w^-_i)^{\frac{2}{3}}\bigg]
\end{equation}
where $J_\alpha(z_i,w_i^+,w_i^-)$ is the Jastrow factor with exponent $\alpha$ formed from the coordinates $z_i,w_i^+,w_i^-$ in that ordering i.e.~$J_\alpha(z_i,w_i^+,w_i^-) = \prod_{i<j} (z_i-z_j)^\alpha$ with $w_i^+ = z_{i+n}$ and $w_i^-=z_{i+2n}$. Notice that the $\prod_i (w^+_i-w^-_i)^{\frac{2}{3}}$ factor will cancel out with the singular terms in $J_{\frac{2}{3}}(z_i,w_i^+,w_i^-)$, so the only terms surviving in the $w_i^\pm \rightarrow w_i$ limit come from those terms in $\Psi_{k=3}^\text{RR}({z_i},w_i^+,w_i^-)$ which are non-singular. In other words, only the partitions where $w_i^+$ and $w_i^-$ for all $i$ are in different clusters will contribute. Let $\mathcal{S}'$ denote the sum over this subset of partitions, and $    \tilde{\Psi}_{k=3}^\text{RR}({z_i},w_i^+,w_i^-)$ contain only the terms coming from this restricted symmetrization. Moreover, let $\tilde{J}$ be the modified Jastrow factor that contains no $(w^+-w^-)^\alpha$ terms. Then:
\begin{align}
     \langle \psi_1 (z_1)...\psi_1(z_n)\psi_2 (w_1)...\psi_2 (w_n)\rangle   = \frac{\tilde{\Psi}_{k=3}^\text{RR}({z_i},w_i,w_i)}{\tilde{J}_{\frac{2}{3}}(z_i,w_i,w_i)}
\end{align}

One may wonder how to relate the CFT construction we have just presented with the fermionic topological order of the Fibonacci state. Indeed, $\frac{\text{SU(2)}_3}{\text{U(1)}_6} \times \text{U(1)}_6 \times \text{U(1)}_6$ has $216$ primary fields, which does not seem to match the number of anyons in the theory. Firstly, one must extend the chiral algebra by the boson $B(z)=\psi_1(z) :e^{2i\phi_s(z)/\sqrt{6}}:$ in $\frac{\text{SU(2)}_3}{\text{U(1)}_6} \times \text{U(1)}_6$ to recover $\text{SU(2)}_3$ in the neutral sector. The full theory post-condensation is now $\text{SU(2)}_3 \times \text{U(1)}_6$ with 24 primaries. This reduces to 6 primaries if we extend the chiral algebra yet again, this time by the fermion $F(z) = \Phi_{j=\frac{3}{2}}(z) :e^{3i\phi_c(z)/\sqrt{6}}:$ where $\Phi_{2j+1}$ is the spin-$j$ multiplet in $\text{SU(2)}_3$. This final chiral extension leads to $\frac{\text{U(6)}_1}{\text{SU(3)}_2}$ which has 6 primaries (modulo the fermion), each associated to a topological sector in the corresponding spin-TQFT. The modular $S$ and $T$ matrices for the bosonic sector of the theory in the $\{V_0, V_1, V_2, \tau V_0, \tau V_1, \tau V_2\}$ basis of Ref.~\onlinecite{Vaezi2014} is:
\begin{align}
&S=
\frac{1}{\sqrt{3(1+\varphi^2)}}
\begin{pmatrix}
1&1&1&\varphi&\varphi&\varphi\\
1&\omega&\omega^2&\varphi&\varphi\omega&\varphi\omega^2\\
1&\omega^2&\omega&\varphi&\varphi\omega^2&\varphi\omega\\
\varphi&\varphi&\varphi&-1&-1&-1\\
\varphi&\varphi\omega&\varphi\omega^2&-1&-\omega&-\omega^2\\
\varphi&\varphi\omega^2&\varphi\omega&-1&-\omega^2&-\omega
\end{pmatrix},\\
&T
=
e^{-2\pi ic/24}\operatorname{diag}
\left(
1,\,
\omega,\,
\omega,\,
\zeta,\,
\omega\zeta,\,
\omega\zeta
\right).
\end{align}
where $c=-\frac{16}{5},\ \varphi=\frac{1+\sqrt{5}}{2}, \ \omega=e^{2\pi i/3},\ \zeta=e^{4\pi i/5}$. The fermionic sector then contributes topological central charge $c=6$ and shifts the anyon spins by $\frac{1}{2}$ if the anyon is chosen to be fused with the fermion. In the spin-TQFT language \cite{PhysRevB.94.155113,wen2016theory}, this theory is $\mathcal{F}_{12} \boxtimes 3^B_{2}\boxtimes 2^B_{14/5} = \mathcal{F}_{12} \boxtimes \overline{\text{SU(3)}_2}$. The invertible Fermionic phases $\mathcal{F}_m$ are defined $m \mod 16$, therefore this theory is equivalent to $\mathcal{F}_{-4} \boxtimes \overline{\text{SU(3)}_2}\equiv \mathcal{C}_{N_{\ell}=2}$ quoted in the main text.

The wavefunction for the SU$(N)_2$ bilayer Fibonacci state at $\nu=\frac{1}{N}+\frac{1}{N}$ can be constructed in an entirely analogous way, keeping in mind that there is an even-odd effect where for $N$ even, the state is bosonic while for $N$ odd the state is fermionic. One starts from the edge theory, $\text{U(1)}_{2N}/\text{SU}(N)_2 = \frac{\text{SU}(2)_N}{\text{U(1)}_{2N}} \times \text{U(1)}_{2N} \times \text{U(1)}_{2N}$ and identifies the coset $ \frac{\text{SU}(2)_N}{\text{U(1)}_{2N}}$ as the $\mathbb{Z}_N$ parafermionic theory. The latter has $N$ parafermionic fields $\psi_l$ with $\mathbb{Z}_N$ fusion rule $\psi_l \times \psi_{l'} = \psi_{l+l' \text{ (mod }N)}$ and conformal weight $h = \frac{l(N-l)}{n}$. Each parafermion has a conjugate partner $\psi^\dagger_l \equiv \psi_{N-l}$ with which it fuses to the identity, and thus with the same conformal weight. The $\text{U(1)}_{2N} \times \text{U(1)}_{2N}$ theory instead has vertex operators $V_{\alpha,\beta}$ with conformal weight $h=\frac{\alpha^2+\beta^2}{4N}$ with $\alpha,\beta\in \mathbb{Z}_{2N}$. To construct the electron operators, we use the fundamental parafermion $\psi_1$ and its conjugate $\psi^\dagger_1$:
\begin{align}
    \Psi_\uparrow(z) = \psi_1(z) V_{N,N-2}(z) = \psi_1(z) :\text{exp}\bigg[\frac{i}{\sqrt{2N}}(N\phi_c(z) + (N-2) \phi_s(z))\bigg]:\\
    \Psi_\downarrow(z) = \psi_1^\dagger(z) V_{N,2-N}(z) = \psi^\dagger_1(z) :\text{exp}\bigg[\frac{i}{\sqrt{2N}}(N\phi_c(z) - (N-2) \phi_s(z))\bigg]:
\end{align}
with conformal weight $h=\frac{N}{2}$. Notice that these operators are bosonic/fermionic for $N$ even/odd respectively, as expected. We construct a wavefunction as in the $N=3$ case:
\begin{align}
    \Psi(z_1,...,z_n;w_1,...,w_n) =&\langle \Psi_\uparrow (z_1)...\Psi_\uparrow (z_n) \Psi_\downarrow (w_1)...\Psi_\downarrow (w_n) \mathcal{O}_\text{bg}(\infty)\rangle \\
    =&\langle \psi_1 (z_1)...\psi_1(z_n)\psi^\dagger_1 (w_1)...\psi^\dagger_1 (w_n)\rangle\\
    &\quad \times \prod_{i<j}(z_i-z_j)^{N-2+\frac{2}{N}}(w_i-w_j)^{N-2+\frac{2}{N}}\times \prod_{ij}(z_i-w_j)^{2-\frac{2}{N}}
\end{align}
One can verify that $N_\phi = \frac{N}{2}N_\mathrm{e} - N$ indicating that the state has total filling $\nu=\frac{2}{N}$ and shift $\mathcal{S}=N$ as expected. The parafermionic correlator can be computed by point-splitting regularization, where now each $w_i$ coordinate is split into $N-1$ coordinates $w_i^a$ ($a=1,2,...,N-1)$. We get
\begin{align}
    \langle \psi_1 (z_1)...\psi_1(z_n)\psi^\dagger_1 (w_1)...\psi^\dagger_1 (w_n)\rangle = \lim_{w_i^a \rightarrow w_i}\langle  \psi_1 (z_1)...\psi_1(z_n)\prod_{a=1}^{N-1}{\psi}_1 (w_1^a)...{\psi}_1 (w_n^{a})\rangle \times \prod_{i=1}^n \prod_{a=1}^{N-2} (w_i^a-w_i^{a+1})^{\frac{2a}{N}}
\end{align}
We note that the order of the pairing in the $(w_i^a-w_i^{a+1})^{-\frac{2a}{N}}$ is not important as that depends on the order in which the fields $\prod_a\psi_1(w_i^a)$ are fused, and will not affect the final form of the wavefunction. Next, we identify $\langle  \psi_1 (z_1)...\psi_1(z_n)\prod_{a=1}^{N-1}{\psi}_1 (w_1^a)...{\psi}_1 (w_n^{a})\rangle$ as the parafermionic correlator giving rise to the $k=N$ Read-Rezayi state. Thus:
\begin{align}
     \langle \psi_1 (z_1)...\psi_1(z_n)\psi^\dagger_1 (w_1)...\psi^\dagger_1 (w_n)\rangle &= \lim_{w_i^a \rightarrow w_i}\bigg[\frac{\Psi_{k=N}^\text{RR}(z_i,w_i^1,...,w_i^{N-1})}{J_{\frac{2}{N}}(z_i,w_i^1,...,w_i^{N-1})} \times \prod_{i=1}^{n} \prod_{a=1}^{N-2} (w_i^a-w_i^{a+1})^{\frac{2a}{N}}\bigg]
\end{align}
where
\begin{align}
    \Psi_{k=N}^\text{RR}(z_1,...,z_{nN}) = \mathcal{S}\bigg[\prod_{p=1}^N\prod_{i<j}(z_i^p-z_j^p)^2\bigg]
\end{align}
Here $\mathcal{S}$ symmetrizes over ways to partition $nN$ particles into $N$ clusters of $n$ particles each. Notice that for each $i$, the Jastrow factor contains $(N-1)(N-2)$ factors of the form $(w_i^a-w_i^b)^{2/N}$, so it is singular with order $-\frac{2}{N}(N-1)(N-2)$. These singular factors cancel out with $\prod_{a=1}^{N-2} (w_i^a-w_i^{a+1})^{\frac{2a}{N}}$, which is also of order $\frac{2}{N}(N-1)(N-2)$ in the $w_i^a$ coordinates. We are left with all those terms in the numerator which do not have any two $w_i^a, w_i^b$ in the same cluster, for each $i$. Finally, we arrive at
\begin{equation}
     \langle \psi_1 (z_1)...\psi_1(z_n)\psi^\dagger_1 (w_1)...\psi^\dagger_1 (w_n)\rangle = \frac{  \tilde{\Psi}_{k=N}^\text{RR}(z_i, \overbrace{w_i, ..., w_i}^{N-1}) }{\tilde{J}_{\frac{2}{N}}(z_i,\underbrace{w_i, ..., w_i}_{N-1})}
\end{equation}
where $\tilde{\Psi}_{k=N}^\text{RR}(z_i, \overbrace{w_i, ..., w_i}^{N-1})$ contains only the non-vanishing terms in ${\Psi}_{k=N}^\text{RR}(z_i, \overbrace{w_i, ..., w_i}^{N-1})$ and $\tilde{J}_\alpha$ is the Jastrow factor with all singular terms removed. 
\subsection{Multilayer Fibonacci states}
In this subsection, we generalize our model wavefunction construction to $\nl$-layered non-Abelian states at filling $\nu=\frac{\nl}{k}$ and shift $\mathcal{S} = k$. We split the construction into three parts. In the first part, we start from the effective theory on the edge, which is a generalized parafermion CFT combined with $k$ free chiral bosons, and construct the electron operators for the edge on each layer, ensuring that they are all mutually local. We then show that the (chiral) conformal correlator of these fields can be interpreted as an electronic wavefunction at the desired shift and filling factor. In the second part, we explicitly compute this conformal correlator using a point-splitting regularization procedure. Finally, in the third part we relate the CFT construction to the topological order in the bulk. 

\subsubsection{Electron operator construction}
Let us now consider the base where we have $\nl$ layers, each at filling $\frac{1}{k}$ so the total filling fraction is $\nu=\frac{\nl}{k}$. The edge theory is now $\frac{\text{U}(k\nl)_{1}}{\text{SU}(k)_\nl}  = \text{SU}(\nl)_k \times \text{U(1)}$. However, we wish to make the $[\text{U(1)}]^\nl$ symmetry explicit, this we can do by performing an appropriate coset:
\begin{equation}
    \frac{\text{SU}(\nl)_k}{[\text{U(1)}]^{k-1}} \times [\text{U(1)}]^k = \text{GPf}[\text{SU}(k)_\nl] \times  [\text{U}(1)]^k 
\end{equation}
where $ \text{GPf}[\text{SU}(k)_\nl] = \frac{\text{SU}(k)_\nl}{[\text{U(1)}_{k\nl}]^{k-1}} $ is the $\text{SU}(k)_\nl$ Gepner parafermion theory \cite{gepner1987new} which has previously appeared in the construction of the non-Abelian spin-singlet state \cite{ardonne1999new, ardonne2001non}. To construct a model wavefunction, we must specify one electron operator for each layer $l$:
\begin{equation}
    \Psi_l(z) = \psi_\bl(z) V_{(\al,\Delta)} = \psi_\bl(z) :\mathrm{exp}\big[i (\al \cdot \boldsymbol{\phi}_s(z) + \Delta \phi_c(z))\big]:
\end{equation}
where $\boldsymbol{\phi}_s$ is a $N_\ell-1$ component boson coming from the neutral sector and $\phi_c$ is the charge boson (single-component). Moreover $\psi_\bl$  are chosen to be root parafermions in $\mathrm{GPf}[\text{SU}(\nl)_k]$, that is, $\bl$ are proportional to roots of $\text{SU}(\nl)$. The reason for this, as first stated in Ref.~\onlinecite{Read1999paired}, is that root parafermions have the lowest conformal weight, and thus form the densest clusters, as we would expect from an FQH wavefunction, especially one obeying a set of generalized exclusion principles. It is easy to verify that using non-root parafermions would lead to a wavefunction that is incompatible with our filling factor. We can then form a valid wavefunction by computing the correlator
\begin{equation}\label{eq:wf1}
    \Psi(\{z_i^l\}) = \Big\langle \prod_l \prod_i\Psi_l(z_1^l) \mathcal{O}_\mathrm{bg,c}\Big\rangle = \Big\langle \prod_i \prod_l \psi_\bl(z_i^l) V_\al(z_i^l)\Big\rangle \times \Big\langle \prod_i \prod_l V_\Delta (z_i^l) \mathcal{O}_\mathrm{bg,c}\Big\rangle
\end{equation}
where $z_i^l$ indicates the coordinate of the $i$th particle in layer $l$. Since we have only a background operator in the charge sector, this correlator evaluates to a finite quantity only if
\begin{equation}
    \sum_l \bl = 0 \text{ mod } kQ \ \text{ and } \ \sum_l \al = 0
\end{equation}
where $Q$ is the charge lattice of $\text{SU}(\nl)$. From this constraint it is natural to define the roots $\bl$ as:
\begin{align}
    &\bl = \frac{1}{\sqrt{k}}(\textbf{e}_l-\textbf{e}_{l+1}), \ l = 1,...,\nl-1,\\
    &\boldsymbol{\beta}_{\nl} = -\sum_{l=1}^{\nl-1}\bl = \frac{1}{\sqrt{k}}(\textbf{e}_\nl-\textbf{e}_1),
\end{align}
where $\textbf{e}_l$ form the standard basis of $\mathbb{R}^{\nl}$ i.e.~$[\textbf{e}_l]_j=\delta_{l,j}$. In other words, $\bl$ are the $\text{SU}(\nl)$ simple roots for $l=1,...,\nl-1$ and $\mathbf{\beta}_{\nl}$ is the negative highest root. These roots sum to zero by construction, moreover they are normalized so that $\bl^2 = \frac{2}{k}$ and $\boldsymbol{\beta}_l \cdot \boldsymbol{\beta}_{l'} = -\frac{1}{k}\delta_{l',l\pm1}$. Since $\bl$ are all roots, we can associate a current in $\text{SU}(\nl)_k$ to each of them:
\begin{align}\label{currents}
    &J_\bl^\pm(z) =  \sqrt{k}\psi_{\pm\bl}(z) :e^{\pm i \bl \cdot \boldsymbol{\phi}_s(z)}:,\\
    &J^3_\bl(z) = \frac{k}{2}i\bl \cdot \partial \boldsymbol{\phi}_s(z).
\end{align}
These currents form an SU$(2)_k$ sub-algebra inside $\text{SU}(\nl)_k$. It is apparent from these relations that:
\begin{equation}
    h(\psi_\bl) = 1 - \frac{\bl^2}{2} = 1-\frac{1}{k},
\end{equation}
and along this root direction, the conformal weights take the same form as the usual $\mathbb{Z}_k$ parafermions, that is $h(\psi_{p\bl}) = \frac{p(k-p)}{k}$ where $p$ is defined mod $k$. More generally, for a root-lattice vector $\boldsymbol{\mu} \in Q$, we have that
\begin{equation}
    h(\psi_{\boldsymbol{\mu}}) = n(\boldsymbol{\mu}) - \frac{\boldsymbol{\mu}^2}{2}
\end{equation}
where $n(\boldsymbol{\mu})$ is the \textit{lowest} affine-current grade at which $\boldsymbol{\mu}$ appears inside the affine vacuum module i.e.~the $\Lambda=0$ representation of $\text{SU}(\nl)_k$. That is, $n(\boldsymbol{\mu})$ is the minimum number of roots needed such that their sum is $\boldsymbol{\mu}$. For example, $h(\psi_{\bl + \boldsymbol{\beta}_{l'}})=2-\frac{2}{k}$ if $|l-l'|>1$ since $\bl + \boldsymbol{\beta}_{l'}$ sits at least at grade $n(\bl + \boldsymbol{\beta}_{l'})=2$ inside the vacuum module. If instead $|l-l'|=1$ then $n(\bl + \boldsymbol{\beta}_{l'})=1$, since $\bl + \boldsymbol{\beta_{l'}}$ is itself a root. From these conformal weights, we deduce the standard parafermionic OPEs:
\begin{align}\label{paraOPE}
    &\psi_\bl(z) \psi_{p\bl}(w) \sim (z-w)^{-\frac{2p}{k}} \psi_{{(p+1)\bl}}(w),
\end{align}
so the monodromy of $\psi_\bl$ and $\psi_{p\bl}$ is $\Delta_{1,p}^{p+1} = \frac{2p}{k}$. Consequently, the operator $\Psi_l(z)$ has conformal weight
\begin{equation}
    h(\Psi_l) = 1-\frac{1}{k}+\frac{\al^2+\Delta^2}{2}.
\end{equation}
We can directly compute the vertex operator correlation functions in Eq.~\eqref{eq:wf1} and obtain
\begin{equation}
    \Psi(\{z_i^l\}) = \Big\langle \prod_i \prod_l \psi_\bl(z_i^l) \Big\rangle \times \prod_l \prod_{i<j}(z_i^l-z_j^l)^{\al^2+\Delta^2} \times\prod_{l<l'} \prod_{ij}(z_i^l-z_j^{l'})^{\al \cdot \boldsymbol{\alpha}_{l'}+\Delta^2}
\end{equation}
Counting the largest power of $z_i^l$ in the wavefunction, we obtain:
\begin{equation}
    N_\phi^{(l)} = \Delta_{1,-1}^0 + \bigg(\frac{N_p}{N_l}-1\bigg)(\al^2+\Delta^2) + \frac{N_p}{N_l}\sum_{l'\neq l}(\alpha_l \cdot \alpha_l{'}+\Delta^2),
\end{equation}
where $\Delta_{1,-1}^0$ is the monodromy between $\psi_\bl$ and $\psi^\dagger_\bl$, which we know from Eq.~\eqref{paraOPE} is $\frac{2(k-1)}{k}$. Moreover, we also have that $\sum_l \al = 0$, hence
\begin{equation}
 N_\phi^{(l)} = \Delta^2 N_p - \bigg[\al^2+\Delta^2 + \frac{2(k-1)}{k}\bigg] \overset{!}{=} \nu^{-1}N_p-\mathcal{S}.
\end{equation}
We thus see that 
\begin{equation}
    \Delta^2 = \frac{k}{\nl}, \  \al^2 = k-2+\frac{2}{k}-\frac{k}{\nl},
\end{equation}
This implies that $h(\Psi_l)=\frac{k}{2}$ which is indeed fermionic (bosonic) for odd (even) $k$, as desired. To confirm that our chosen $\Psi_l(z)$ operators produce a valid wavefunction, we must also check that they are all local with respect to each other. In particular, we find that
\begin{equation}
    \Psi_l \times \Psi_l = \psi_{2\bl} V_{(2\al,2\Delta)} \ \implies\  \Delta_{l,l} = \al^2-\bl^2+\Delta^2 = k-2
\end{equation}
so electrons in the same layer are local, and the wavefunction vanishes to the $k-2$ power as any two of their coordinates are set to be equal each other. Similarly, for electrons in adjacent layers we find that
\begin{equation}
    \Psi_l \times \Psi_{l\pm1} = \psi_{\bl + \boldsymbol{\beta_{l\pm1}}} V_{(\al + \boldsymbol{\alpha}_{l\pm1},2\Delta)} \ \implies \  \Delta_{l,l\pm1} = \al \cdot \boldsymbol{\alpha}_{l\pm1} + \Delta^2 - \frac{k-1}{k},
\end{equation}
where we used the fact that $ \psi_{\bl + \boldsymbol{\beta}_{l\pm1}}$ is also a root parafermion (as $\bl + \boldsymbol{\beta}_{l\pm1}$ is also a $\text{SU}(\nl)$ root). We expect $\Delta_{l,l\pm1} = 0$ by locality, hence
\begin{equation}
   \al \cdot \boldsymbol{\alpha}_{l\pm1} = 1- \frac{1}{k}-\frac{k}{\nl}
\end{equation}
Finally, for non-adjacent layer electrons we get
\begin{equation}
    \Psi_l \times \Psi_{l'} =\psi_{\bl + \boldsymbol{\beta_{l'}}} V_{(\al + \boldsymbol{\alpha}_{l'},2\Delta)} \ \implies  \ \Delta_{l,l'} = \al \cdot \boldsymbol{\alpha}_{l'} + \Delta^2
\end{equation}
where we used the fact that $h(\psi_{\bl + \boldsymbol{\beta}_{l'}}) = 2-\frac{2}{k}$. Again locality forces this monodromy to vanish, yielding
\begin{equation}
    \al \cdot \boldsymbol{\alpha}_{l'} = -\frac{k}{N_l}, \text{ for } |l-l'|>1
\end{equation}
We thus obtain the Gram matrix for the $\al$ vectors
\begin{equation}
    G^{(\alpha)}_{l,l'} \equiv \al \cdot \boldsymbol{\alpha}_{l'}= -\frac{k}{N_l}+\bigg(1-\frac{1}{k}\bigg)\delta_{l,l'\pm1} + \bigg(k-2+\frac{2}{k}\bigg)\delta_{l,l'}
\end{equation}
if we want the corresponding $\Psi_l$ operators to generate a non-singular wavefunction. We also note that $G^{(\alpha)} \boldsymbol{1} = 0$ (as a matrix-vector product), so this Gram matrix automatically enforces the neutrality condition $\sum_l \al = 0$ in the spin-boson charges. Moreover, due to the circulant form of $G^{(\alpha)}$, it can be readily diagonalised using Fourier modes. This leads to a solution for $\al$ which is unique up to orthogonal transformation in the $N_l$ layers. Letting $\boldsymbol{\varepsilon}_l = \textbf{e}_l-\frac{1}{\nl}\sum_{j=1}^\nl \textbf{e}_j$ then one finds that
\begin{equation}\label{eq:alphas}
    \al = \frac{(k-1)\boldsymbol{\varepsilon}_l+\boldsymbol{\varepsilon}_{l+1}}{\sqrt{k}} = \frac{1}{\sqrt{k}}\bigg[(k-1)\textbf{e}_l+\textbf{e}_{l+1}-\frac{k}{\nl}\sum_{j=1}^{\nl} \textbf{e}_j\bigg]
\end{equation}
Nevertheless, the precise form of $\al$ is not so important for now, as the only physical information that enters our wavefunction construction is the Gram matrix itself. We thus obtain
\begin{equation}
    \Psi(\{z_i^l\}) = \Big\langle \prod_{i=1}^{n} \prod_{l=1}^{\nl} \psi_\bl(z_i^l) \Big\rangle \times \prod_l \prod_{i<j}(z_i^l-z_j^l)^{k-2+\frac{2}{k}} \times\prod_{\langle l, l'\rangle} \prod_{ij}(z_i^l-z_j^{l'})^{1-\frac{1}{k}}
\end{equation}
where $\langle l, l'\rangle$ denotes unordered pairs of adjacent layer indices. 
\subsubsection{Point-splitting procedure}
The only remaining task is to compute the parafermionic correlator, which we do by a point-splitting regularization procedure. In particular, we would like to reduce this correlator to $\Big\langle \prod_{i=1}^{N} \prod_{l=1}^{\nl-1} \psi_\bl(z_i^l) \Big\rangle $ which is known to produce the $\text{SU}(\nl)$ singlet wavefunction \cite{PhysRevB.87.205137,PhysRevB.95.125130}, up to some additional Jastrow factors. This is completely analogous to the construction of the $\text{SU(2)}_3$ Fibonacci state, where we wanted to reduce the correlator $\langle \psi_1(z_1)...\psi_1(z_n)\psi_2(w_1)...\psi_2(w_n)\rangle$ to $\langle \psi_1(z_1)...\psi_1(z_n)\psi_1(z_{n+1})...\psi_1(z_{3n})\rangle$, which we know is the $\mathbb{Z}_3$ Read-Rezayi wavefunction (again up to a Jastrow factor). 

We thus have to re-write $\psi_{\bbeta_\nl}(z_i^\nl)$ in terms of $\psi_{\bl}(z)$ for $l=1,...,\nl-1$. Since the parafermionic roots are only defined modulo $kQ$, we can write $\psi_{\bbeta_\nl} = \psi_{-(\bbeta_1+...+\bbeta_{\nl-1})} = \psi_{(k-1)(\bbeta_1+...\bbeta_{\nl-1})}$, which we can split using OPEs. Firstly, let's fuse $\prod_{l=1}^{\nl-1} \psi_\bl$ into $\psi_{\bbeta_1+...+\bbeta_{\nl-1}}$. We find that:
\begin{equation}
    \psi_{\bbeta_1+...+\bbeta_{nl-1}}(z) = \lim_{\{w_a\} \rightarrow z}\bigg[\prod_{a=1}^{\nl-2}(w_a-w_{a+1})^{1-\frac{1}{k}} \times \prod_{a=1}^{\nl-1}\psi_{\bbeta_a}(w_a)\bigg]
\end{equation}
If we assemble $k-1$ of these we obtain:
\begin{equation}
     \prod_{p=1}^{k-1}\psi_{\bbeta_1+...+\bbeta_{nl-1}}(z_p) = \lim_{\{w_{a,p}\} \rightarrow z_p}\bigg[\prod_{p=1}^{k-1}\bigg(\prod_{a=1}^{\nl-2}(w_{a,p}-w_{a+1,p})^{1-\frac{1}{k}} \times\prod_{a=1}^{\nl-1}\psi_{\bbeta_a}(w_{a,p})\bigg)\bigg]
\end{equation}
but at the same time we have that
\begin{equation}
     \psi_{\bbeta_{\nl-1}}(z) = \psi_{(k-1)(\bbeta_0+...+\bbeta_{\nl-1})}(z) = \lim_{\{z_p\}\rightarrow z}\bigg[\prod_{p=1}^{k-2}(z_p-z_{p+1})^{2p/k}\times\prod_{p=1}^{k-1}\psi_{\bbeta_1+...+\bbeta_{\nl-1}}(z_p)\bigg]
\end{equation}
Consequently one finds that
\begin{align}
    \psi_{\bbeta_{\nl-1}}(z_i^\nl) =  \lim_{\substack{{\{w_{i,a,p}\} \rightarrow z_{i,p}}\\{\{z_{i,p}\}\rightarrow z_i^\nl}}}\bigg[\prod_{p=1}^{k-2}(z_{i,p}-z_{i,p+1})^{2p/k}\times\prod_{p=1}^{k-1}\prod_{a=1}^{\nl-2}(w_{i,a,p}-w_{i,a+1,p})^{1-\frac{1}{k}}\times\prod_{p=1}^{k-1}\prod_{a=1}^{\nl-1}\psi_{\bbeta_a}(w_{i,a,p})\bigg]
\end{align}
and thus
\begin{equation}\label{eq:P_split}
    \prod_{i=1}^n \psi_{\bbeta_{\nl-1}}(z_i^\nl)    =\lim_{\substack{{\{w_{i,a,p}\} \rightarrow z_{i,p}}\\{\{z_{i,p}\}\rightarrow z_i^\nl}}}\bigg[P_\text{split}(\{z_{i,p}\},\{w_{i,a,p}\}) \times \prod_{i=1}^n\prod_{p=1}^{k-1}\prod_{a=1}^{\nl-1}\psi_{\bbeta_a}(w_{i,a,p})\bigg]
\end{equation}
where we defined the point-splitting polynomial:
\begin{equation}
    P_\text{split}(\{z_{i,p}\},\{w_{i,a,p}\})  = \prod_{i=1}^n\bigg(\prod_{p=1}^{k-2}(z_{i,p}-z_{i,p+1})^{2p/k}\times\prod_{p=1}^{k-1}\prod_{a=1}^{\nl-2}(w_{i,a,p}-w_{i,a+1,p})^{1-\frac{1}{k}}\bigg)
\end{equation}
Substituting Eq.~\eqref{eq:P_split} into the parafermionic correlator we get:
\begin{align}
    \Big\langle \prod_{i=1}^{n} \prod_{l=1}^{\nl} \psi_\bl(z_i^l) \Big\rangle &=  \lim_{\substack{{\{w_{i,a,p}\} \rightarrow z_{i,p}}\\{\{z_{i,p}\}\rightarrow z_i^\nl}}} \bigg[P_\text{split}(\{z_{i,p}\},\{w_{i,a,p}\}) \times\Big\langle \prod_{i=1}^n\prod_{p=1}^{k-1}\prod_{a=1}^{\nl-1}\psi_{\bbeta_a}(w_{i,a,p}) \times \prod_{i=1}^{n} \prod_{l=1}^{\nl-1} \psi_\bl(z_i^l) \Big\rangle \bigg]\\
    &= \lim_{\substack{{\{w_{i,a,p}\} \rightarrow z_{i,p}}\\{\{z_{i,p}\}\rightarrow z_i^\nl}}} \bigg[P_\text{split}(\{z_{i,p}\},\{w_{i,a,p}\}) \times\Big\langle\prod_{I=1}^{kn} \prod_{l=1}^{\nl-1} \psi_\bl(z_I^l) \Big\rangle \bigg]
\end{align}
where in the second line we merged the $i=1,...,n$ and $p=1,...,k$ indices into one index: $(i,p)\rightarrow I$, with $z_{(i,a)}^l = w_{i,a,l}$ for $p=1,...,k-1$ and  $z_{(i,k)}^l = z_i^k$. The correlator $\Big\langle\prod_{I=1}^{kn} \prod_{l=1}^{\nl-1} \psi_\bl(z_I^l) \Big\rangle$ is known in closed form \cite{ardonne1999new,regnault2008bridge, PhysRevB.95.125130}. More precisely, using the $\text{SU}(\nl)_k$ the current operators defined in Eq.~\eqref{currents}, one has that
\begin{equation}
   \Phi_{\text{SU}(\nl)_k}^{(k)}  = \Big\langle \prod_{I=1}^{kn}\prod_{l=1}^{\nl-1} J_\bl^+(z_I^l)\Big \rangle = S\bigg[\prod_{p=1}^k \Phi_{\text{SU}(\nl)}^\mathrm{Halp}(\{z_{i,p}^l\})\bigg].
\end{equation}
Here, we take the $kn$ particles, and partition them into $k$ groups of $n$ particles each. $S$ symmetrizes over the different ways to partition the $nk$ particles into $k$ clusters. $\Phi_{\text{SU}(\nl)}^\mathrm{Halp}$ is the Halperin state whose $K$ matrix matches the Cartan matrix of $\text{SU}(\nl)$ \cite{regnault2008bridge}:
\begin{equation}
    \Phi_{\text{SU}(\nl)}^\mathrm{Halp} = \prod_{l}\prod_{i<j}(z_i^l-z_j^l)^2  \times \prod_{\langle l,l'\rangle}\prod_{ij}(z_i^l-z_j^{l'})^{-1}
\end{equation}
Note the negative exponent in the interlayer terms. Therefore, for each way of partitioning the particles into $k$ groups, we compute this $\text{SU}(\nl)$ Halperin state for each group of particles, and then take their product. Finally, we symmetrize over all possible partitionings. This is very closely related to the construction of the non-Abelian spin singlet (NASS) state, where the Halperin wavefunction that is used, is instead $\prod_{l}\prod_{i<j}(z_i^l-z_j^l)^2 \times\prod_{\langle l,l'\rangle}\prod_{ij}(z_i^l-z_j^{l'})$ (with positive interlayer exponent now). Finally, since $J^+_\bl = \psi_\bl :\mathrm{exp}(i\bl \cdot \boldsymbol{\phi}_s):$, and 
\begin{equation}
    \Big\langle \prod_{I=1}^{kn} V_{\bl}(z^l_i) \Big\rangle = \prod_{l}\prod_{I<J}(z_I^l-z_J^l)^{\bbeta_l^2}  \times \prod_{\langle l,l'\rangle}\prod_{IJ}(z_I^l-z_J^{l'})^{\bbeta_l\cdot \bbeta_{l'}} = \big[\Phi_{\text{SU}(\nl)}^\mathrm{Halp}(\{z_I^l\})\big]^{1/k} 
\end{equation}
where this time the Halperin state is computed over all $nk$ particles rather than the $n$ particles in a specific cluster. We find that
\begin{equation}
    \Big\langle\prod_{I=1}^{kn} \prod_{l=1}^{\nl-1} \psi_\bl(z_I^l) \Big\rangle = \frac{ \Phi_{\text{SU}(\nl)_k}^{(k)}(\{z_I^l\}) }{\big[\Phi_{\text{SU}(\nl)}^\mathrm{Halp}(\{z_I^l\}) \big]^{1/k} }
\end{equation}
In conclusion, the complete $\text{SU}(\nl)_k$ wavefunction at filling $\nu = \frac{\nl}{k}$ and shift $\mathcal{S}=k$ is given by 
\begin{align}
    \Psi(\{z_i^l\}) = \lim_{\substack{{\{w_{i,a,p}\} \rightarrow z_{i,p}}\\{\{z_{i,p}\}\rightarrow z_i^\nl}}} \bigg[P_\text{split}(\{z_{i,p}\},&\{w_{i,a,p}\}) \times \frac{ \Phi_{\text{SU}(\nl)_k}^{(k)}(\{z_I^l\}) }{\big[\Phi_{\text{SU}(\nl)}^\mathrm{Halp}(\{z_I^l\}) \big]^{1/k} } \bigg] \nonumber\\
    &\times \prod_l \prod_{i<j}(z_i^l-z_j^l)^{k-2+\frac{2}{k}} \times\prod_{\langle l, l'\rangle} \prod_{ij}(z_i^l-z_j^{l'})^{1-\frac{1}{k}}
\end{align}
where $z_{(i,a)}^l = w_{i,a,l}$ for $p=1,...,k-1$ and  $z_{(i,k)}^l = z_i^k$. One can verify that the singular terms all cancel out, and that the final wavefunction is well-defined and finite with no branch-cuts or discontinuities. 
\subsubsection{TQFT construction}
Using our CFT construction we can recover the TQFT description of these states. Firstly, by condensing the condensable algebra in $\text{GPf}[\text{SU}(k)_N] \times [\text{U(1)}_{Nk}]^{k-1}$ one obtains $\text{SU}(k)_N$ in the neutral sector. The condensable algebra here corresponds to the root lattice of $\text{SU}(k)_N$, so it suffices to condense all the simple-root currents $J^\pm_\beta(z) = \psi_{\pm \bl}(z) :e^{\pm i \bl \cdot \boldsymbol{\phi}_s}:$ for $l=1,...,\nl-1$. We are left with $\text{SU}(\nl)_k \times \text{U(1)}_{k\nl}$ which has $k\nl {{\nl+k-1}\choose k}$ primaries. We will write these primaries in the form $\Phi_\lambda(z) e^{im\phi_c(z)/\sqrt{k\nl}}$ where $\lambda$ is the Dynkin label of an integrable highest weight in $\text{SU}(\nl)_k$ and $m\in \mathbb{Z}_{k\nl}$. Inside this theory is the physical fermion operator which we identify as $F(z) = \Phi_{k\boldsymbol{\Lambda_1}}(z) : e^{i\sqrt{\frac{k}{\nl}}\phi_c(z)}:$. Indeed the $\sqrt{\frac{k}{\nl}}$ charge in the vertex operator is fixed by the filling of the state, and $\Phi_{k\boldsymbol{\Lambda_1}}(z)$ is a simple current generator, where $\boldsymbol{\Lambda}_p$ for $p=1,...,\nl-1$ are the fundamental weights of $\text{SU}(\nl)$. Note that $\Phi_{k\boldsymbol{\Lambda_p}}$ are simple currents as they have quantum dimension 1. It can be verified that with our choice of $\al$ and $\bl$, the electron operators $\Psi_l$ are components of the $\Phi_{k\boldsymbol{\Lambda_1}}$ multiplet. 

We therefore proceed to extending the chiral algebra by $F(z)$. For even $k$, $F(z)$ is bosonic and thus this amounts to condensation process in TQFT language. However, for odd $k$, $F(z)$ is fermionic, and the resulting theory will be a spin-TQFT. When we perform such a simple-current extension, we must only keep those fields that are local with respect to $F(z)$. This imposes a selection rule, if we consider the fusion of $\Phi_\lambda(z) e^{im\phi_c(z)/\sqrt{k\nl}}$ with $F(z)$ then we get $\Phi_{\lambda \times k\Lambda_1}(z)e^{i(m+k)\phi_c(z)/\sqrt{k\nl}}$, the monodromy is given by:
\begin{equation} 
    h(\Phi_{\lambda \times k\Lambda_1}) - h(\Phi_\lambda) -h(\Phi_{k\Lambda_1}) -\frac{m}{\nl} 
\end{equation}
The monodromy in the $\text{SU}(\nl)_k$ sector is known to be $\frac{|\lambda|}{\nl}$ where $|\lambda| = \sum_{j=1}^{\nl-1} j\lambda_j$ is the $N$-alith of $\lambda$ \cite{francesco2012conformal}. For this monodromy to be trivial, we must require:
\begin{equation}
    \frac{|\lambda|}{\nl} - \frac{m}{\nl} = 0 \mod 1 \implies m = |\lambda| \mod \nl
\end{equation}
It follows that for a given $\lambda$, there are $k$ possible values of $m$ in $\mathbb{Z}_{k\nl}$ that yield an operator $\Phi_\lambda(z)e^{im\phi_c(z)/\sqrt{k\nl}}$ that is local with $F(z)$. Thus, there are $k {{\nl+k-1}\choose{k}}$ local primaries. Furthermore, local fields that differ by $F(z)$ in fusion must now be identified. Since $F(z)$ is a simple current of order $\nl$, this will lead to $\frac{k}{\nl} {{\nl+k-1}\choose{k}}$ distinct primaries in the end. After a bit of shuffling, we are left with:
\begin{equation}
    \# \mathrm{ primaries} = {{k+\nl-1}\choose{\nl}}
\end{equation}
which are precisely the number of integrable highest weights in $\text{SU}(k)_\nl$. This is not a coincidence, indeed by the level-rank duality \cite{nakanishi1992level,hsin2016level} we know that such a simple-current extension of $\text{SU}(\nl)_k \times \text{U(1)}_{k\nl}$ leads to a theory whose fusion rules are those of $\overline{\text{SU}(k)}_\nl$ (the time-reversed conjugate of $\text{SU}(k)_\nl$). The topological spins are instead shifted by $\frac{m}{2}$ relative to those of $\overline{\text{SU}(k)}_\nl$. Moreover, note that the central charge of our edge theory is
\begin{equation}
    c\bigg(\frac{\text{U}(k\nl)_1}{\text{SU}(k)_\nl}\bigg) = 1 +\frac{k(\nl^2-1)}{\nl+k}
\end{equation}
while for $\overline{\text{SU}(k)}_\nl$ it is
\begin{equation}
    c\big(\overline{\text{SU}(k)}_\nl\big) = - \frac{\nl(k^2-1)}{\nl+k}
\end{equation}
so they differ by $\delta c = k\nl$. Therefore, the TQFT for our state is not exactly $\overline{\text{SU}(k)}_\nl$. This is also highlighted by the fact that the spins differ by $\frac{m}{2}$. 

In the fermionic case, both of these shifts can be absorbed by stacking an invertible fermionic theory which corrects the central charge and shifts the spins by $\frac{1}{2}$, as desired, making it possible to choose anyons from $ \overline{\text{SU}(k)}_\nl$ and attach if necessary a fermion to it to make the topological spin match. We find that the spin-TQFT describing our state is then $\mathcal{F}_{2k\nl} \boxtimes \overline{\text{SU}(k)}_\nl$, where $\mathcal{F}_\nu$ is an invertible fermionic phase with central charge $\frac{\nu}{2}$. Physically, this corresponds to time-reversal conjugating a bosonic bulk theory of the type $\text{SU}(k)_\nl$, and then stacking it with $2k\nl$-component chiral Majorana fermion (or equivalently $k\nl$ component complex fermion) to turn it into a fermionic topological order with the correct central charge and spins. However, it does not affect the charge of the anyons in the theory, as the Majorana fermion is neutral. In the bosonic case, we instead have $\frac{\text{SU}(\nl)_k\boxtimes \text{U(1)}_{k\nl}}{\mathbb{Z}_\nl}$ where the coset by $\mathbb{Z}_\nl$ arises from condensing the $F(z)$ boson. In other conventions, this is denoted as $\text{U}(\nl)_{k,k}$ \cite{hsin2016level}. To conclude, we have that
\begin{equation}
\mathrm{TQFT}=
\begin{cases}
     \mathcal{F}_{2k\nl} \boxtimes \overline{\text{SU}(k)}_\nl, \ &\text{ for $k$ odd (fermionic)}\\[4pt]
     \frac{\text{SU}(\nl)_k\boxtimes \text{U(1)}_{k\nl}}{\mathbb{Z}_\nl}, \ &\text{ for $k$ even (bosonic)}
\end{cases}
\end{equation}

\subsection{Alternative wavefunctions for bilayer Fibonacci state}
\label{sec:interlayer_Pfaffian}

A possible wavefunction for the bilayer Fibonacci state was proposed in Ref.~\onlinecite{Vaezi2014} via a parton construction, where each electron is split into three partons that inherit the layer degree of freedom of the electrons. The proposed wavefunction is $P_\mathrm{LLL}(\Phi_{\nu=2})^3$ where $\Phi_{\nu=2}$ describes the two lowest layer-symmetrized Landau levels of the partons being completely filled. We now work out the shift of this wavefunction. The partons feel magnetic field $B/3$. If the electrons feel flux $N_\phi$, then the partons feel flux $N_\phi/3$ and the LLL and SLL together can hold $2N_\phi/3+4$ partons. So $2N_\mathrm{e}=2N_\phi/3+4$ hence $\mathcal{S}=6$. Another way to derive this shift is to use the general expression $\mathcal{S}=\sum_\alpha n_\alpha=2+2+2=6$, where $n_\alpha$ are the filling factors of the partons \cite{Balram2021b}. So this wavefunction does not have the same shift as the bilayer Fibonacci state (which has $\mathcal{S}=3$).

\begin{figure}
        \centering
        \includegraphics[width=\linewidth]{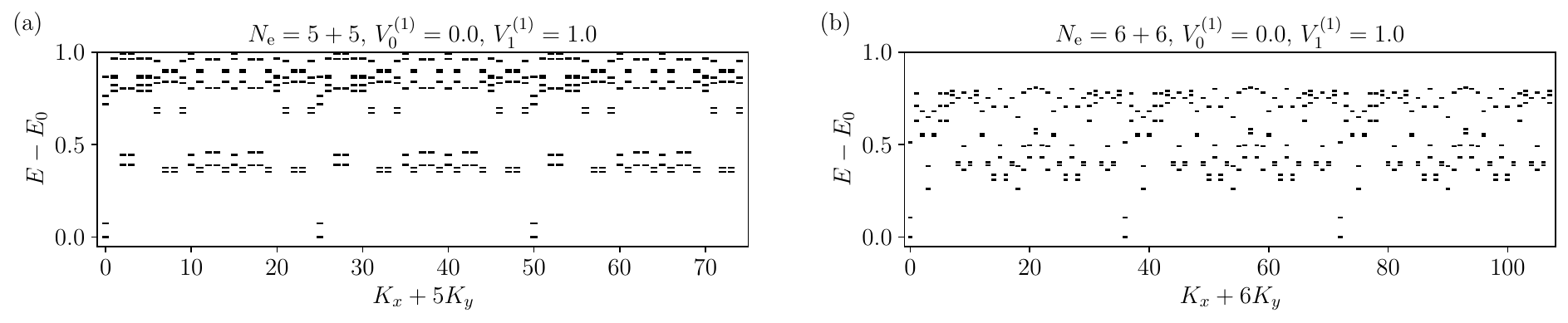}
        \caption{\textbf{Energy spectra on the torus for Fibonacci phase.} We compute the energy gap for (a) $N_\mathrm{e}=5+5$ and (b) $N_\mathrm{e}=6+6$ particles at $V_0^{(1)}=0$ and $V_1^{(1)} = 1$. The ground state degeneracy is 6, as expected for the Fibonacci state, and in contrast to the ground state degeneracy of 9 for the interlayer Pfaffian phase.}
        \label{fig:app_torus_Fib}
    \end{figure}

Another alternative wavefunction for the bilayer Fibonacci state is the paired spin-singlet state (PSS) (also referred to in the literature as the interlayer Pfaffian state \cite{Geraedts2015}) which was proposed in Ref.~\onlinecite{PhysRevB.65.041305} and investigated numerically in Ref.~\onlinecite{PhysRevA.79.033609}. It takes the form
\begin{equation}\label{PSS}
    \Psi_\mathrm{PSS}^{(m)}(\{z_i^\uparrow\},\{z_i^\downarrow\}) = \text{Pf}\bigg(\frac{1}{z_i-z_j}\bigg)\prod_{i<j}(z^\uparrow_i-z^\uparrow_j)^{m+1}\prod_{i<j}(z^\downarrow_i-z^\downarrow_j)^{m+1}\prod_{ij}(z_i^\uparrow-z_j^\downarrow)^m,
\end{equation}
where $\text{Pf}\big(\frac{1}{z_i-z_j}\big)$ is the Pfaffian of the anti-symmetric matrix $A_{ij} = \frac{1}{z_i-z_j}$ and $z_i$ denotes all particle's coordinates in both layers i.e.~if we have $n$ spin up and $n$ spin down particles then $A_{ij}$ is a $2n\times 2n$ matrix. It can be readily verified that this wavefunction has filling factor $\nu=\frac{2}{2m+1}$ and shift $\mathcal{S}=m+2$, which for $m=1$ makes Eq.~\eqref{PSS} a candidate wavefunction for the Fibonacci state on the sphere. Nevertheless, as we show below, the topological order that Eq.~\eqref{PSS} describes is inconsistent with the numerically observed ground state degeneracy on the torus. Following Ref.~\onlinecite{PhysRevB.65.041305}, the presence of the Pfaffian term implies the existence of a Majorana particle in the underlying theory, and strongly suggests that the edge theory is of the form $\mathrm{Ising} \times U(1)_s \times U(1)_c$. Indeed, we can identify the electron operators in each layer as:
\begin{align}
    &\Psi_\uparrow(z) = \psi(z) :\text{exp}[i(\sqrt{2m+1} \phi_c(z)+\phi_s(z))/\sqrt{2}]:,\\
    &\Psi_\downarrow(z) = \psi(z) :\text{exp}[i(\sqrt{2m+1} \phi_c(z)-\phi_s(z))/\sqrt{2}]:,
\end{align}
where $\psi(z)$ is the Majorana fermion in the Ising theory, $\phi_s$ and $\phi_c$ are the spin and charge bosons respectively. The wavefunction in Eq.~\eqref{PSS} can then be expressed as $\Psi_\mathrm{PSS}^{(m)} = \big\langle \prod_i \prod_{\sigma=\uparrow,\downarrow} \Psi_\alpha(z_i^\sigma)\mathcal{O}_\mathrm{bg}\big\rangle$. Note $h(\Psi_m) = \frac{m+2}{2}$ so the state is fermionic (bosonic) for $m$ odd (even), which means the $m=1$ wavefunction is indeed fermionic. With this definition of the electron operators, we see that the microscopic edge theory is $\text{Ising}\times U(1)_2 \times U(1)_{4m+2}$, with $12(2m+1)$ primaries. After extending the chiral algebra by $\Psi_{\sigma}(z)$ however, half of these primaries become confined as they are not local with respect the electron operators. The remaining $6(2m+1)$ primaries local with respect to the electron operators then organize into orbits, since $\Psi_{\sigma}(z)$ are simple currents of order 2. We are thus left with $3(2m+1)$ distinct topological sectors, thus this state should have a ground state degeneracy of $9$ for $m=1$ on the torus. This does not agree with the observed ground state degeneracy of $6$ (see energy spectrum in Fig.~\ref{fig:app_torus_Fib}, thus disqualifying it from describing the bilayer Fibonacci phase.
Furthermore, Ref.~\onlinecite{PhysRevB.84.205134} derives the exclusion principles for the Interlayer Pfaffian. Following these rules, the predicted quasihole counting for a state in the Interlayer Pfaffian phase is $D_0\oplus 2D_1\oplus D_2\oplus 2D_3\oplus D_4\oplus D_5$ as shown in Tab.~\ref{tab:ipf_roots}, This is inconsistent with the numerically observed counting shown in Fig.~\ref{fig:QHspectra}. 
\begin{table}[h]
\centering
\begin{tabular}{c|c|c}
\hline \hline
Root & $L_z$ & \# \\
\hline
$20020020020020$ $(\times 1)$
& $5$ & $1$ \\
\hline
$20020020020011$ $(\times 2)$
& $4$ & $2$ \\
\hline
$20020020020002$ $(\times 2)$
& \multirow{2}{*}{$3$} & \multirow{2}{*}{$4$} \\
$20020020011011$ $(\times 2)$
& & \\
\hline
$20020020011002$ $(\times 2)$
& \multirow{2}{*}{$2$} & \multirow{2}{*}{$5$} \\
$20020011011011$ $(\times 3)$
& & \\
\hline
$20020011011002$ $(\times 3)$
& \multirow{3}{*}{$1$} & \multirow{3}{*}{$7$} \\
$20020020002002$ $(\times 2)$
& & \\
$20011011011011$ $(\times 2)$
& & \\
\hline
$20011011011002$ $(\times 3)$
& \multirow{3}{*}{$0$} & \multirow{3}{*}{$8$} \\
$20020011002002$ $(\times 3)$
& & \\
$11011011011011$ $(\times 2)$
& & \\
\hline \hline
\end{tabular}
\caption{\textbf{Quasihole counting for the interlayer Pfaffian state in the fixed $S_z=0$ sector.}
We show the layer-summed root configurations for $N_{\rm e}=10$ electrons and $N_\phi=13$ flux quanta consistent with the exclusion principles of the Interlayer Pfaffian phase. Some of the layer-summed root configurations have non-trivial multiplicity $m$, which we denote by $(\times m)$. The quasihole manifold is predicted to decompose into angular momentum multiplets $D_0\oplus 2D_1\oplus D_2\oplus 2D_3\oplus D_4\oplus D_5$.}
\label{tab:ipf_roots}
\end{table}

\subsection{Stereographic projection}

In this subsection, we explain the standard procedure by which one can relate quantum Hall wavefunctions in the disk and in the spherical geometry using stereographic projections, as first done in Ref.~\onlinecite{fano1986configuration}. In the disk geometry, we employ complex coordinates $z_i=x_i+iy_i$. The LLL single-particle wavefunctions take the form $z_i^me^{-|z_i|^2/4}$, where we are using units in which the magnetic length $l_B=1$. Any many-body wavefunction in the LLL will take the form
\begin{equation}
    \Psi_\mathrm{Disk}(\{z_i\})=f(\{z_i\})\prod_ie^{-|z_i|^2/4},
    \label{eq:MB_disk}
\end{equation}
where $f(\{z_i\})$ is a holomorphic polynomial. Note that usually we drop the Gaussian factor for conciseness and only write the polynomial factor, however the Gaussian factor is always implicitly present. 

While the wavefunctions appear simpler in the disk geometry, when performing exact diagonalization it is preferable to use the spherical geometry due to its lack of a boundary. To translate a wavefunction written in the disk geometry to the spherical geometry, we can use the stereographic projection. A particle's position on the sphere is specified in terms of the usual polar coordinates $(\theta_i,\phi_i)$ and in terms of those it is useful to define the so-called spinor coordinates 
\begin{equation}
    u_i=e^{i\phi_i/2}\cos\theta_i/2, \quad v_i=e^{-i\phi_i/2}\sin\theta_i/2.
\end{equation}
The stereographic projection relates the spinor coordinates to a coordinate in the plane via
\begin{equation}
    z_i=2R\frac{v_i}{u_i}=2R\ e^{-i\phi_i}\tan\theta_i/2,
    \label{eq:stereo}
\end{equation}
where $R$ is the radius of the sphere. The LLL single-particle orbitals in the spherical geometry take the form \cite{Read1999paired}
\begin{equation}
    \frac{z_i^{m}}{\left(1 + |z_i|^2 / 4R^2\right)^{1 + N_\phi/2}},
\end{equation}
and hence the many-particle wavefunction on the sphere corresponding to Eq.~\eqref{eq:MB_disk} is 
\begin{equation}
\Psi_\mathrm{Sphere}(\{\theta_i,\phi_i\}) = f(\{z_i\}) \prod_i
\left(1 + |z_i|^2 / 4R^2\right)^{-(1 + N_\phi/2)} .
\end{equation}
keeping in mind the stereographic projection Eq.~\eqref{eq:stereo}.

As an example let us consider the Jastrow factor. On the disk we have
\begin{equation}
    \prod_{i<j}(z_i-z_j)^q,
\end{equation}
again dropping the Gaussian factors. Using Eq.~\eqref{eq:stereo} we have
\begin{equation}
    \prod_{i<j}[2R(u_iv_j-v_iu_j)]^q,
\end{equation}
where $2R|u_iv_j-v_iu_j|$ is precisely the chord distance between the two particles on the sphere.

\section{Monte-Carlo overlap of model wavefunctions on the sphere}\label{app:overlaps}
\begin{figure}[h]
\centering
\begin{minipage}[c]{0.49\linewidth}
\centering
\scriptsize
\renewcommand{\arraystretch}{1.1}
\resizebox{\linewidth}{!}{
\begin{tabular}{ccccc} \hline\hline 
$N_{\mathrm e}$ & $N_\phi$ & $\dim(L=0)$ & $\dim(L=S=0)$ & Overlap \\ \hline 
$6$ & $6$ & $5$ & $4$ & \begin{tabular}{@{}c@{}} 
$|\langle \Psi_{\mathrm H}|\Psi_{\mathrm{ED}}\rangle|^2 = 0.6900 \pm 0.0021$ \\ 
$|\langle \Psi_{\mathrm F}|\Psi_{\mathrm{ED}}\rangle|^2 = 0.9377 \pm 0.0017$ \\ 
$|\langle \Psi_{\mathrm{IPf}}|\Psi_{\mathrm{ED}}\rangle|^2 = 0.9689 \pm 0.0010$ \\
$|\langle \Psi_{\mathrm{Fib}}|\Psi_{\mathrm{IPf}}\rangle|^2 = 0.9411 \pm 0.0030$
\end{tabular} \\[3.0ex] \hline 
$8$ & $9$ & $36$ & $16$ & \begin{tabular}{@{}c@{}} 
$|\langle \Psi_{\mathrm H}|\Psi_{\mathrm{ED}}\rangle|^2 = 0.5525 \pm 0.0008$ \\ 
$|\langle \Psi_{\mathrm F}|\Psi_{\mathrm{ED}}\rangle|^2 = 0.8702 \pm 0.0042$ \\ 
$|\langle \Psi_{\mathrm{IPf}}|\Psi_{\mathrm{ED}}\rangle|^2 = 0.9110 \pm 0.0010$ \\
$|\langle \Psi_{\mathrm{Fib}}|\Psi_{\mathrm{IPf}}\rangle|^2 = 0.8986 \pm 0.0034$
\end{tabular} \\[3.0ex] \hline 
$10$ & $12$ & $345$ & $105$ & \begin{tabular}{@{}c@{}} 
$|\langle \Psi_{\mathrm H}|\Psi_{\mathrm{ED}}\rangle|^2 = 0.4405 \pm 0.0006$ \\ 
$|\langle \Psi_{\mathrm F}|\Psi_{\mathrm{ED}}\rangle|^2 = 0.8039 \pm 0.0068$ \\ 
$|\langle \Psi_{\mathrm{IPf}}|\Psi_{\mathrm{ED}}\rangle|^2 = 0.8500 \pm 0.0011$ \\
$|\langle \Psi_{\mathrm{Fib}}|\Psi_{\mathrm{IPf}}\rangle|^2 = 0.8516 \pm 0.0044$
\end{tabular} \\[3.0ex] \hline 
$12$ & $15$ & $5466$ & $1210$ & \begin{tabular}{@{}c@{}} 
$|\langle \Psi_{\mathrm H}|\Psi_{\mathrm{ED}}\rangle|^2 = 0.3470 \pm 0.0004$ \\ 
$|\langle \Psi_{\mathrm F}|\Psi_{\mathrm{ED}}\rangle|^2 = 0.7349 \pm 0.0073$ \\ 
$|\langle \Psi_{\mathrm{IPf}}|\Psi_{\mathrm{ED}}\rangle|^2 = 0.7801 \pm 0.0009$ \\
$|\langle \Psi_{\mathrm{Fib}}|\Psi_{\mathrm{IPf}}\rangle|^2 = 0.8001 \pm 0.0129$
\end{tabular} \\[3.0ex] \hline 
$14$ & $18$ & $101266$ & $18607$ & \begin{tabular}{@{}c@{}} 
$|\langle \Psi_{\mathrm H}|\Psi_{\mathrm{ED}}\rangle|^2 = 0.2730 \pm 0.0009$ \\ 
$|\langle \Psi_{\mathrm F}|\Psi_{\mathrm{ED}}\rangle|^2 = 0.6701 \pm 0.0162$ \\ 
$|\langle \Psi_{\mathrm{IPf}}|\Psi_{\mathrm{ED}}\rangle|^2 = 0.7150 \pm 0.0016$ \\
$|\langle \Psi_{\mathrm{Fib}}|\Psi_{\mathrm{IPf}}\rangle|^2 = 0.7488 \pm 0.0086$
\end{tabular} \\ \hline\hline
\end{tabular}

}
\end{minipage}
\hfill
\begin{minipage}[c]{0.5\linewidth}
\centering
\includegraphics[width=\linewidth]{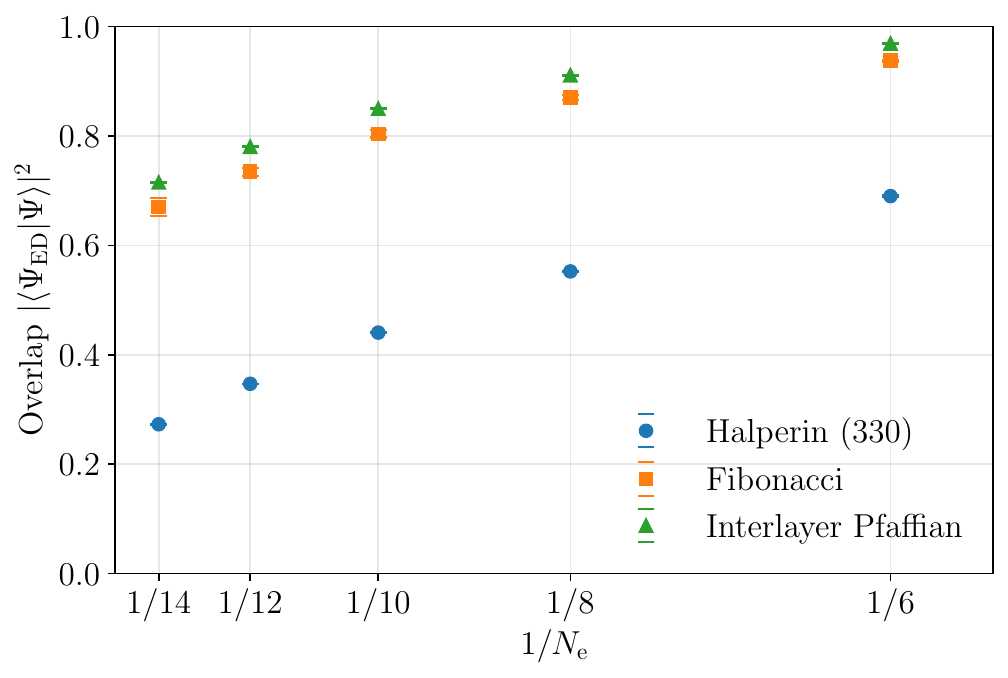}
\end{minipage}

\caption{Overlaps of the ED ground state $\Psi_\text{ED}$ at $V_0^{(1)}=0$, $V_1^{(1)}=1$ with the Fibonacci state $\Psi_\mathrm{Fib}\equiv\Psi_\text{F}$, interlayer Pfaffian state $\Psi_\mathrm{PSS}^{(3)}\equiv\Psi_\text{IPf}$, and decoupled Laughlin state $\Psi_{330}\equiv\Psi_\text{H}$, obtained via Monte Carlo sampling with $10^8$ samples. We also list the overlap of the interlayer Pfaffian with the Fibonacci state.}
\label{fig:MC_overlaps_and_scaling}
\end{figure}

We compute the overlap of the ED ground state at $V_0^{(1)}=0$, $V_1^{(1)}=1$ with several model wavefunctions at filling $\nu=\frac{1}{3}+\frac{1}{3}$ and shift $\mathcal{S}=3$: the Fibonacci state $\Psi_\text{Fib}$ in Eq.~\eqref{eq:Fib_WF}, the interlayer Pfaffian/PSS state $\Psi_\text{PSS}$ in Eq.~\eqref{PSS} and the decoupled Laughlin state $\Psi_{330}$ in Eq.~\eqref{eq:halperin_mmn}. We compute the overlaps for various system sizes using Monte Carlo sampling with $N_\text{mc}=10^8$ samples, the results are shown in Fig.~\ref{fig:MC_overlaps_and_scaling}. 

Firstly, we note that even for systems as large as $N_\mathrm{e}=7+7$ where the $L=0$ Hilbert space dimension is 101266, the overlap of the ED state with both the Fibonacci and interlayer Pfaffian wavefunctions is $\sim 0.7$. Moreover, the interlayer Pfaffian performs marginally better than the Fibonacci state for the system sizes we considered. However, it is not possible to conclude which wavefunction is a better description of the phase based on this data alone. Indeed, the ED ground state for the chosen SU(2)-symmetric interactions is a layer pseudospin singlet, as is the interlayer Pfaffian by construction. In contrast, the Fibonacci state is not SU(2) symmetric, although it can be made so. Thus, the Fibonacci overlap would probably increase if the wavefunction was first projected to the $S=0$ sector via Young symmetrization procedure, as has been done in Ref.~\onlinecite{PhysRevB.87.045310} for the spin-singlet Gaffnian state. Given the substantial reduction of the Hilbert space dimension from the $L=0$ sector to the $L=S=0$ sector, the fact that the Fibonacci overlaps without projection are this large is reassuring. Furthermore, the Fibonacci overlaps improve as one goes away from the SU(2) symmetric interactions by increasing $V_1^{(1)}$, while the interlayer Pfaffian overlaps deteriorate relative to it. For example, at $V_0^{(1)}=0$ and $V_1^{(1)} = 2$, the Fibonacci overlap is $\sim 0.823$ while the interlayer Pfaffian overlap is $\sim 0.780$ for $N_\mathrm{e}=6+6$. 

To probe how similar the Fibonacci and interlayer Pfaffian wavefunctions are, we also computed the overlap between these two candidate model states. We see that for $N_\mathrm{e}=7+7$ particles, their overlap is $\sim 0.75$, so it is unsurprising that their overlaps with the ED ground states are also very similar. Indeed, the two wavefunctions actually share a larger overlap with each other than with the ED ground state. Again, these overlaps would also likely increase after projecting the Fibonacci wavefunction to the singlet sector (and normalizing the projected Fibonacci wavefunction). 

Finally, we remark that while it is not possible to ascertain which of the two model wavefunctions agrees best with the numerical data based on the overlaps alone, topological data such as ground state degeneracy on the torus and quasihole counting on the sphere all point to the Fibonacci state as being the correct description. Indeed, wavefunction overlaps on their own are not a robust metric for topological order. For example, it is well known that the Gaffnian state \cite{PhysRevB.75.075317} and the Jain $\nu=\frac{2}{5}$ composite fermion state have a high overlap \cite{regnault2008bridge}, despite the two states describing completely different physics. Thus, one should rely on more robust, topological metrics such as the ground state degeneracy and quasihole counting.

\section{Phase diagrams for different layer systems}\label{app:multilayer}

\begin{figure}[H]
    \centering
    \includegraphics[width=0.95\columnwidth]{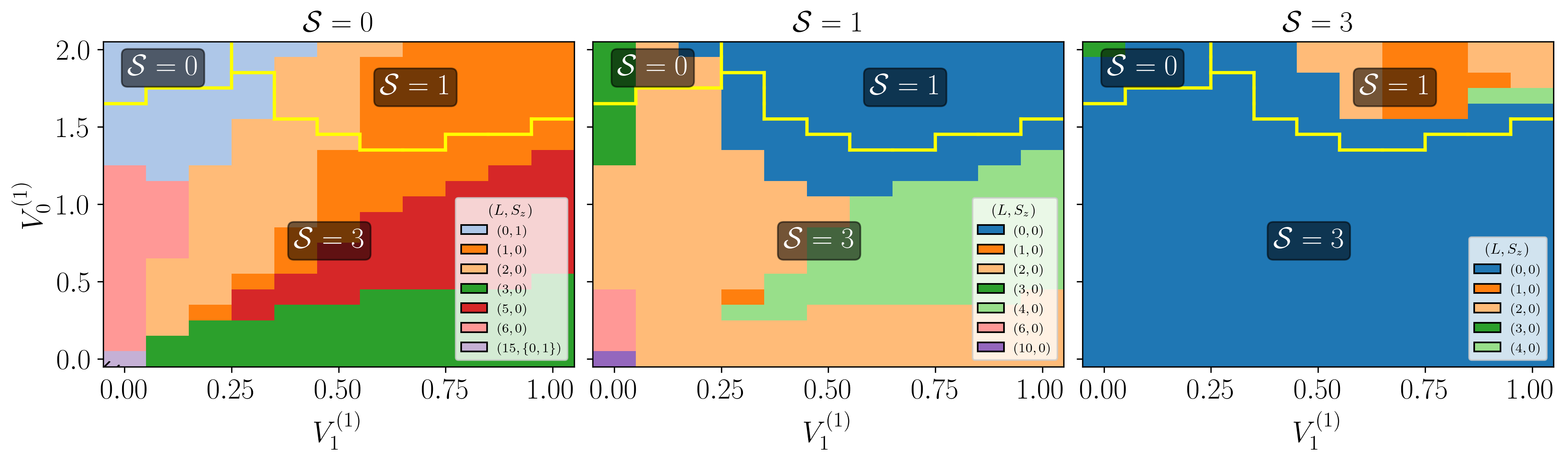}
    \caption{\textbf{Layer polarization in the bilayer on the sphere.} We show the $L$ and $S_z$ quantum number of the ground state of the bilayer system on the sphere with $N_\mathrm{e}=10$ particles and shift $\mathcal{S}=0,1$ and 3, for varying $V_1^{(1)}$ and $V_0^{(1)}$. Note since we have spin inversion symmetry, we only look at the $S_z\geq 0$ sectors, with the understanding that there are corresponding degenerate states at $-S_z$. For the decoupled point $V_1^{(1)} = V_0^{(1)}=0$, we expect exact zero modes in both $S_z=0$ and $S_z=1$ with $L$ up to 15, hence we label it by $(L=15, S_z=\{0,1\})$. We also show in yellow the phase boundaries between regions where each shift sector is the true ground state of the system (determined by choosing the shift sector with the largest gap). We see that regions where $\mathcal{S}=1$ or $\mathcal{S}=3$ are the ground states are spin-unpolarized. The phase where the $\mathcal{S}=0$ sector is the ground state has a small $S_z=\pm1$ spin polarization, but this could be due to finite-size effects.}
    \label{fig:spin_polarization}
\end{figure}

\begin{figure}[H]
    \centering
    \includegraphics[width=0.95\columnwidth]{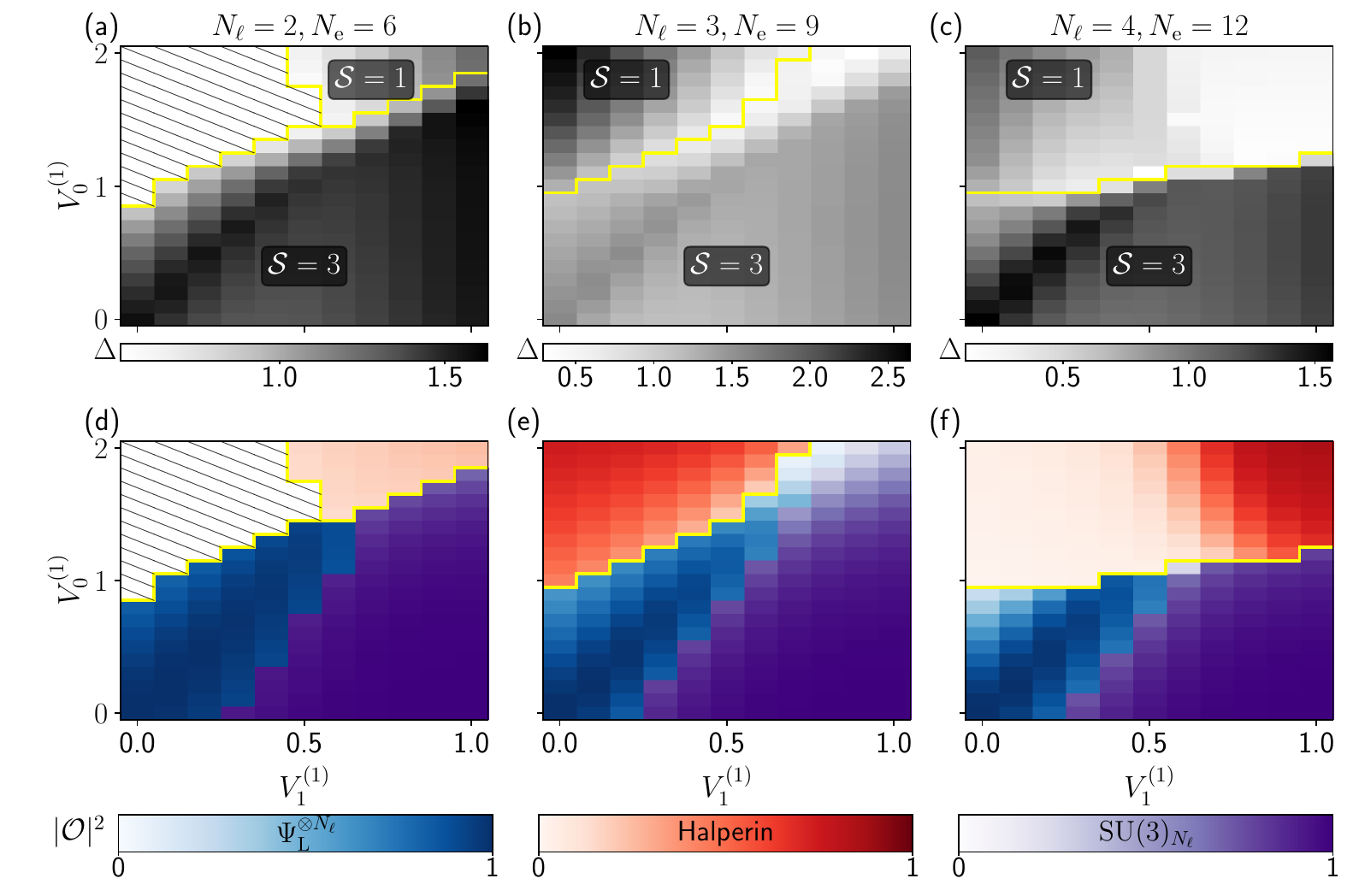}
    \caption{Phase diagrams for the quantum Hall (a) and (d) bilayer, (b) and (e) trilayer, and (c) and (f) tetralayer in the spherical geometry. In all phase diagrams we choose the intralayer interaction to consist of a single non-zero pseudopotential $V_1^\mathrm{(0)}=1$. We tune the interlayer onsite and ``nearest-neighbor" pseudopotentials $V_0^\mathrm{(1)}$ and $V_1^\mathrm{(1)}$. The pseudopotentials satisfy ``periodic boundary conditions" in the $z$-direction, i.e.~the top and bottom layers in the stack interact, for the bilayer we do not double count the interlayer interactions.  At each point, the upper row shows the shift $\mathcal{S}$ with the largest computed gap, with intensity proportional to that gap; the lower row shows the model state with the largest corresponding overlap, with intensity proportional to $|\mathcal{O}|^2$. For all the cases, we have taken $N_\mathrm{e}=3$ per each layer. For the bilayer, we have stripped out the $\mathcal{S}=0$ region since the ground state there has finite $L$ rather than $L=0$ (hashed region). For the Fibonacci state overlaps, we have plotted the overlap with the ground state at $V_1^\mathrm{(0)}=1$ and $V_1^\mathrm{(1)}=1$, rather than using the model wavefunctions from the main text. We know from Fig.~\ref{fig:QHspectra} that the quasihole counting at $V_1^\mathrm{(0)}=1, V_1^\mathrm{(1)}=1$ agrees with the SU(3)$_\nl$ state. For the Halperin state overlaps, we have used the generalized Halperin state from $K_{ij}^{(\nl)}$, which reduces to the ordinary Halperin~(112) state at $\nl=2$ (see \ref{app:genHalperin}).}
    \label{fig:placeholder}
\end{figure}

\begin{figure}[H]
    \centering
    \includegraphics[width=0.35\columnwidth]{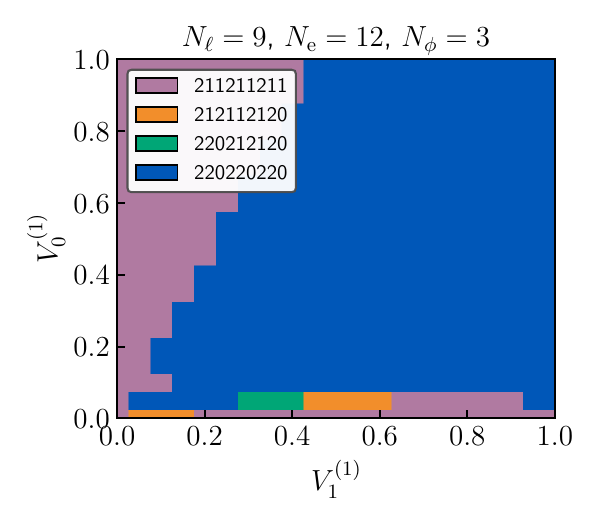}
    \caption{\textbf{Ground state particle configuration for $\nl=9$ layers.} Minimum energy for different spin sector with $V_1^{(0)}=1$ gives different spin polarization sector over a large region of $V_0^{(1)}$ and $V_1^{(1)}$ and confirms full polarization for strong interlayer pseudopotentials. Different labeling provides different particle configurations.}
    \label{fig:particle_configuration}
\end{figure}

We provide additional phase diagrams for $N_{\ell}=2,3,4$, and $9$ layer systems, supplementing the results in the main text. Figure~\ref{fig:spin_polarization} shows the spin polarization of the bilayer ground state as a function of the interlayer pseudopotentials $V_{0}^{(1)},$ and $V_{1}^{(1)}$. Different shift sectors, $\mathcal{S}=0,1,3$, are separated via a marked phase boundary in the plots. By comparing across all shift sectors, we find that the ground state has a small layer polarization ($S_z=\pm1$) at large $V_0^{(1)}/V_{1}^{(0)}\sim 2$, while the system remains fully unpolarized $S_{z}=0$ in regions characterized by shift sectors $\mathcal{S}=1$ and $\mathcal{S}=3$. Note however that in systems with $N_\mathrm{e}=8$ and $N_\mathrm{e}=12$ particles, the ground state of the $\mathcal{S}=0$ phase was found to be pseudospin-unpolarized but with finite $L$ quantum number.  
In Figure \ref{fig:placeholder} 
 we show phase diagram of the exact diagonalization ground state on a sphere with $N_{\ell}=2,3$, and $4$ layers. The phases are identified by the shift sectors with the largest gap, and their corresponding overlap calculations with the analytical Fibonacci wavefunction (Eq.~\eqref{eq:Fib_WF}) for the bilayer, and the ground state at $V_0^{(1)}=0$, $V_1^{(1)} = 1$ for the $3$ and $4$-layer systems. 
 Finally, in Fig.~\ref{fig:particle_configuration} we identify which particle-number-per-layer sector contains the global ground state of $N_{\ell}=9$ system as a function of the interlayer pseudopotentials. We compute the energy over all particle configurations and report the configuration with minimum energy. The plot thus makes it explicit the competition between layer-occupation configurations, and shows that the pair-foliated charge order survives as the ground state sector over an extended portion of the interaction plane. 
 

\section{Energy spectra for SU(2) symmetric interactions}\label{app:su2symmetric_spectra}
\begin{figure*}[h!]
    \centering
    \includegraphics[width=\linewidth]{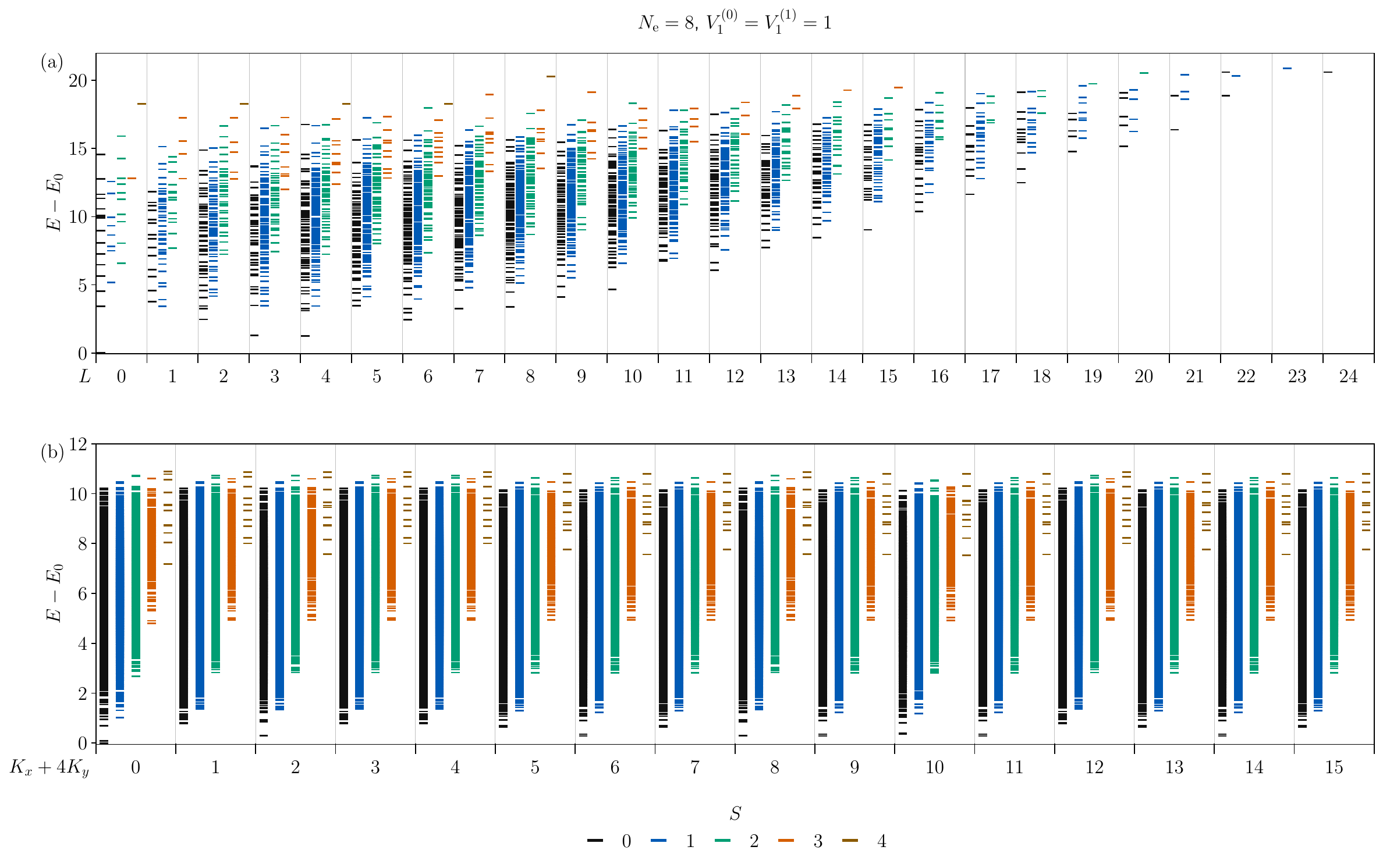}
    \caption{\textbf{Full energy spectra for the bilayer system with SU(2) symmetric interactions.} We plot the energy spectrum for a bilayer with $N_\mathrm{e}=8$ particles with SU(2)-symmetric interactions $V_1^{(0)}=V_1^{(1)}=1$ sorted by (a) $(K_x+4K_y,S)$ sectors, where $K_x,K_y$ are momenta corresponding to the magnetic translation symmetries, and $S$ labels the layer SU(2)$_S$ pseudospin irreducible representation. (b) $(L,S)$ sectors corresponding to multiplets of $\mathrm{Spin}(4) = \text{SU}(2)_L \times \text{SU}(2)_S$. Note that the ground state is a singlet $(L=0,S=0)$ as expected from a singlet FQH state. Note the spectrum looks less dense on the sphere because there we also decompose into orbital angular momentum multiplets.}
    \label{fig:SU2_spectra}
\end{figure*}

In this section we show the energy spectra on both torus and sphere geometries for SU(2)-symmetric interactions where $V_1^{(0)}=V_1^{(1)}=1$. Due to the additional layer SU(2) symmetry, we can label eigenstates by spin multiplets $S$ in addition to already existing quantum numbers ($K_x,K_y$ on the torus and $L$ on the sphere). We see that in both cases the ground state and low energy excited states are all spin-singlets (Fig.~\ref{fig:SU2_spectra}).

\section{Spin-resolved low-energy torus spectrum from decoupled 330 to Fibonacci}
\label{app:torus_sz_scan_330_fib}

\begin{figure}
    \centering
    \includegraphics[width=0.8\linewidth]{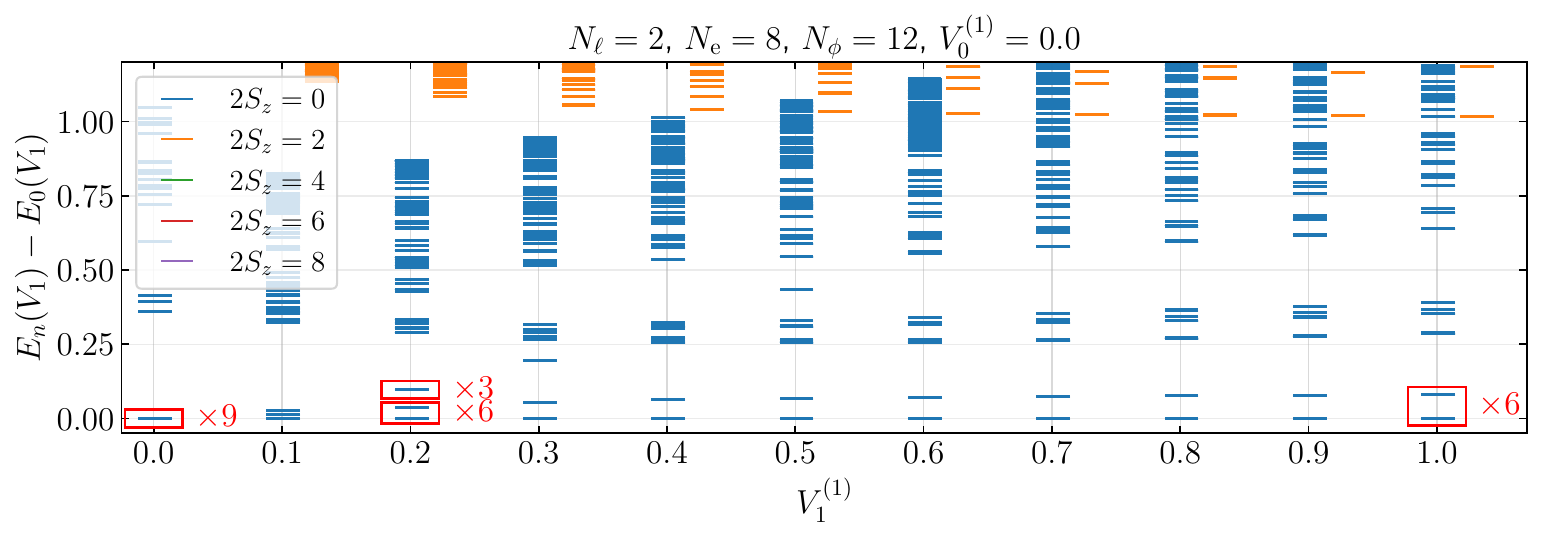}
    \caption{\textbf{Low-energy spectrum on the torus along $V_1^{(1)}$.}
    Low-energy spectrum on the bilayer square torus along a one-parameter cut from the decoupled Laughlin state toward the Fibonacci phase. We fix $V_1^{(0)}=1$, $V_0^{(1)}=0$ and vary $V_1^{(1)}$ for $N_\mathrm{e}=8$ and $N_\phi=12$. Different colors denote different $S_z$ sectors. For each value of $V_1^{(1)}$, the energy is shifted by the lowest energy among all checked $S_z$ sectors. The lowest states remain in the $S_z=0$ sector along the scan, indicating that the interpolation is not dominated by a change of spin polarization. For the low-energy states, we indicate the actual degeneracies where relevant. 
    }
    \label{fig:torus_sz_scan_330_fib}
\end{figure}

In this appendix, we examine the low-energy torus spectrum along a one-parameter cut from the decoupled Halperin~(330) limit toward the Fibonacci region. We fix $V_1^{(0)}=1$ and $V_0^{(1)}=0$, and vary the interlayer pseudopotential $V_1^{(1)}$. The calculation is performed for $N_\mathrm{e}=8$ and $N_\phi=12$, corresponding to total filling $\nu=2/3$. For each value of $V_1^{(1)}$, we diagonalize the Hamiltonian in all spin sectors $2S_z=0,2,4,6,8$ (up to spin-inversion symmetry $S_z \rightarrow -S_z$), and shift the spectrum by the lowest energy at that value of $V_1^{(1)}$.

Figure~\ref{fig:torus_sz_scan_330_fib} shows that the lowest states remain in the unpolarized sector throughout the scan. Starting from the decoupled Laughlin limit at $V_1^{(1)}=0$, the low-energy levels evolve continuously as $V_1^{(1)}$ is increased, while the partially polarized and polarized sectors stay at higher energy. This indicates that, along this cut, the low-energy evolution occurs within the unpolarized sector rather than through a competition with partially polarized or fully polarized states. While we do not explicitly see a collapse of the many-body energy gap at the transition between the decoupled Laughlin and Fibonacci phases, there is a visible softening of the first excited energy band above the decoupled Laughlin's ground state manifold at $E-E_0 \approx 0.3-0.4$. In the thermodynamic limit, we expect this band to collapse and close the energy gap at some critical $V_1^{(1)}$ leading to a critical point with diverging correlation length, thus signalling the topological phase transition between the two phases. We note that a very similar transition, but with inverted energy ordering due to the presence of attractive interactions, was presented in Ref.~\onlinecite{crepel2024attractive}.

\section{Overlap of the PH(1/3) state in the torus three-fold degenerate region}
\label{app:ph13test}
\begin{figure}[h]
    \centering
    \includegraphics[width=0.4\columnwidth]{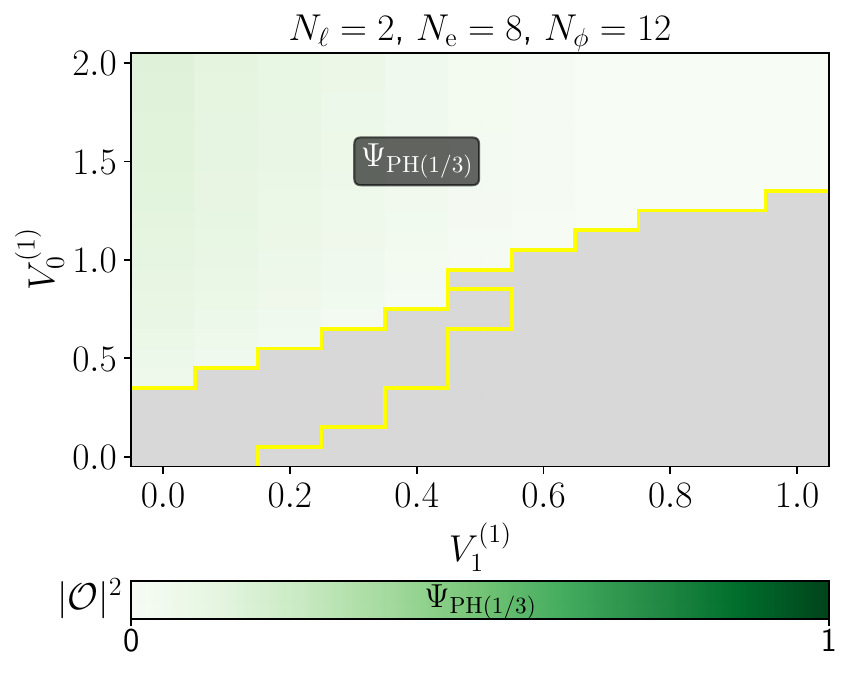}
    \caption{\textbf{Overlap with the PH(1/3) state on the square torus.} Squared overlap $|\mathcal{O}|^2$ of the $S_z=0$-rotated $\Psi_{\text{PH}(1/3)}$ with the 3-fold degenerate ground-state manifold on the square torus for bilayer with $N_\mathrm{e}=8$, $N_\phi=12$ with $V_1^{(0)}=1$. The yellow lines mark the boundary of the different degenerate regions [same boundary as in Fig.~\ref{fig:bilayer_torus}]. The overlap remains small throughout the 3-fold degenerate region.}
    \label{fig:ph13_overlap}
\end{figure}
As discussed in Sec.~\ref{sec:bilayerthintorus}, the torus phase diagram contains a three-fold degenerate ground-state manifold ($D=3$) for which Table~\ref{tab:FQHphases} lists two candidate phases sharing the same degeneracy: the Halperin~(112) state and the layer-polarized particle--hole conjugate of the Laughlin $1/3$ state, $\Psi_{\text{PH}(1/3)}$ [Eq.~\eqref{eq:PH1/3}]. The gap $\Delta$ above this manifold for $N_\mathrm{e}=8$, $N_\phi=12$ is shown in Fig.~\ref{fig:bilayer_torus}(a) in the main text. On the torus, the ground state remains pseudospin-unpolarized ($S_z=0$) throughout this entire region. We nonetheless test the identification with $\Psi_{\text{PH}(1/3)}$ directly, using the same construction as in Sec.~\ref{sec:bilayerthintorus}: we take the SU(2)-symmetric Coulomb ground state in the fully spin-polarized $2S_z=N_\mathrm{e}$ sector as a proxy for $\Psi_{\text{PH}(1/3)}$, rotate it to $S_z=0$ by repeated application of the spin-lowering operator $S^-$, and compute its overlap with the $D=3$ degenerate ground-state manifold at $S_z=0$.

As Fig.~\ref{fig:ph13_overlap} shows, the overlap stays small across the entire $D=3$ region. This is consistent with the sphere-geometry analysis already presented in App.~\ref{app:multilayer} (Fig.~\ref{fig:spin_polarization}): In the corresponding shift $\mathcal{S}=0$ sector on the sphere, the $(S_z,L)$ quantum numbers of the lowest-energy state are not stable across system size, appearing as $(S_z,L)=(1,0)$ for $N_\mathrm{e}=10$ but $(S_z,L)=(0,L\neq0)$ for other system sizes. Since a finite-size-consistent $(S_z,L)=(0,0)$ ground state is not available in this region of the phase diagram, there is no stable numerical state on the sphere to compare against the $S_z=0$-rotated $\Psi_{\text{PH}(1/3)}$, independent of the small overlap value found above.
The unpolarized torus ground state and small overlap with $\Psi_{\text{PH}(1/3)}$ rule out this identification for the $D=3$ region. The gap remains sizable and grows with $V_0^{(1)}$ at small $V_1^{(1)}$ [Fig.~\ref{fig:bilayer_torus}(a)], excluding a gapless or finite-size origin. Interpolating between the model Hamiltonian and a bare Coulomb Hamiltonian known to realize the Halperin~(112) state, we find the gap above the three-fold manifold closes and a different manifold reopens partway along the path, signaling a phase transition and ruling out (112) as well. The isotropic structure factor at this point (App.~\ref{appendix:strucfac}) further disfavors charge-density-wave order. With $\Psi_{\text{PH}(1/3)}$, (112), and charge-density-wave order all excluded, the topological character of this robustly gapped $D=3$ region remains an open question for future work.

\section{Structure factor on the torus geometry}\label{appendix:strucfac}
\begin{figure}[h!]
    \centering
    \includegraphics[width=0.99\linewidth]{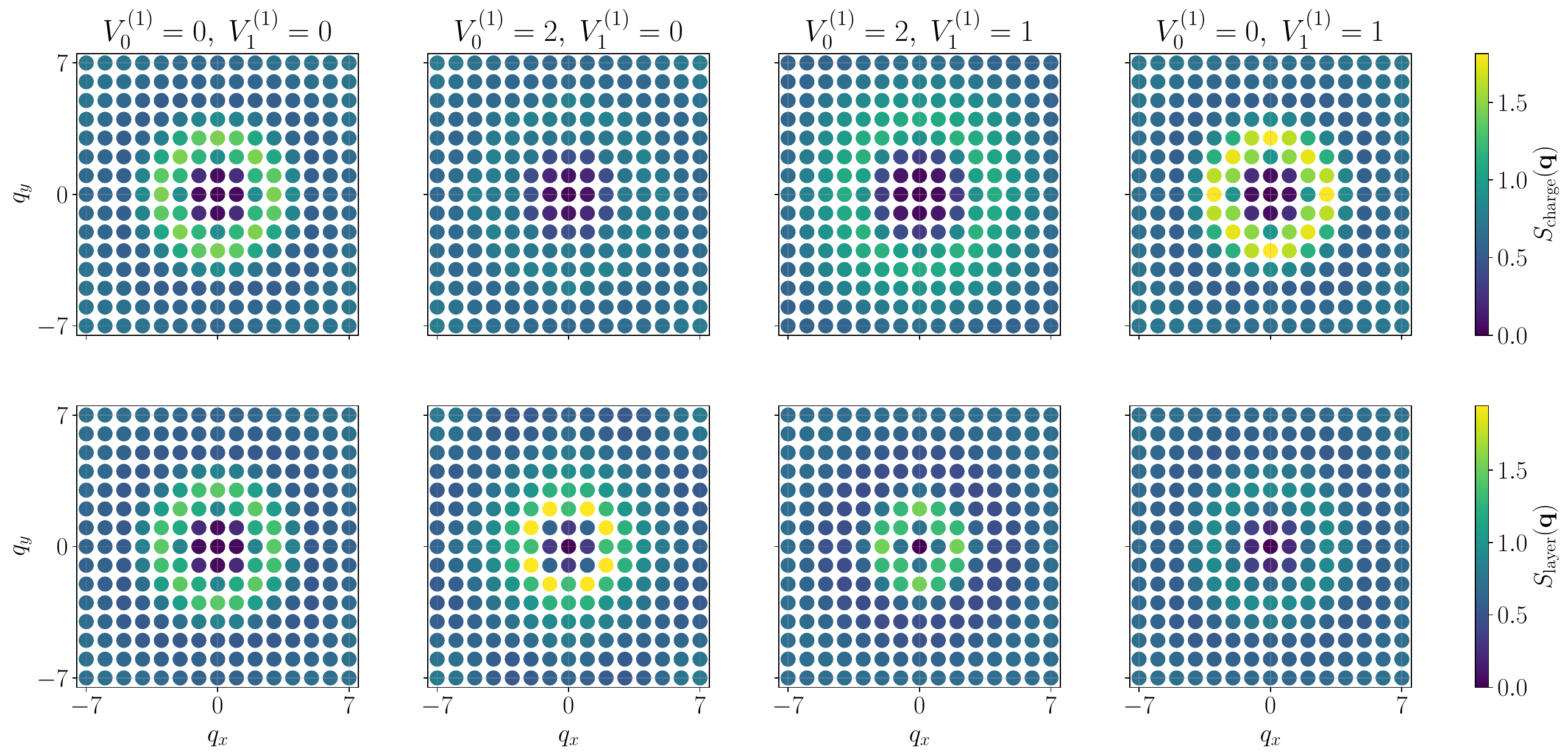}
    \caption{\textbf{Bilayer structure factors on the square torus.} We compute $S_\text{charge}(\textbf{q})$ (top row) and $S_\text{layer}(\textbf{q})$ (bottom row) for a bilayer with $N_\mathrm{e}=10$ particles at the four corners of the bilayer phase diagram similar to Fig.~\ref{fig:bilayer_torus}. At the origin, namely $q_x=q_y=0$, we subtract the particle number $N_\mathrm{e}$ and get a vanishing structure factor.}
    \label{fig:strucfactorus}
\end{figure}

Here we compute the structure factors at the four corners of the phase diagram on the square torus in Fig.~\ref{fig:bilayer_torus}, that is, at the points $(V_0^{(1)}, V_1^{(1)}) = (0.0,0.0),(0.0,1.0),(2.0,0.0),(2.0,1.0)$. At each of these four points, we take the ground state manifold $\{\ket{\Psi_i}\}_{i=1,...,D}$ where $D$ is the ground state degeneracy we find modulo center of mass translations. Thus we have $D=2$ and $D=3$ for the Fibonacci and decoupled Laughlin corners respectively, while $D=1$ for the other two corners. We then compute the structure factors for each state and average over the ground state manifold:
\begin{align}
    &S_\text{charge}(\textbf{q}) = \frac{1}{D}\sum_{i=1}^D \frac{1}{N_\mathrm{e}}\langle\Psi_i|\bar\rho_c(\textbf{q}) \bar \rho_c(-\textbf{q})|\Psi_i\rangle\\
    &S_\text{layer}(\textbf{q}) = \frac{1}{D}\sum_{i=1}^D  \frac{1}{N_\mathrm{e}} \langle\Psi_i|\bar\rho_s(\textbf{q}) \bar \rho_s(-\textbf{q})|\Psi_i\rangle
\end{align}
where we defined the lowest Landau level projected charge and pseudo-spin density operators:
\begin{equation}
    \bar\rho_c(\textbf{q}) = \bar\rho_\uparrow(\textbf{q}) + \bar\rho_\downarrow(\textbf{q}), \   \bar\rho_s(\textbf{q}) = \bar\rho_\uparrow(\textbf{q}) - \bar\rho_\downarrow(\textbf{q})
\end{equation}
and $\bar\rho_\uparrow(\textbf{q})$ ($\bar\rho_\downarrow(\textbf{q})$) are the projected density operators in the top (bottom) layer. At the origin, one gets $S_\text{charge}(\textbf{q}=0) = S_\text{layer}(\textbf{q}) = N_\mathrm{e}$, so we subtract $N_\mathrm{e}$ at the point $\textbf{q}=0$ to be able to clearly see peaks at finite $\textbf{q}$ if there are any. We plot the structure factor in Fig.~\ref{fig:strucfactorus}. For a translation symmetry breaking state, we expect to observe crystal peaks at finite $\textbf{q}=\textbf{q}_0$ consistent with the observed ground state degeneracy, and furthermore the peak in the structure factor should scale as $S(\textbf{q}_0)\sim O(N_\text{e})$ in the thermodynamic limit. We see that the structure factors in Fig.~\ref{fig:strucfactorus} are all isotropic without any sharp peaks that would be on the order of magnitude of $N_\mathrm{e}$ (we also computed the structure factor for $N_\mathrm{e}=8$ and found that the maxima are very similar in magnitude to $N_\mathrm{e}=10$)
, which are all consistent with FQH liquids. 

We also note that for the decoupled Laughlin point, the charge and layer structure factors are identical. This is expected, as with our definitions $S_\text{charge}(\textbf{q})-S_\text{layer}(\textbf{q})$ measures the inter-layer density correlations. The other three corners are therefore states with appreciable inter-layer density correlations, especially $V_0^{(1)}=0, V_1^{(1)}=1$ and $V_0^{(1)}=2, V_1^{(1)}=1$.

\section{Thin-torus analysis}\label{appendix:thintorus}

In this appendix, we derive the effective one-dimensional Hamiltonian for the bilayer quantum Hall system in the thin-torus limit \cite{papic2014solvable}. We begin be reviewing how the Haldane pseudopotential interactions reduce to a tractable second-quantized form on a torus, before carrying out the systematic expansion in the parameter $\epsilon$ as performed in Sec. \ref{sec:bilayerthintorus}. We consider $N_{\rm e}$ electrons in $N_{\ell}$ layers on a rectangular torus with dimensions $L_x$ and $L_y$, threaded by $N_{\phi}=L_xL_y/2\pi l_B^2$ magnetic flux quanta. We work in the Landau gauge ${\bf A}=-Bx\hat{y}$, in which the lowest Landau level single-particle orbitals take the form,
\begin{align}
    \psi_m(x,y)=\frac{1}{\pi^{1/4}\sqrt{L_y l_B}} \exp\left( -\frac{(x-k_m l_B^2)^2}{2l_B^2} \right) \exp\left( ik_m y \right), ~~~~~~~\text{with}~~~~~~~ k_m=\frac{2\pi m}{L_y}, ~~m=0,1,\dots,(N_{\phi}-1) .
\end{align}
Each orbital is a Gaussian strip of width $\sim l_B$ in the $x$-direction, centered at the guiding-center position $x_m=2\pi m l_B^2/L_y=m\Delta x$, where $\Delta x=2\pi l_B^2/L_y$. In the thin-torus limit $L_y \ll l_B$, one has $\Delta x \gg l_B$, so that the overlap between neighboring orbitals is exponentially suppressed. This motivates defining the dimensionless parameters, 
\begin{align}
    \kappa \equiv \frac{2\pi l_B}{L_y}=\frac{\Delta x}{l_B}, ~~~~~\epsilon\equiv e^{-\kappa^2/2} 
\end{align}
which controls the perturbative expansion. Indeed, any two-body matrix element involving a momentum transfer of $n$ orbital spacings carries a Gaussian factor $\sim \epsilon^{n^2}$. 

A generic translationally invariant two-body interaction $V({\bf r}_1 - {\bf r}_2)$, projected onto the lowest Landau level takes the form,
\begin{align}
    H = \frac{1}{2} \sum_{i,k,r} V_{k,r} c_{i+k+r}^{\dagger} c_{i}^{\dagger} c_{i+r}^{} c_{i+k}^{}
\end{align}
where $i$ is summed over all orbitals, and $k$ and $r$ are integers. We impose periodic boundary conditions along $x$ so that the orbital index is understood modulo $N_{\phi}$. Specific to the case of Haldane pseudopotentials, the matrix element $V_{k,r}$ factorizes into a polynomial prefactor $f(k,r;\kappa)$ times a Gaussian suppression $ V_{k,r} =f(k,r;\kappa) \exp\left( -\frac{\kappa^2}{2} (k^2+r^2) \right) = f(k,r;\kappa) \epsilon^{k^2+r^2}$.
Contributions at order $\mathcal{O}(\epsilon^n)$ involve only integer pairs with $k^2+r^2=n$. 
The pseudopotential $V_m$ is fixed by the requirement that it penalize pairs whose relative wavefunction vanishes as the $m$th power as the two particles approach each other. Equivalently, the Haldane pseudopotentials correspond to the form factors $L_m(q^2 l_B^2)$, the Laguerre polynomials being orthonormal under the Gaussian weight, in the Landau gauge it determines the polynomial $f(k,r;\kappa)$.

For the two pseudopotentials relevant to our bilayer system, 
the $m=0$ pseudopotential gives a constant prefactor $f(k,r;\kappa)=1$, while $m=1$ gives $f(k,r) = \kappa^2(k^2 - r^2)$. 
Note that for the intralayer interaction, $m = 0$ is forbidden by the Pauli principle, so $V_{0}$ contributes only to the interlayer interaction. As a reminder, a superscript $V_{m}^{(l-l')}$ will be used later to distinguish between the intralayer ($l=l'$) and interlayer ($l\neq l'$) pseudopotentials. This notation is consistent with the one used in the main text. 

Restoring the layer indices $l\in \{1,2\}$, the bilayer Hamiltonian with on-site and NN interactions, take the form $H=H_1^{(0)}+ H_{0}^{(1)} + H_1^{(1)}$, where,
\begin{align}
    H_1^{(0)} &= V_1^{(0)} \sum_{i,k,r,l} \kappa^2 (k^2-r^2) \epsilon^{k^2+r^2} c_{i+k+r,l}^{\dagger} c_{i,l}^{\dagger} c_{i+r,l}^{} c_{i+k,l}^{} \\
    H_0^{(1)} &= V_0^{(1)} \sum_{i,k,r,l} \epsilon^{k^2+r^2} c_{i+k+r,l}^{\dagger} c_{i,\bar l}^{\dagger} c_{i+r,\bar l}^{} c_{i+k,l}^{} \\
    H_1^{(1)} &= V_1^{(1)} \sum_{i,k,r,l} \kappa^2 (k^2-r^2) \epsilon^{k^2+r^2} c_{i+k+r,l}^{\dagger} c_{i,\bar l}^{\dagger} c_{i+r,\bar l}^{} c_{i+k,l}^{}
\end{align}
where we now introduced the superscript to indicate the intralayer ($l-l'=0$) and the interlayer (NN $l-l'=1$) interactions, and $l$ is the layer index with $\bar l=-l$. The first term $H_{1}^{(0)}$ therefore is the intralayer interaction, while the next two terms $H_{0}^{(1)}, H_{1}^{(1)}$ are the interaction between the two layers. We now expand this Hamiltonian around the thin-torus limit $\epsilon\rightarrow 0$. This is not strictly speaking a power series in one parameter but rather an expansion in the range of the interaction: A term connecting orbitals $(k,r)$ apart is suppressed by $\epsilon^{k^2+r^2}$, so truncating in $\epsilon$ retains terms up to a fixed range. 

We look at all contributions up to fourth order $\epsilon^4$. At $\epsilon^0$ order only $(k,r)=(0,0)$ contribute, therefore an interlayer density-density interaction survives at this order from $H_{0}^{(1)}$. Similarly, at $\epsilon^{1}$ order we have contributions from four pairs $(k,r)=(\pm 1,0),(0,\pm 1)$. At $\epsilon^2$, we have $(k,r)=(\pm 1,\pm 1)$. There are no terms at $\epsilon^3$, since $k^2+r^2=3$ does not have any integer solutions. Finally, at $\epsilon^4$ we have $(k,r)=(\pm 2,0),(0,\pm 2)$. After collecting all these terms, we obtain,
\begin{align}
    H &= V_0^\perp \sum_{i,l} n_{i,l}\, n_{i,\bar l} + V_1^{\parallel} \sum_{i,l} n_{i+1,l} n_{i,l} + V_1^{\perp} \sum_{i,l} \left[n_{i+1,l} n_{i,\bar l} + n_{i,l} n_{i+1,\bar l}\right] \nonumber\\ 
    &\quad + J_1 \sum_{i,l} 
    \left[c^\dagger_{i+1,l} c^\dagger_{i,\bar l} c_{i+1,\bar l} c_{i,l} + \mathrm{H.c.}\right] + \tilde{J} \sum_{i,l}
    \left[c^\dagger_{i+1,l} c^\dagger_{i-1,\bar l} c_{i,\bar l} c_{i,l} + c^\dagger_{i-1,l}\ c^\dagger_{i+1,\bar l} c_{i,\bar l} c_{i,l}+ \mathrm{H.c.}\right] \nonumber\\
  &\quad + V_2^{\parallel} \sum_{i,l} n_{i+2,l} n_{i,l} + V_2^{\perp} \sum_{i,l} \left[n_{i+2,l} n_{i,\bar l}  + n_{i-2,l} n_{i,\bar l}\right]+ J_2 \sum_{i,l} \left[c^\dagger_{i+2,l} c^\dagger_{i,\bar l} c_{i+2,\bar l} c_{i,l} + \mathrm{H.c.}\right] + \mathcal{O}(\epsilon^5)
\end{align}
where the interaction coefficients are defined in terms of the intra/interlayer pseudopotentials in the following way
\begin{align}
    \mathcal{O}(\epsilon^0)&:~~~~~~~~~~V_0^\perp=V_0^{(1)},\\
    \mathcal{O}(\epsilon^1)&:~~~~~~~~~~V_1^{\parallel}=4\epsilon \kappa^2V_1^{(0)},~~V_1^{\perp}=\epsilon \left( V_0^{(1)} + \kappa^2 V_1^{(1)} \right),~~J_1=\epsilon\left( V_0^{(1)} - \kappa^2 V_1^{(1)} \right),\\
    \mathcal{O}(\epsilon^2)&:~~~~~~~~~~\tilde{J} = \epsilon^2 V_0^{(1)},\\
    \mathcal{O}(\epsilon^4)&:~~~~~~~~~~V_2^{\parallel}=16\epsilon^4 \kappa^2V_1^{(0)},~~V_2^{\perp}=\epsilon^4 \left( V_0^{(1)} + 4\kappa^2 V_1^{(1)} \right),~~J_2=\epsilon^4\left( V_0^{(1)} - 4\kappa^2 V_1^{(1)} \right).
\end{align}

We focus on the regime where $V_1^\parallel$ is the largest energy scale and $V_2^\parallel$ is appreciable. 
We further consider $V_{0}^\perp, V_{1}^\perp$ and $J_1$ to be of the same order as $V_2^\parallel$. Hence we have $\tilde{J} = \epsilon^2 V_0^\perp$ and $V_2^\perp \sim \epsilon^3 V_1^\perp$, $J_2 \sim \epsilon^3 J_1$ suppressed with powers of $\epsilon$, and we will ignore these terms. We note that this choice would correspond to a $V_1^{(0)}$ much larger than other pseudo-potentials. However, we do not expect the exact matching between the energetics in the thin-torus regime and the thermodynamic limit. But we will demonstrate that the trends of stability of the various phases match with the numerical simulations. 

We are then left with the following Hamiltonian,
\begin{align}
    \tilde{H} &= \sum_{r=1,2} V_r^{\parallel} \sum_{i,l} n_{i+r,l} n_{i,l}  + \sum_{r=0,1} V_r^\perp \sum_{i,l} n_{i+r,l}  n_{i,\bar l} + J_1 \sum_{i,l} 
    \left[c^\dagger_{i+1,l} c^\dagger_{i,\bar l} c_{i+1,\bar l} c_{i,l} + \mathrm{H.c.}\right].
    \label{eq:app:Heff-torus}
\end{align}
Now let us consider filling $1/3$ in each layer. With the above energetics, the Hilbert space can be decomposed into disjoint sectors. To see this, let us denote the configuration in the occupation basis with a shorthand notation of total particle number on each site. For example $|200\dots\rangle$ denotes a state $\left| \begin{matrix}
100\dots\\100\dots
\end{matrix} \right\rangle$, and $|110\dots\rangle$ denotes a sub-Hilbert space with two states $\left| \begin{matrix}
100\dots\\010\dots
\end{matrix} \right\rangle$ and $\left| \begin{matrix}
010\dots\\100\dots
\end{matrix} \right\rangle$. 

It is readily seen that states with only $2$'s are eigenstates of the Hamiltonian since these states are annihilated by the pair-hopping terms $J_1$. Within this class of states, the ground states are charge density waves with periodicity three, i.e.~$|200200\dots\rangle$, with energy $E_{D_0} = N V_0^\perp/2$, where $N$ is the total particle number.
We call these states `diagonal' in the main text, and they are threefold degenerate.
The excited states have energies $E_{\rm D_{\rm{ex}}} = N V_0^\perp/2 + m V_2^\perp$, where $m$ is the pair of second nearest neighbor pairs. 

Next, we consider states with only $1$'s. We can view the pair hopping terms as only changing the layer labels. Hence sub-Hilbert spaces with different shorthand notations do not mix. Within this class of sub-Hilbert spaces, the states with lowest energies are ``maximally" sparse, such as $|110110\dots \rangle$. By maximally sparse, we mean that there are least numbers of consecutive $1$'s, and in turn least number of $V_1^{\perp}$ interaction pairs. A state in the maximally sparse subspace have lower $V_1^\perp$ and can gain more exchange energy from $J_1$ compared. There are three of such maximally sparse sub-Hilbert space, related by lattice translation. The dimension of each sub-Hilbert space is $2^{N/2}$. Within such sub-Hilbert space, the effective Hamiltonian can be mapped to the transverse field Ising model, by grouping three sites into one and define locally two orbitals,
\begin{align}
\ket{\uparrow}\equiv  \left| \begin{matrix}
100\\010
\end{matrix} \right\rangle, \ket{\downarrow}\equiv \left| \begin{matrix}
010\\100
\end{matrix} \right\rangle.
\end{align}
The Hamiltonian Eq.~\eqref{eq:app:Heff-torus} can be written as,
\begin{align}\label{eq:model_spinchain}
\tilde{H}_{\text{Ising}} = \sum_{i} J S_i^x - \sum_{i}\Delta S_{i}^{z} S_{i+1}^z + V_1^\perp N/2 + V_2^\parallel N /4,
\end{align}
where $S^{x,z}$ are Pauli matrices, $J = J_1$ and $\Delta = V_2^\parallel/2$. The above Hamiltonian was also obtained in Ref.~\onlinecite{Vaezi2014} with a different microscopic Hamiltonian. $\tilde{H}_{\text{Ising}}$ has two phases: ferromagnetic phase when $\Delta > |J|$ and paramagnetic phase otherwise. In the ferromagnetic phase, the ground states are two-fold degenerate, and are smoothly connected to the polarized states $|\uparrow \uparrow \dots \rangle$ and $|\downarrow \downarrow \dots \rangle$. Together with the center of mass translations, this leads to a sixfold degeneracy, which becomes a total 9-fold degeneracy if we include the diagonal states, as expected for the decoupled Laughlin state. In the paramagnetic phase, the ground state is smoothly connected to a product state. When $J>0$, the product state is $\ket{s}=\frac{1}{\sqrt{2}}(|\uparrow\rangle - |\downarrow\rangle)$ locally, which we call ``layer anti-symmetric" in the main text. When $J<0$, the product state is $\frac{1}{\sqrt{2}}(|\uparrow\rangle + |\downarrow\rangle)$ locally, which we call ``layer symmetric". The center of mass translations make this state 3-fold degenerate, which together with the diagonal states, constitute the 6-fold degenerate ground state manifold of the Fibonacci state.

So far, we have only considered the ground states in the sectors with all $2$'s or all $1$'s in our shorthand notation. In general, the sectors with mixed $2$'s and $1$'s lies above the ground state. 
In the regime where $V_1^{\parallel}$ dominates, each layer forms a period 3 CDW and every orbital makes a single interlayer contact, so the classical part of the energy for any such configuration is $E\sim V_0^{\perp} N_{200} + V_1^{\perp} N_{110} + V_2^{\parallel} N_{\text{intercell}}$, with $N_{110}+N_{200}=N_{\text{cell}}$ and $N_{\text{intercell}}$ being the same-layer next-nearest-neighbor orbital contacts at domain boundaries. As all coefficients are non-negative, $E\ge \min(V_0^{\perp},V_1^{\perp})N_{\text{cell}}$, i.e., no mixed configuration falls below the lower of the two uniform patterns. When the two are nearly degenerate the residual splitting is set by the $J_1$ pair-hopping, which lowers the sparse sector but annihilates every doubly-occupied orbital, so any admixture of $\ket{200}$ units forfeits this gain. The mixed sectors are therefore separated from the ground state manifold, and we don't expect them to be smoothly connected to the topologically degenerate ground states on the torus in the thermodynamic limit.

\textbf{Optimized CDW roots in multilayer system.} We now address the problem of finding the optimized CDW roots in a multilayer system such that for each filled layer at a given filling $\nu$, the CDW root is able to minimize its classical density-density interaction energy by `shifting' its root relative to the adjacent layers. Once again, since we are interested in the configurations that correspond to filling of $\nu=1/3,2/3,$ and $2/9$ in each filled layer, their roots are $\dots 100100100 \dots$ of period 3 for $\nu=1/3$, $\dots 110110110\dots$ of period 3 for $\nu=2/3$, and $\dots 100010000\dots$ of period 9 for $\nu=2/9$. We now label different CDW root configurations in each layer by $\alpha_l$ for a given filling. For example, in the 9-orbital unit cell at $\nu=2/9$, $\alpha_l=0$ has occupied positions at $\{0,4\}$, represented by $100010000$, while for $\nu=1/3$, the root $\alpha_l=0$ has occupied positions at $\{0,3,6\}$ represented by $100100100$. The $\nu=2/9$ CDW is 9-fold degenerate per layer, while $\nu=1/3$ is 3-fold degenerate, as $\alpha_l=0,3,6$ produce the same pattern.

Since in our thin-torus Hamiltonian Eq.~(\ref{eq:thintorus_N6}), we have density-density interaction terms between NN and NNN layers, the ground states would minimize NN and NNN contacts between filled orbitals in the CDW by staggering the roots between different layers. For a large number of layer $N_{\ell}$, we find the most optimal relative CDW root pattern that minimizes NN and NNN contacts between occupied orbitals. Moreover, since $V_0^{\perp}=V_0^{(1)}$ is at order $\mathcal{O}(\epsilon^0)$, configurations with energies proportional to $V^{\perp}_0$ will not be energetically favourable. The $\nu = 2/9$ root places only 2 electrons per 9 orbitals, providing substantial freedom to stagger the CDW patterns across layers. For large $N_\ell$, a period-2 alternating shift sequence $\alpha_{l+1} - \alpha_l \in \{2, 6\}$ achieves $1/2$ NN contact and 2 NNN contacts per layer while satisfying all on-site constraints. The generated $\alpha_l$ values have period~18, with the NNN shift $2+6 \equiv 8 \pmod 9$ lying outside the forbidden set $\{0,4,5\}$ (that result in configurations with energy $\sim V_0^{\perp}$) throughout. Therefore, for the $\nu=2/9$ uniformly filled system, $[\frac{2}{9}\frac{2}{9}\frac{2}{9}]^{N_{\ell}/3}$, the most optimal CDW root configuration is given by the sequence $\alpha_l=(0,2,8,1,7,0,6,...)$, such that $\alpha_{l+1}-\alpha_l$ alternates between $2$ and $6$.

\section{Non-Abelian states}\label{app:nonabelian_states}
\subsection{Fibonacci anyon content}
 The full quasiparticle content is summarised in
Table~\ref{tab:qps}.
 
\begin{table}[ht!]
\centering
\begin{tabular*}{0.5\columnwidth}{@{\extracolsep{\fill}}ccc}
\hline\hline
Species & Charge & Quantum dimension \\
\hline
$a_0 \equiv \mathbf{1}$ & $0$       & $1$ \\
$a_1$                   & $2e/3$    & $1$ \\
$a_2$                   & $4e/3$    & $1$ \\
$\tau$                  & $0$       & $\varphi$ \\
$a_1\tau$               & $2e/3$    & $\varphi$ \\
$a_2\tau$               & $4e/3$    & $\varphi$ \\
\hline\hline
\end{tabular*}
\caption{\textbf{Anyon content of the Fibonacci state.} The three
Abelian sectors $a_{\alpha}$ are inherited from the $(330)$ state; the three
non-Abelian sectors are obtained by dressing with the Fibonacci anyon
$\tau$.}
\label{tab:qps}
\end{table}

\subsection{$\text{SU(3)}_{N_\ell}$ quasihole counting}\label{app:QHcount}
Here we generalize the exclusion principles of the bilayer Fibonacci state to the case of $\text{SU(3)}_{N_\ell}$ states for arbitrary $N_\ell$. The $\text{SU(3)}_{N_\ell}$ exclusion principles can be derived in a similar fashion as was done in Ref.~\onlinecite{Liu2015} for $N_\ell=2$. The torus Hamiltonian with $V_1^{(0)}$, $V_0^{(1)}$ and $V_1^{(1)}$ pseudopotentials takes the form:
\begin{align}
    &H = H_1^{(0)} + H_0^{(1)} + H_1^{(1)},\\
    &H_1^{(0)} = V_1^{(0)} \sum_l \sum_{j,r,s} U^{r,s}_{1} 
    c^\dagger_{j+r,l}c^\dagger_{j+s,l} c_{j,l}c_{j+r+s,l},\\
    &H_0^{(1)} = V_0^{(1)} \sum_{\braket{ll'}} \sum_{j,r,s} U^{r,s}_0
    (A^{ll'}_{j+r,j+s})^\dagger A^{ll'}_{j,j+r+s},\\
    &H_1^{(1)} = V_1^{(1)} \sum_{\braket{ll'}} \sum_{j,r,s} U^{r,s}_1
    (S^{ll'}_{j+r,j+s})^\dagger S^{ll'}_{j,j+r+s},
\end{align}
where $c_{j,l}$ annihilates an electron in the LLL orbital in the Landau gauge with momentum $k_y = 2\pi j/L_y$ in layer $l$, $U^{r,s}_0 = \frac{\kappa}{\sqrt{2\pi}}e^{-(r^2+s^2)\kappa^2/2}$, $U^{r,s}_1 = \frac{\kappa^3}{\sqrt{2\pi}}(r^2-s^2)e^{-(r^2+s^2)\kappa^2/2}$ with $\kappa = \frac{2\pi l_B}{L_y}$, and we defined the following two-particle annihilation operators:
\begin{align}
    &A_{ij}^{ll'} = \frac{1}{\sqrt{2}}(c_{i,l}c_{j,l'} - c_{i,l'}c_{j,l}), \\ 
    &S_{ij}^{ll'} = \frac{1}{\sqrt{2}}(c_{i,l} c_{j,l'} + c_{i,l'} c_{j,l}),
\end{align}
which annihilate two particles in orbitals $i$ and $j$ and layers $l$ and $l'$ in a layer-antisymmetric or layer-symmetric superposition respectively. In the thin-torus limit where $\kappa \rightarrow \infty$, we can truncate the sum over $r,s$ and obtain the thin-torus Hamiltonian obtained in Ref.~\onlinecite{Liu2015}:
\begin{align}
    &H_1^{(0)} \rightarrow V_1^{(0)}\sum_l \sum_j [U^{1,0}_1  n_{j,l} n_{j+1,l} + U_1^{2,0} n_{j,l}n_{j+2,l}],\\
    &H_0^{(1)} \rightarrow V_0^{(1)} \sum_{\braket{ll'}} \sum_j [U^{0,0}_0 n_{j,l} n_{j,l'} + U^{1,0}_0 (A_{j,j+1}^{ll'})^\dagger  A_{j,j+1}^{ll'}],\\
    &H_1^{(1)} \rightarrow V_1^{(1)} \sum_{\braket{ll'}} \sum_j U^{1,0}_0 (S_{j,j+1}^{ll'})^\dagger S_{j,j+1}^{ll'}.
\end{align}
In the limit where $V_0^{(1)}\ll V_1^{(1)}, V_1^{(0)}$, the Hamiltonian penalizes pairs of particles in adjacent layers forming layer-symmetric states, assuming their intralayer separation is sufficiently small. Since we truncated the Hamiltonian to NNN intralayer terms, sufficiently small will mean that the particles are separated by at most one orbital (e.g.~the thin-torus patterns $[...11...]$ and $[...101...]$ could get penalized, but not $[...1001...]$). 

More precisely, $V_1^{(0)}$ imposes a $(k,r) = (1,3)$ exclusion principle \cite{bernevig2008properties} in each layer, that is:
\begin{equation}
    n_{i,l} + n_{i+1,l} + n_{i+2,l} \leq 1 \text{ for }l=1,2,...,N_\ell,
\end{equation}
where $n_{j,l}$ is the occupation number at site $j$ and layer $l$ of the thin-torus pattern. Summing over the layers we thus find:
\begin{equation}
    n_{i} + n_{i+1} + n_{i+2} \leq N_\ell,
\end{equation}
where $n_i$ is the occupation number at site $j$ of the layer-summed thin-torus pattern. The intralayer interaction thus imposes a $(k,r)=(N_\ell,3)$ exclusion principle of no more than $N_\ell$ particles in any three consecutive orbitals on the layer-summed thin-torus patterns. 

However, a given layer-summed thin-torus pattern may have multiple corresponding layer-resolved thin-torus patterns, which of these are not penalized by the interlayer interaction $V_1^{(1)}$? From our previous discussion, we must remove those patterns which have sufficiently close (i.e.~at most NNN intralayer separation) pairs of particles in adjacent layers forming layer-symmetric states.

Let us focus on $N_\ell=3$, and suppose we have found a layer-summed thin-torus state $[\lambda_1,...,\lambda_{N_\mathrm{e}}]$ of $N_\mathrm{e}$ particles that obeys the $(k,l)=(N_\ell,3)$ exclusion principle, where $\lambda_n$ is the orbital occupied by particle $n$. For each layer-resolved thin-torus pattern $[(\lambda_1,l_1),...,(\lambda_{N_\mathrm{e}},l_{N_\mathrm{e}})]$ with particle $n$ in layer $l_n$, we can associate a weight $\psi_{l_1,...,l_{N_\mathrm{e}}}$, forming the thin-torus state $\sum_{\{l_i\}} \psi_{l_1,...,l_{N_\mathrm{e}}} [(\lambda_1,l_1),...,(\lambda_{N_\mathrm{e}},l_{N_\mathrm{e}})]$. This state should obey the interlayer anti-symmetrization exclusion principle, which effectively anti-symmetrizes the tensor $\psi$ in some of its indices, depending on the intralayer separation of the corresponding occupied orbitals. The number of remaining degrees of freedom in $\psi$ indicates the multiplicity of the layer-summed thin-torus pattern. For a single particle thin-torus pattern, the tensor $\psi_l$ transforms under the fundamental representation $\mathbf{3}$ of $\text{SU(3)}$. For a two-particle thin-torus pattern, the tensor $\psi_{l,l'}$ transforms as $\mathbf{3} \otimes \mathbf{3} = \mathbf{6} \oplus \bar{\mathbf{3}}$, since it can be decomposed into a symmetric part (transforming under $\mathbf{6}$) and anti-symmetric part (transforming under $\bar{\mathbf{3}}$). If the two particles are close, only $\bar{\mathbf{3}}$ will survive, while if they are sufficiently separated the state can live in both representations. Finally, for a three particle thin-torus pattern, the tensor $\psi_{l_1,l_2,l_3}$ transforms as $\mathbf{3} \otimes \mathbf{3} \otimes \mathbf{3}= \mathbf{10} \oplus \mathbf{8} \oplus \mathbf{8} \oplus \mathbf{1}$. The $\mathbf{10}$ is the fully symmetric representation, that is the space of fully symmetric rank-3 tensors $\psi_{(l_1,l_2,l_3)}$. Similarly, $\mathbf{1}$ is the fully anti-symmetric singlet $\psi_{[l_1,l_2,l_3]} \equiv \epsilon_{l_1,l_2,l_3}$. Finally the two copies of the adjoint representation $\mathbf{8}$ are mixed, for example if we imagine $\psi_{l_1,l_2,l_3}$ being antisymmetric in $l_1,l_2$ and free in $l_3$, then it transforms in $\bar{\mathbf{3}}\otimes \mathbf{3} = \mathbf{1} \oplus \mathbf{8}$. 

Let us look at some concrete examples. $\ket{300}$, or equivalently $[0,0,0]$, has multiplicity one, as anti-symmetry forces $\psi_{l_1,l_2,l_3}$ in the singlet representation $\mathbf{1}$. Likewise $\ket{210}, \ket{120}$ and $\ket{111}$ also must be singlets. Together with their translated counterparts, these thin-torus patterns account for the 10-fold ground state degeneracy of the $\text{SU(3)}_3$ state on the torus. This also agrees with the number of anyon species in the $\text{SU(3)}_3$ Chern-Simons theory. 

For $\ket{2003001}$ the first two particles must transform in the $\bar{\mathbf{3}}$ representation, the next three in $\mathbf{1}$ and the last in $\mathbf{3}$, leading to $\mathbf{1}\oplus \mathbf{8}$ and thus a multiplicity of nine of which three come from the layer-balanced sector (same number of particles in each layer). Note that this last example was particularly easy as it is clear how to partition the state into clusters on which the exclusion principles act independently (by cluster we mean ordered blocks of particles whose positions fit inside a length-three orbital window). Let us instead consider $\ket{1110111}$ where several options are available. We can have two clusters of three particles, $(111)(111)$, each in the singlet representation. Alternatively, we can form three pairs of particles $(11)(101)(11)$ transforming under the representation $\bar{\mathbf{3}} \otimes \bar{\mathbf{3}} \otimes \bar{\mathbf{3}}$. Clearly, the first option is favourable as it minimizes the number of unhappy interlayer bonds across which the state is not projected to the anti-symmetric sector. Thus this state has a multiplicity of one.

In general, one should form the clusters of 3 consecutive orbitals in order of the contained particle number, first clusters of 3, then 2 and finally 1. For clusters of 2, prioritize $110$ clusters over $101$ clusters. This ensures that the density of layer-singlets is maximal, as required by the thin-torus Hamiltonian. Once a decomposition into clusters has been found, each cluster will transform under a $\mathbf{1}$, $\bar{\mathbf{3}}$ or $\mathbf{3}$ depending on if it is a cluster of 3, 2 or 1 particles respectively. This gives the quasihole counting resolved into $\text{SU(3)}$ layer-multiplets. For a given distribution of particles in each layer, say $(N_1,N_2,N_3)$, the multiplicity of a root decomposed into $m$ clusters with $q_j$ particles in cluster $j$ can be obtained by computing the coefficient of the monomial $x_1^{N_1} x_2^{N_2} x_3^{N_3}$ in the generating functional
\begin{equation}
    Z(x_1,x_2,x_3) = \prod_{i=1}^m e_{q_j}(x_1,x_2,x_3)
\end{equation}
where $e_{q_j}(x_1,x_2,x_3)$ is the elementary symmetric polynomial in three variables of order $q_j$:
\begin{align}
    &e_{1}(x_1,x_2,x_3) = x_1+x_2+x_3\\
    &e_2(x_1,x_2,x_3) = x_1x_2+x_2x_3+x_3x_1\\
    &e_3(x_1,x_2,x_3) = x_1x_2x_3
\end{align}
Generalizing to $N_\ell$-layers, we have the following $\text{SU(3)}_{N_\ell}$ exclusion principles:
\begin{enumerate}
    \item[(i)] no more than $N_\ell$ particle in 3 consecutive orbitals,
    \item[(ii)] decompose the layer-summed thin-torus pattern into clusters of particles so as to maximize the layer-singlet density. Hence, start with $N_\ell$ particle clusters, then $N_\ell-1$ particle clusters and so on. For each class of particle cluster, prioritize formation of clusters where the particles are closer together i.e.~for $m$ particle cluster, $m00...$ then $(m-1)10...$ and so on.
    \item[(iii)] for each layer-summed thin-torus pattern, with a decomposition into $m$ clusters of particles, with $q_j$ being the number of particles in cluster $j$, compute the coefficient of the monomial $\prod_{i=1}^{N_\ell} x_i^{N_i}$ in $\prod_{j=1}^m e_{q_j}(x_1,...,x_{N_\ell})$, where $e_q(x_1,...,x_{N}) = \sum_{1\leq a_1\leq...\leq a_q\leq N}x_{a_1}...x_{a_q}$ are the elementary symmetric polynomials. This coefficient is the multiplicity of the thin-torus pattern in the sector with layer-occupation numbers $N_1,...,N_{N_\ell}$,
\end{enumerate}

\begin{figure}
    \centering
    \includegraphics[width=0.99\linewidth]{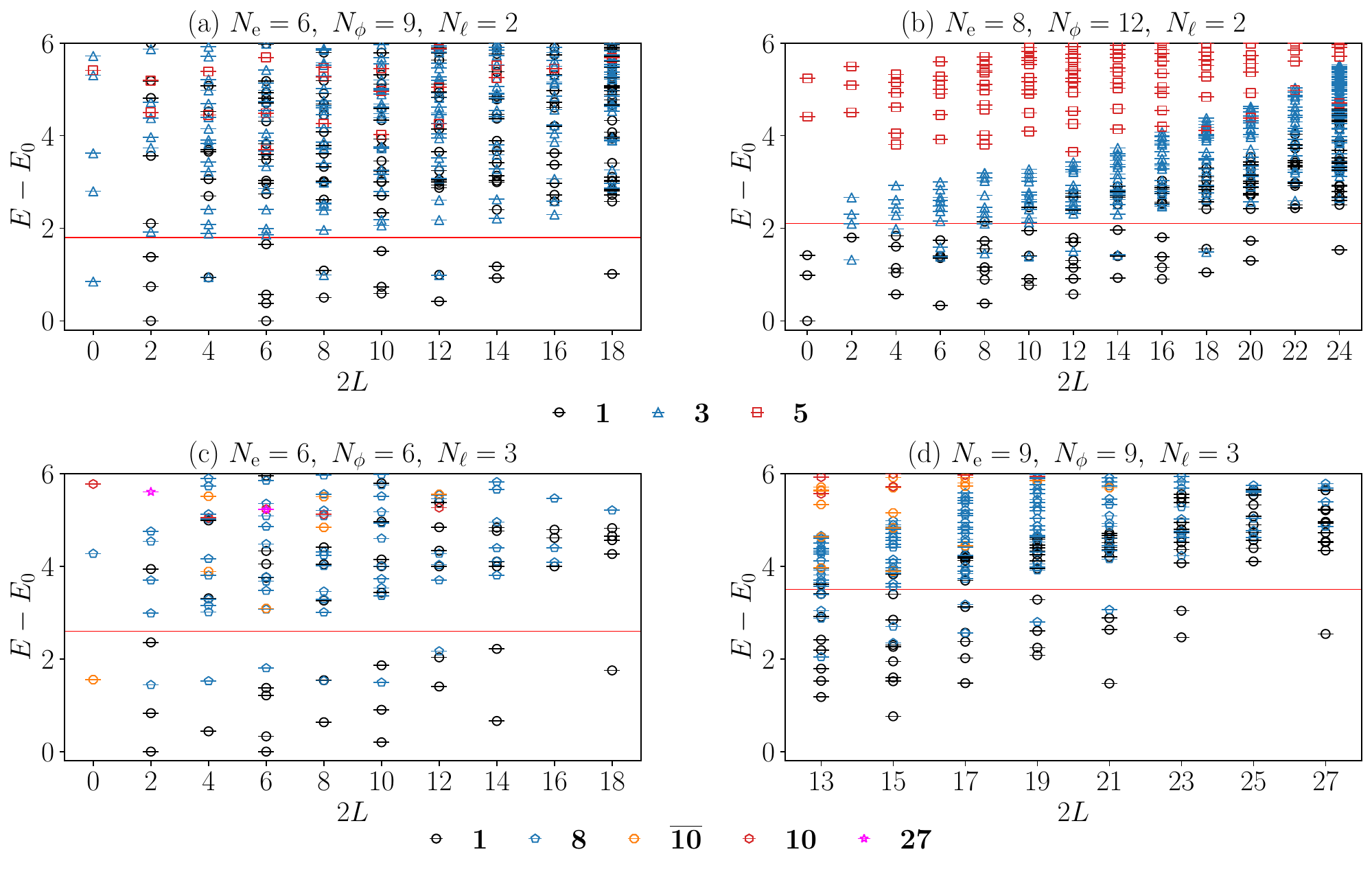}
    \caption{Quasihole spectra in the homogeneous particle-per-layer sectors for $N_\mathrm{e}=6$ and $N_\mathrm{e}=8$ particle systems with $N_\ell=2$, and $N_\mathrm{e}=6$ and $N_\mathrm{e}=9$ for $\nl=3$, with 3 flux quanta added on top of the $\mathcal{S}=3$ ground state and $\text{SU}(\nl)$-symmetric interactions $V_1^{(0)}=V_1^{(1)}=1$. Due to the SU(2) rotation symmetry and $\text{SU}(\nl)$ layer symmetry, we can label the eigenstates by their respective irreps. The red line indicates where the quasihole manifold terminates. We only plot the spectra in those $L$ sectors where we computed enough eigenstates to determine the multiplet decomposition up to the red line, so there may be additional states above it (but not below). This is why panel (d) is cut off below $2L = 13$.
    }
    \label{fig:QH_bilayer_trilayer}
\end{figure}

We have computed the quasihole counting obtained from these exclusion principles for $N_\ell = 2, 3$, and compared it with the numerically observed counting in ED in Tab.~\ref{tab:bilayer-trilayer-multiplets}. Owing to the combined orbital SU(2)$_\mathrm{orb}$ and layer $\text{SU}(N_\ell)_\mathrm{layer}$ symmetries when $V_1^{(1)} = V_1^{(0)}$, the quasihole manifold can be simultaneously decomposed into SU(2)$_\mathrm{orb}$ and $\text{SU}(N_\ell)_\mathrm{layer}$ multiplets. As we can see, the predicted counting agrees with the observed counting down to a certain value of $2L$, below which finite-size effects begin to affect the counting. For example, in the trilayer system with $N_\mathrm{e}=9$, the counting agrees up to $2L=17$, below which the quasihole gap becomes negligibly small (see Fig.~\ref{fig:QH_bilayer_trilayer}), and there's no reason to  expect the quasihole manifold to remain separate from other, non-universal excitations in the system anymore. 

\newcommand{\match}{\textcolor{blue}{\checkmark}}

\begin{table}[ht]
\centering
\renewcommand{\arraystretch}{1.05}
\setlength{\tabcolsep}{2pt}

\begin{adjustbox}{max width=\textwidth, max totalheight=0.82\textheight, center}

\begin{tabular}{|c|cc|cc|cc|cc|}
\hline
& \multicolumn{2}{c|}{Bilayer, $N_\mathrm{e}=6$}
& \multicolumn{2}{c|}{Bilayer, $N_\mathrm{e}=8$}
& \multicolumn{2}{c|}{Trilayer, $N_\mathrm{e}=6$}
& \multicolumn{2}{c|}{Trilayer, $N_\mathrm{e}=9$} \\
\hline
$2L$
& Pred. & ED
& Pred. & ED
& Pred. & ED
& Pred. & ED \\
\hline

$0$  &
$\mathbf{3}$ & \match &
$\mathbf{1}^{\oplus 4}$ & $\mathbf{1}^{\oplus 3}$ &
$\overline{\mathbf{10}} \oplus \mathbf{8}$ & $\overline{\mathbf{10}}$ &
& \\

$1$  &
& &
& &
& &
$\mathbf{8} \oplus \mathbf{1}^{\oplus 2}$ & -- \\

$2$  &
$\mathbf{1}^{\oplus 3}$ & \match &
$\mathbf{3} \oplus \mathbf{1}$ & \match &
$\mathbf{8} \oplus \mathbf{1}^{\oplus 3}$ & \match &
& \\

$3$  &
& &
& &
& &
$\mathbf{8}^{\oplus 3} \oplus \mathbf{1}^{\oplus 8}$ & -- \\

$4$  &
$\mathbf{3} \oplus \mathbf{1}$ & \match &
$\mathbf{1}^{\oplus 6}$ & $\mathbf{3} \oplus \mathbf{1}^{\oplus 5}$ &
$\mathbf{1}$ & $\mathbf{8} \oplus \mathbf{1}$ &
& \\

$5$  &
& &
& &
& &
$\mathbf{8}^{\oplus 4} \oplus \mathbf{1}^{\oplus 9}$ & -- \\

$6$  &
$\mathbf{1}^{\oplus 5}$ & $\mathbf{1}^{\oplus 4}$ &
$\mathbf{3}^{\oplus 2} \oplus \mathbf{1}^{\oplus 4}$ & \match &
$\mathbf{8}^{\oplus 2} \oplus \mathbf{1}^{\oplus 5}$ & $\mathbf{8} \oplus \mathbf{1}^{\oplus 4}$ &
& \\

$7$  &
& &
& &
& &
$\mathbf{8}^{\oplus 4} \oplus \mathbf{1}^{\oplus 10}$ & -- \\

$8$  &
$\mathbf{3} \oplus \mathbf{1}^{\oplus 2}$ & \match &
$\mathbf{3} \oplus \mathbf{1}^{\oplus 8}$ & $\mathbf{3} \oplus \mathbf{1}^{\oplus 6}$ &
$\mathbf{8} \oplus \mathbf{1}^{\oplus 2}$ & \match &
& \\

$9$  &
& &
& &
& &
$\overline{\mathbf{10}} \oplus \mathbf{8}^{\oplus 5} \oplus \mathbf{1}^{\oplus 11}$ & -- \\

$10$ &
$\mathbf{1}^{\oplus 3}$ & \match &
$\mathbf{3} \oplus \mathbf{1}^{\oplus 4}$ & \match &
$\mathbf{8} \oplus \mathbf{1}^{\oplus 3}$ & \match &
& \\

$11$ &
& &
& &
& &
$\mathbf{8}^{\oplus 4} \oplus \mathbf{1}^{\oplus 11}$ & -- \\

$12$ &
$\mathbf{3} \oplus \mathbf{1}^{\oplus 2}$ & \match &
$\mathbf{3} \oplus \mathbf{1}^{\oplus 7}$ & $\mathbf{3} \oplus \mathbf{1}^{\oplus 6}$ &
$\mathbf{8} \oplus \mathbf{1}^{\oplus 2}$ & \match &
& \\

$13$ &
& &
& &
& &
$\mathbf{8}^{\oplus 3} \oplus \mathbf{1}^{\oplus 8}$ &
$\mathbf{8}^{\oplus 4} \oplus \mathbf{1}^{\oplus 7}$ \\

$14$ &
$\mathbf{1}^{\oplus 2}$ & \match &
$\mathbf{3} \oplus \mathbf{1}^{\oplus 3}$ & \match &
$\mathbf{1}^{\oplus 2}$ & \match &
& \\

$15$ &
& &
& &
& &
$\mathbf{8}^{\oplus 3} \oplus \mathbf{1}^{\oplus 9}$ &
$\mathbf{8}^{\oplus 2} \oplus \mathbf{1}^{\oplus 8}$ \\

$16$ &
$\mathrm{None}$ & \match &
$\mathbf{1}^{\oplus 4}$ & \match &
$\mathrm{None}$ & \match &
& \\

$17$ &
& &
& &
& &
$\mathbf{8}^{\oplus 2} \oplus \mathbf{1}^{\oplus 5}$ & \match \\

$18$ &
$\mathbf{1}$ & \match &
$\mathbf{3} \oplus \mathbf{1}^{\oplus 2}$ & \match &
$\mathbf{1}$ & \match &
& \\

$19$ &
& &
& &
& &
$\mathbf{8} \oplus \mathbf{1}^{\oplus 4}$ & \match \\

$20$ &
& &
$\mathbf{1}^{\oplus 2}$ & \match &
& &
& \\

$21$ &
& &
& &
& &
$\mathbf{8} \oplus \mathbf{1}^{\oplus 3}$ & \match \\

$22$ &
& &
$\mathrm{None}$ & \match &
& &
& \\

$23$ &
& &
& &
& &
$\mathbf{1}^{\oplus 2}$ & \match \\

$24$ &
& &
$\mathbf{1}$ & \match &
& &
& \\

$25$ &
& &
& &
& &
$\mathrm{None}$ & \match \\

$26$ &
& &
& &
& &
& \\

$27$ &
& &
& &
& &
$\mathbf{1}$ & \match \\

\hline
\end{tabular}
\end{adjustbox}

\caption{Comparison of the predicted and numerically observed layer multiplet decompositions
of the quasihole spectra for bilayer and trilayer systems. The check mark indicates that the ED decomposition agrees with the prediction.
Entries in the ED column where the Lanczos did not converge a sufficient number of energy values
to get a full count below the quasihole gap are marked with --. For the bilayer, $\mathbf{1}$ and $\mathbf{3}$ are the singlet ($S=0$) and triplet ($S=1$) representations respectively, while for the trilayer $\mathbf{1}$, $\mathbf{3}$, $\mathbf{8}$ and $\mathbf{10}$ are the singlet, fundamental, adjoint and decuplet representations respectively, with $\bar{\mathbf{a}}$ denoting the complex conjugate representation of $\mathbf{a}$ (note $\mathbf{1}$ and $\mathbf{8}$ are self-conjugate).}
\label{tab:bilayer-trilayer-multiplets}
\end{table}

We close by remarking that our quasihole counting also works if the symmetries in the Hamiltonian are broken. For example, if the Hamiltonian lacks SU$(\nl)$ symmetry, then the total quasihole count in each $L$ sector will still be the same, but we won't be able to label the quasihole states by SU$(\nl)$ irreps. For example, suppose we want to obtain the quasihole counting in the $L=6$ sector for the bilayer with $N_\mathrm{e}=6$ particles using a Hamiltonian that has spin U(1)$_S$ symmetry rather than SU(2)$_S$. We see from Tab.~\ref{tab:bilayer-trilayer-multiplets} that in the presence of SU(2)$_S$ we expect one $S=1$ triplet and two $S=0$ singlets. With U(1)$_S$ symmetry, $S_z$ is a good quantum number, the triplet and the singlets will both contribute to the $S_z=0$ sector, so we expect three quasihole states there. Meanwhile, the $S_z=1$ receives contributions from the triplet, so we expect only 1 state. If we further added impurities at the poles of the sphere, breaking the SU(2)$_L$ rotation symmetry down to U(1)$_L$, we would have to go through a similar procedure and decompose $L$-multiplets into individual $L_z$ sectors. Note that throughout this process the total  dimension of the quasihole manifold is unchanged. Therefore, even if we break all of the symmetries in the system, we will still expect the same total number of quasihole states.

\section{Finite-size gap scaling for the foliated Laughlin phase}\label{app:finitesize_scaling_extra}
In Tab.~\ref{tab:gaps_flaughlin}, we report the values of the energy gap for the foliated Laughlin ground state at different system sizes, as a function of the number of particles in each layer $N$ and the number of layers $\nl$. The physically motivated way to extrapolate to the thermodynamic limit would be to first extrapolate the energy gaps at fixed $\nl$ in the $N\rightarrow \infty$ limit:
\begin{equation}
    \Delta(N,\nl) = \Delta_\infty(\nl) + \frac{a(\nl)}{N}
\end{equation}
and then extrapolate in the $\nl \rightarrow \infty$ limit:
\begin{equation}
    \Delta(\nl) = \Delta_\infty +  \frac{b^{(\Delta)}}{\nl}, \ a(\nl) = a_\infty + \frac{b^{(a)}}{\nl}
\end{equation}
Since we have to fit the systems with different $\nl$ separately, this method is not reliable in our case where we only have a few system sizes for the larger $\nl$ systems. We find $\Delta_\infty(\nl=2) = 0.596$ and $a(\nl=2) = 2.724$ for $\nl=2$ where we have 5 data points, and $\Delta_\infty(\nl=3) = 0.461$ and $a(\nl=3) = 2.921$ for $\nl=3$ where we have 4 data points. 

Due to the limited number of system sizes that are amenable to exact diagonalization, we also perform a simple linear fit
\begin{equation}
    \Delta = \Delta_\infty + \frac{a}{N} + \frac{b}{\nl}
\end{equation}
and obtain $\Delta_\infty = 0.366$, $a = 3.06$ and $b = 0.218$. We see that the gap is mainly controlled by $\frac{1}{N}$, so the dominant finite-size effects come from a small number of particles per layer. Note that $\Delta_\infty$, the energy gap that we extrapolate to the infinite system size limit $N,\nl \rightarrow \infty$, is finite, supporting the claim that the foliated Laughlin phase is a robust phase of matter.
\begin{table}[h]
\centering
\begin{tabular}{c|ccccc}
\hline\hline
$N_\ell$ & $N=2$ & $N=3$ & $N=4$ & $N=5$ & $N=6$ \\
\hline
$2$  & $1.97154028$ & $1.49201530$ & $1.23894471$ & $1.16732509$ & $1.06131984$ \\
$3$  & $1.94259992$ & $1.41219558$ & $1.12326927$ & $1.11406951$ & -- \\
$4$  & $1.94187616$ & $1.40544644$ & $1.08068397$ & -- & -- \\
$5$  & $1.94186121$ & $1.40521067$ & -- & -- & -- \\
$6$  & $1.94186056$ & $1.40515068$ & -- & -- & -- \\
$7$  & $1.94186054$ & -- & -- & -- & -- \\
$8$  & $1.94186054$ & -- & -- & -- & -- \\
$9$  & $1.94186054$ & -- & -- & -- & -- \\
$10$ & $1.94186054$ & -- & -- & -- & -- \\
$11$ & $1.94186054$ & -- & -- & -- & -- \\
$12$ & $1.94186054$ & -- & -- & -- & -- \\
$13$ & $1.94186054$ & -- & -- & -- & -- \\
\hline\hline
\end{tabular}
\caption{Energy gaps for the ground state at $V_0^{(1)}=0.6$ and $V_1^{(1)}=0.3$ as a function of the number of layers $N_\ell$ and particle number per layer $N$.} 
\label{tab:gaps_flaughlin}
\end{table}

\section{Non-Abelian Chern-Simons Ginzburg Landau theory}
\label{sec:NA_GL}
Infinite-component Chern-Simons theories with $\text{U(1)}$ gauge fields have been discussed in the layered three-dimensional quantum Hall systems \cite{Levin2009,Ma2022,wu2023two}. So far, the studies have focused on Abelian states. Here, we consider a non-Abelian version of such Chern-Simons theories. A natural question to ask is whether for infinite layered systems, there can be a transition from the decoupled Laughlin state to a non-Abelian state, as we have seen in the few-layer numerical studies in the main text. However, in contrast to the foliated Fibonacci state studied numerically in the main text which is accompanied by a layer charge density wave, in this section we consider a foliated Fibonacci state without this charge density modulation. To address this question, we consider an infinite-component Chern-Simons theory with non-Abelian gauge fields. For the filling $\nu = 1/3$ per layer of interest, we introduce an $\text{SU(3)}$ gauge field $a_z$ for the $z$-th layer. The gauge field in different layers are independent fields, resulting in an ``infinite-component" theory.

The $\text{SU(3)}$ gauge field naturally arises in the following parton construction \cite{Jain1989, blok1990}: $c_{i,z} = d^1_{i,z} d^2_{i,z} d^3_{i,z}$ with physical fermion $c_{i,z}$ at site $i$ of $z$-th layer and parton fermions $d^\alpha_{i,z}$ of three species. The parton fermions transform as an $\text{SU(3)}$ fundamental representation, which leaves the physical fermion untouched. At filling $1/3$, the partons of each layer experience $2\pi$ magnetic flux. Ignoring the layer coupling, the partons of each layer can form an integer quantum Hall state, which produces a Laughlin state per layer (foliated Laughlin state), corresponding to an $\text{SU(3)}_1$ Chern-Simons theory for each $a_z$-component. However, the interlayer interaction between the electrons can introduce an ``exciton" condensation of the parton fields, namely $\langle (d^\alpha_{i,z})^\dag  d^\beta_{i, z+1} \rangle \neq 0$, reducing the gauge symmetry via the Higgs mechanism. In the bilayer case, this induces the transition from the decoupled Laughlin state, described by $\text{SU(3)}_1 \times \text{SU(3)}_1$ CS theory, to a Fibonacci state, described by $\text{SU(3)}_2$ CS theory \cite{Vaezi2014}. In the infinite layer case, the Higgs condensation of the partons of neighboring layers can also break the discrete translation symmetry along the $z$-direction. In the extremal case, the interlayer interaction might even prefer a dimerization, effectively resulting in one Fibonacci state for two layers, which is the so-called ``foliated Fibonacci state". We note that this is a different foliated Fibonacci state to the one considered in the main text, since there is no layer charge modulation. 

To describe the foliated Laughlin and Fibonacci states, we introduce a bi-fundamental Higgs boson field $\Phi_{z, z+1}$ that couples to the two $\text{SU(3)}$ gauge fields in layer $z$ and $z+1$, and consider the following Chern-Simons Ginzburg Landau action,
\begin{equation}
    \begin{split}
        S[\Phi_{z,z+1}, a_z] = S_{\text{iCS}}[a_z] +  S_{\text{int}}[\Phi_{z,z+1},a_z] + S_{\text{ec}}[\Phi_{z,z+1}],
    \end{split}
\end{equation}
where
\begin{equation}
    S_{\text{iCS}}[a_z] = \frac{i}{4\pi}\sum_z \int \mathrm{d}^3 x\ \text{Tr}[a_z \wedge d a_z + \frac{2}{3} a_z\wedge a_z \wedge a_z]
\end{equation}
denotes the CS theory of $\text{SU(3)}_1$ for each layer, with $z \in \mathbb{Z}$ and the integration being over $x=(t,\mathbf{r})$, where $\mathbf{r}$ is a two-dimensional coordinate in the plane. The second term is 
\begin{equation}
    S_{\text{int}}[\Phi_{z,z+1},a_z] = \sum_{z} \int \mathrm{d}^3 x\ \text{Tr}[(D_\mu \Phi_{z,z+1})^\dag D_\mu \Phi_{z,z+1}],
    \label{eq:Sint}
\end{equation}
where
\begin{equation}
    D_\mu \Phi_{z,z+1} = \partial_\mu \Phi_{z,z+1} - i a_{z,\mu} \Phi_{z,z+1} + i \Phi_{z,z+1} a_{z+1,\mu}
\end{equation}
being the gauge covariant derivative that couples $\Phi_{z,z+1}$ to the two adjacent layers in an alternating fashion, with the sum over $\mu=0,1,2$ being implicit in Eq.~\eqref{eq:Sint}. The last term is
\begin{equation}
\begin{split}
S_{\text{ec}}[\Phi_{z,z+1}] =& \sum_{z} \int\mathrm{d}^3 x\ r  \text{Tr}[\Phi_{z,z+1}^\dag \Phi_{z,z+1}] \\
&+ u \left\{\text{Tr}\left[\Phi_{z,z+1}^\dag \Phi_{z,z+1}\right]\right\}^2 \\
&+ \kappa  \text{Tr}\left\{\left[\Phi_{z,z+1}^\dag \Phi_{z,z+1} - \frac{\text{Tr}[\Phi_{z,z+1}^\dag \Phi_{z,z+1} ]}{3}\right]^2\right\}\\
&+ 2 v~ \text{Tr}[\Phi_{z-1,z}^\dag \Phi_{z-1,z}]\text{Tr}[\Phi_{z,z+1}^\dag \Phi_{z,z+1}]
\end{split}
\end{equation}
and denotes the interaction between the $\Phi_{z,z+1}$ fields.

The fields $\Phi_{z,z+1}$ should be understood as the order parameter of the exciton condensate of the partons in the adjacent layers. When $r>0$, all $\Phi_{z,z+1}$ are gapped. Integrating out the $\Phi_{z,z+1}$ fields yields the foliated Laughlin state. When $r<0$,  $\Phi_{z,z+1}$ condenses. Here we let $ v >u >0$ and $\kappa > 0$. Due to the interaction between $\Phi_{z,z+1}$ and $\Phi_{z-1,z}$, there are two degenerate configurations: $\langle \Phi_{2m-1, 2m} \rangle = \phi_z \mathbf{1}$, $\langle \Phi_{2m,2m+1}\rangle = 0$ and vice versa, where $\phi_z$ is a complex number and $m$ is an integer. The condensate breaks the discrete translation symmetry $z \rightarrow z+1$. Furthermore, due to the Higgs mechanism, only the diagonal $\text{SU(3)}$ part of the $\text{SU(3)} \times \text{SU(3)}$ gauge fields of the neighboring layers remains. Hence, the effective action can be written as,
\begin{equation}
    \begin{split}
        S_{\text{eff,~Fib}}&[a_m] 
        = 2\frac{ i}{4\pi}\sum_m \int \mathrm{d}^3 x\ \text{Tr}[a_{2 m} \wedge d a_{2 m} + \frac{2}{3} a_{2 m}\wedge a_{2 m} \wedge a_{2 m}],
    \end{split}
\end{equation}
where we let $a_{2m-1} \sim a_{2m}$ for $\langle \Phi_{2m-1,2m} \rangle \neq 0 $. Similarly, we can let $a_{2m+1} \sim a_{2m}$ for the case when $\langle \Phi_{2m,2m+1} \rangle \neq 0 $. Due to the dimerization, one gauge field is shared between adjacent layers and the level of the $\text{SU(3)}$ CS theory is doubled to two, giving rise to the foliated Fibonacci state.  

Here we designed the phenomenological Ginzburg-Landau interaction of the $\Phi_{z,z+1}$ fields in the quest for proposing a foliated non-Abelian state, starting from layered Abelian state. The next important question is whether such a non-Abelian state can be realized in a microscopic Hamiltonian with realistic parameters. However, we have not found such non-Abelian state in the vicinity of the phase space of the layer symmetric, decoupled Laughlin state. The energetic stabilization of this foliated Fibonacci state without layer charge ordering remains an interesting open question.

\end{appendix}

\end{document}